\documentclass[iop]{emulateapj}
\usepackage[colorlinks=true,linkcolor=blue,anchorcolor=blue,citecolor=blue,urlcolor=blue]{hyperref}  % allcolors=blue
\hypersetup{pdfauthor={Ho Seong HWANG}, pdftitle={Local Analogs of DOGs}}
\newcommand{\iras}{\textit{IRAS} }
\newcommand{\akari}{\textit{AKARI} }
\newcommand{\wise}{\textit{WISE} }
\newcommand{\galex}{\textit{GALEX} }

\newcommand{\spitzer}{\textit{Spitzer} }

\slugcomment{Last updated: \today}
\shorttitle{Local Analogs of DOGs}
\shortauthors{Hwang and Geller}

\usepackage{natbib}

\begin{document}

\title{DUST-OBSCURED GALAXIES IN THE LOCAL UNIVERSE}

\author{Ho Seong Hwang and Margaret J. Geller}
\affil{Smithsonian Astrophysical Observatory, 60 Garden St., Cambridge, MA 02138}
\email{hhwang@cfa.harvard.edu, mgeller@cfa.harvard.edu}

%\author{Ho Seong Hwang\altaffilmark{1}}
%\author{Margaret J. Geller\altaffilmark{1}}
%\altaffiltext{1}{Smithsonian Astrophysical Observatory, 60 Garden St., Cambridge, MA 02138;
%hhwang@cfa.harvard.edu, mgeller@cfa.harvard.edu}
%\email{  }

\begin{abstract}
We use {\it Wide-field Infrared Survey Explorer} ({\it WISE}),
  {\it AKARI}, and {\it Galaxy Evolution Explorer} ({\it GALEX}) data
  to select local analogs of high-redshift ($z\sim2$) 
  dust obscured galaxies (DOGs).
We identify 47 local DOGs with
  $S_{12\mu m}/S_{0.22 \mu m}\geq892$
  and $S_{12\mu m}>20$ mJy at $0.05<z<0.08$ 
  in the Sloan Digital Sky Survey data release 7.
The infrared luminosities of these DOGs are in the range
 $3.4\times10^{10}  ({\rm L}_\odot) \lesssim L_{\rm IR} \lesssim 7.0\times10^{11}  ({\rm L}_\odot)$
 with a median $L_{\rm IR}$ of $2.1\times10^{11}$ (${\rm L}_\odot$).
We compare the physical properties of local DOGs
  with a control sample of galaxies
  that have lower $S_{12\mu m}/S_{0.22 \mu m}$
  but have similar redshift, IR luminosity, and stellar mass distributions. 
Both {\it WISE} 12 $\mu$m and {\it GALEX} near-ultraviolet (NUV) flux densities of DOGs
  differ from the control sample of galaxies,
  but the difference is much larger in the NUV. 
Among the 47 DOGs,
  $36\pm7$\% have small axis ratios in the optical (i.e., $b/a<0.6$),
  larger than the fraction among the control sample ($17\pm3$\%).
There is no obvious sign of interaction for many local DOGs.
No local DOGs have companions 
  with comparable optical magnitudes closer than $\sim$50 kpc.
The large- and small-scale environments of DOGs
  are similar to the control sample.
Many physical properties of local DOGs are similar to those of high-$z$ DOGs,
  even though the IR luminosities of local objects
  are an order of magnitude lower than for the high-$z$ objects:
  the presence of two classes (active galactic nuclei- and star formation-dominated) of DOGs,
  abnormal faintness in the UV rather than 
  extreme brightness in the mid-infrared, and
  diverse optical morphology.
These results suggest a common underlying physical origin of local and high-$z$ DOGs.
Both seem to represent the high-end tail of the dust obscuration distribution 
  resulting from various physical mechanisms
  rather than a unique phase of galaxy evolution.
\end{abstract}


\section{Introduction}

Understanding how the star formation activity of galaxies 
 evolves with cosmic time
 is one of the key issues in the study of galaxy formation and evolution.
Redshift $z\sim2$ is an interesting epoch
  because most of the stellar mass in galaxies today formed around this epoch
  \citep{dic03} and 
  because the cosmic star formation rate density also peaked (e.g., \citealt{beh12}).
Interestingly, most star formation at this epoch took place in dusty galaxies, 
  and the infrared luminosity from dust is higher 
  than the observed ultraviolet (UV) luminosity
  even in UV selected galaxies \citep{red08,red12}.
It is thus important to study $z\sim2$ dusty galaxies 
  to understand what drives the intense star formation at this epoch.

One way to identify high-$z$ dusty objects is
  to select optically faint, mid-infrared bright galaxies 
  (e.g., \citealt{yan04,dey08,fio08,lon09,saj12}).
A color selection of ($R-[24]$) $\geq 14$ (mag in Vega, or
  $S_{24\mu m}/S_{0.65 \mu m (R)}\geq982$)
  produces a sample of $z\sim2$ star-forming galaxies %high-$z$ galaxies
  with large dust obscuration: 
  dust-obscured galaxies (DOGs, \citealt{dey08,fio08,pen12,hwa12shels}).
As the name suggests,
  optical spectra reveal that
  DOGs suffer from severe extinction \citep{brand07,mel11}.
The amount of extinction inferred 
  from the Balmer decrement (i.e., H$\alpha$/H$\beta$)
  is $A$(H$\alpha$)$\sim2.4-4.6$,
  much larger than for other galaxies at similar redshift,
  but comparable with submillimeter galaxies \citep{tak06}
  or with extreme local ultraluminous infrared galaxies (ULIRGs; \citealt{vei09}).
  
The rest-frame near-infrared (NIR) spectral energy distributions (SEDs) 
  of these DOGs suggest that there are two types of DOGs: 
  ``bump'' and ``power-law'' DOGs \citep{dey08}.
Bump DOGs are generally fainter than power-law DOGs at 24 $\mu$m 
  (e.g., S$_{\rm 24 \mu m}<0.8$ mJy).
Their SEDs show a rest-frame 1.6 $\mu$m bump
 caused by a minimum in the opacity of the H$^-$ ion 
 present in the atmospheres of cool stars.
Their mid-infrared (MIR) spectra 
  show strong polycyclic aromatic hydrocarbon (PAH) emission \citep{pope08,desai09}, 
  a typical feature of ongoing star formation.
In contrast, the NIR/MIR SEDs of power-law DOGs
  show a rising continuum (i.e., power-law shape),
  and usually do not contain PAH emission \citep{hou05}.
The power-law continuum is usually attributed to a hot dust component,
  indicating the presence of active galactic nuclei (AGN).
The rest-frame UV/optical images reveal that
  bump DOGs are generally larger than power-law DOGs, and
  have more diffuse and irregular morphologies 
  \citep{mel09,don10,buss09hstp,buss11hstb}.
Clustering analysis of these DOGs suggests that
  more luminous ones at 24 $\mu$m 
  tend to reside in richer environments (i.e., strongly clustered)
  than less luminous ones \citep{bro08}.

Recent observational efforts have extended the study of DOGs
  to the far-infrared (FIR) and submillimeter regimes,
  providing a better view of their infrared properties 
  \citep{pope08,tyler09,buss09,mel12,pen12}.
For example,
  \citet{pen12} found that the specific star formation rates 
  (sSFR, i.e., SFR per unit stellar mass)
  of DOGs are similar to the majority of 
  typical star-forming galaxies at similar redshift
  (but see also \citealt{mel12}).
They also suggest that
  the extreme ratios between rest-frame MIR and UV flux densities
  mainly result from abnormal faintness in the UV
  rather than abnormal brightness in the MIR.

Physical models 
  for the formation and evolution of DOGs
  account for several observational features
  including IR luminosity, dust temperature, and stellar mass 
  (e.g., \citealt{nar10}).
Using numerical simulations,  Narayanan et al.
  conclude that DOGs are a diverse population
  ranging from secularly evolving star-forming disk galaxies
  to extreme gas-rich galaxy mergers.
They further suggest that
  some DOGs seem to be a transition phase 
  in the evolutionary sequence of galaxy mergers;
  the sequence progresses from submillimeter galaxies to DOGs 
  to quasars to elliptical galaxies. 
However, because of their extreme distances, 
  it is difficult to compare detailed observational features
  with model predictions.

In this study, 
  we search for local analogs of high-$z$ DOGs.
We study their physical properties in detail
  by taking advantage of their proximity and 
  of the wealth of multiwavelength data.
The local DOGs are useful testbeds 
  for studying the evolutionary significance, and for   
  providing an important hint of the nature of their high-$z$ counterparts.  
We also construct a control galaxy sample
  with distributions of physical parameters
  (e.g., redshift, IR luminosity, stellar mass)
  similar to local DOGs,
  but not satisfying the DOG criterion.
      
Section \ref{data} describes the observational data and
  the method we use to identify both local analogs of DOGs and 
  the control sample of galaxies.
We compute the IR luminosities and the AGN contribution
  of local DOGs in Section \ref{sed}.
We compare several physical parameters of local DOGs
  with the control sample
  and with high-$z$ DOGs in Sections \ref{comp} and \ref{highz}, respectively.
We discuss the results and conclude in Sections
 \ref{discuss} and \ref{sum}, respectively.
Throughout,
  we adopt flat $\Lambda$CDM cosmological parameters:
  $H_0 = 70$ km s$^{-1}$ Mpc$^{-1}$, 
  $\Omega_{\Lambda}=0.7$ and $\Omega_{m}=0.3$.
All magnitudes are on the AB magnitude system. 
SDSS $ugriz$ data are Petrosian magnitudes unless otherwise noted.
When we convert SDSS magnitudes
  into AB magnitudes for the SED fit,
  we adopt the offset
  used in Kcorrect (v4.2) software of \citet{bla07kcorr}:
  $\Delta$m =  m$_{\rm AB}$ $-$ m$_{\rm SDSS}$ 
  = $-0.036$, 0.012, 0.010, 0.028, 0.040
  for $ugriz$ bands.
  
\section{Sample Selection}\label{data}

\subsection{Multiwavelength Data}

We use a spectroscopic sample of galaxies and quasars
   in the Sloan Digital Sky Survey (SDSS; \citealt{york00})
   data release 9 (DR9, \citealt{ahn12}).   
The spectroscopic completeness of the SDSS data is
  poor for bright galaxies with $m_r<14.5$ and
  for galaxies in high-density regions.
Thus, we supplement the galaxy data
  by compiling redshifts
  for the photometric sample of galaxies with $m_r<17.77$
  from the literature (see \citealt{hwa10lirg} for details).

For this SDSS sample,
  we compile the available multiwavelength photometric data 
  from far-UV (FUV) to FIR.
We first add the {\it Galaxy Evolution Explorer} ({\it GALEX}; \citealt{mar05}) 
  UV data from the {\it GALEX} general release 6
 (GR6\footnote{http://galex.stsci.edu/GR6}) that provides
  the cross-matched table (\textbf{xSDSSDR7}) against SDSS 
  DR7\footnote{Because we use the \galex GR6 catalog based on SDSS DR7,
  the SDSS sample in this study is actually equivalent to DR7 rather than DR9,
  even though all the photometric parameters are adopted from DR9
  (see \citealt{ahn12} for details about the difference between DR9 and DR7).}.
The matching tolerance is $5\arcsec$ ($\sim$FWHM of the \galex PSF).
To avoid contamination by nearby sources within the matching tolerance,
  we select only unique matches;
  for a given SDSS object, 
  we choose the {\it GALEX} object closest to the SDSS object
  and vice versa.
  
The  {\it GALEX} GR6 contains several imaging survey databases including 
  the Nearby Galaxy Survey (NGS), Deep (DIS), Medium (MIS), and All Sky Surveys (AIS).
We use all the sources covered by these surveys.
The typical exposure time for the shallowest survey (i.e., AIS) is $\sim$100 s
  corresponding to 5$\sigma$ limiting magnitudes of 
  $m_{\rm FUV}\sim$20 and  $m_{\rm NUV}\sim$20.8. 
We correct for Galactic extinction following \citet{wyder07}.
Apertures for {\it GALEX} photometry 
  are not matched to the bands in other wavelengths, 
  but they are large enough to contain most of galaxy light.
Thus aperture correction is unnecessary \citep{ree07}.

We adopt the NIR data from the extended source catalog \citep{jar00}
  of the Two Micron All Sky Survey (2MASS; \citealt{skr06}).
The matching tolerance is $1.5\arcsec$ \citep{obr06}.
We use $JHK_s$ 20 mag arcsec$^{-2}$ isophotal 
  fiducial elliptical aperture magnitudes (AB).

We include the MIR data from the all-sky survey 
  catalog\footnote{http://wise2.ipac.caltech.edu/docs/release/allsky/expsup/}
  of the {\it Wide-field Infrared Survey Explorer} ({\it WISE}; \citealt{wri10}),
  containing uniform photometric data for over 563 million objects at 4 MIR bands 
  (3.4, 4.6, 12 and 22 $\mu$m).
We identify \wise counterparts of SDSS objects
  with a matching tolerance of 
  3\arcsec~($\sim$ 0.5$\times$FWHM of the \wise PSF at 3.4 $\mu$m).
Because of the high number density of \wise sources, %HH
  we use 0.5$\times$FWHM as a matching tolerance 
  rather than 1$\times$FWHM;
  the expected false detection rate with 3\arcsec~tolerance 
  is only 0.05\% \citep{don12}.
We use the point source profile-fitting magnitudes.
\wise 5$\sigma$ photometric sensitivity is estimated to be better 
  than 0.08, 0.11, 1 and 6 mJy
  at 3.4, 4.6, 12 and 22 $\mu$m
  in unconfused regions on the ecliptic plane \citep{wri10}.

We also add 9 and 18 $\mu$m flux density measurements
  from the all-sky survey Point Source Catalog (PSC ver. 1.0, \citealt{ish10})
  of the {\it AKARI} telescope \citep{mur07}.
The 5$\sigma$ detection limits at 9 and 18 $\mu$m
  are 50 and 90 mJy, respectively.
      
We obtain the FIR data
  by cross-correlating the SDSS sample with
  the {\it Infrared Astronomical Satellite} ({\it IRAS}; \citealt{neu84})
  Faint Sources Catalog ver. 2 (\citealt{mos92}; with 12, 25, 60 and 100 $\mu$m bands)
  and the {\it AKARI}/Far-Infrared Surveyor (FIS; \citealt{kaw07})
  all-sky survey Bright Source Catalog
  (BSC)\footnote{http://www.ir.isas.jaxa.jp/AKARI/Observation/PSC/Public/RN/AKARI-FIS$\_$BSC$\_$V1$\_$RN.pdf}
   ver. 1.0 (with 65, 90, 140 and 160 $\mu$m bands).

\subsection{{\it AKARI} 9 $\mu$m Selected Local DOGs}

To identify local analogs of DOGs,
  we first use {\it GALEX} NUV ($\sim$0.22 $\mu$m) and {\it AKARI} 9 $\mu$m data
  for the SDSS galaxies at $z<0.1$.
These data are roughly equivalent to
  the $R$-band (0.65 $\mu$m) and \spitzer 24 $\mu$m data 
  originally used for selecting $z\sim2$ DOGs \citep{dey08}.
  
\begin{figure}
\center
\includegraphics[width=85mm]{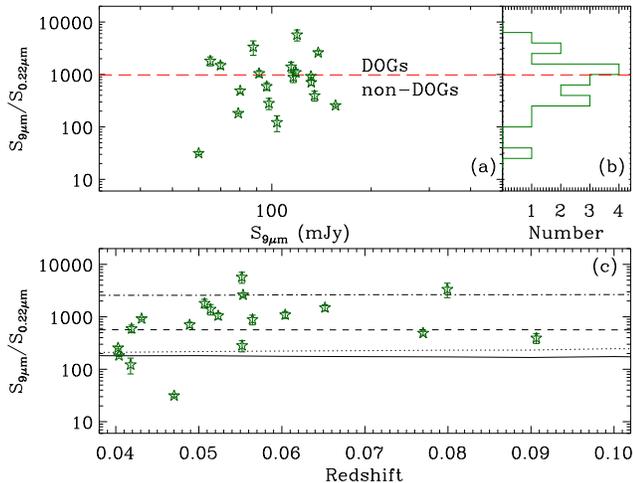}
\caption{Flux density ratio between \akari 9 $\mu$m and \galex NUV
  observations of SDSS galaxies
  as a function of 9 $\mu$m flux density (a), and its histogram (b).
The flux density ratio as a function of redshift (c).
The long dashed line in (a) indicates the selection criterion 
  for local DOGs ($S_{9\mu m}/S_{0.22 \mu m}=982$).
In the bottom panel, 
  we overplot the expected ratios from several SED templates for 
  local star-forming, or AGN-host galaxies 
  \citep[M82: solid, Arp220: dotted,
  Mrk 231: dashed, IRAS 19254-7245 South: dash dot]{pol07}.
}\label{fig-samp9um}
\end{figure}

We plot the flux density ratio between the {\it AKARI} 9 $\mu$m and {\it GALEX} NUV 
  as a function of 9  $\mu$m flux density in the top panel of 
  Figure \ref{fig-samp9um}.
We restrict our analysis to galaxies with $0.04< z<0.1$.
The lower limit is set by the small, fixed-size (3\arcsec)
  aperture for the SDSS spectra covering only a small ($<20\%$) portion 
  of the entire galaxy light at $z<0.04$ \citep{kew05}.
The upper limit selects bright galaxies
  so that visual inspection of galaxy morphology in the SDSS image 
  is reliable.

%Figure 1 %%%%%%%%%%%%%%%%%%%%
%%%%%%%%%%%%%%%%

We show the \citeauthor{dey08} criterion 
  (i.e., rest-frame MIR and UV flux density ratio $\geq 982$)
  as a long dashed line in the top panels of Figure \ref{fig-samp9um}.
Because of the difference in bandpasses between \akari 9 $\mu$m
  and {\it Spitzer} 24 $\mu$m observation at $z\sim2$ (i.e., 8 $\mu$m),
  the \akari 9 $\mu$m flux density for local DOG selection
  could be smaller by $\sim10\%$ 
  (i.e., $S_{9\mu m}/S_{0.22 \mu m}\gtrsim0.9\times982$)
  if we assume the M82 SED at $z\sim0.07$.
However, we retain the $S_{9\mu m}/S_{0.22 \mu m}\geq 982$
  as our selection criterion for local DOGs.
Among the 19 galaxies in the figure, eight satisfy the DOG criterion.

We also show the redshift dependence of the flux density ratio between 
   {\it AKARI} 9 $\mu$m and {\it GALEX} NUV in the bottom panel of 
   Figure \ref{fig-samp9um}.
We overplot the flux density ratios expected from several SED templates of 
  local star-forming, or AGN-host galaxies
  \citep[M82: solid, Arp220: dotted, Mrk 231: dashed, 
  {\it IRAS} 19254-7245 South: dash dot]{pol07}.
Interestingly, 
  none of these templates except IRAS 19254-7245 South\footnote{
  {\it IRAS} 19254-7245 South satisfies
  the DOG criterion, but the relevant UV SED of this galaxy is not well constrained 
  (see \citealt{berta03}).}
  satisfy the DOG criterion (see also Figure 1 in \citealt{dey08}).

%Figure 2 %%%%%%%%%%%%%%%%%%%%
\begin{figure*}
\center
\includegraphics[width=135mm]{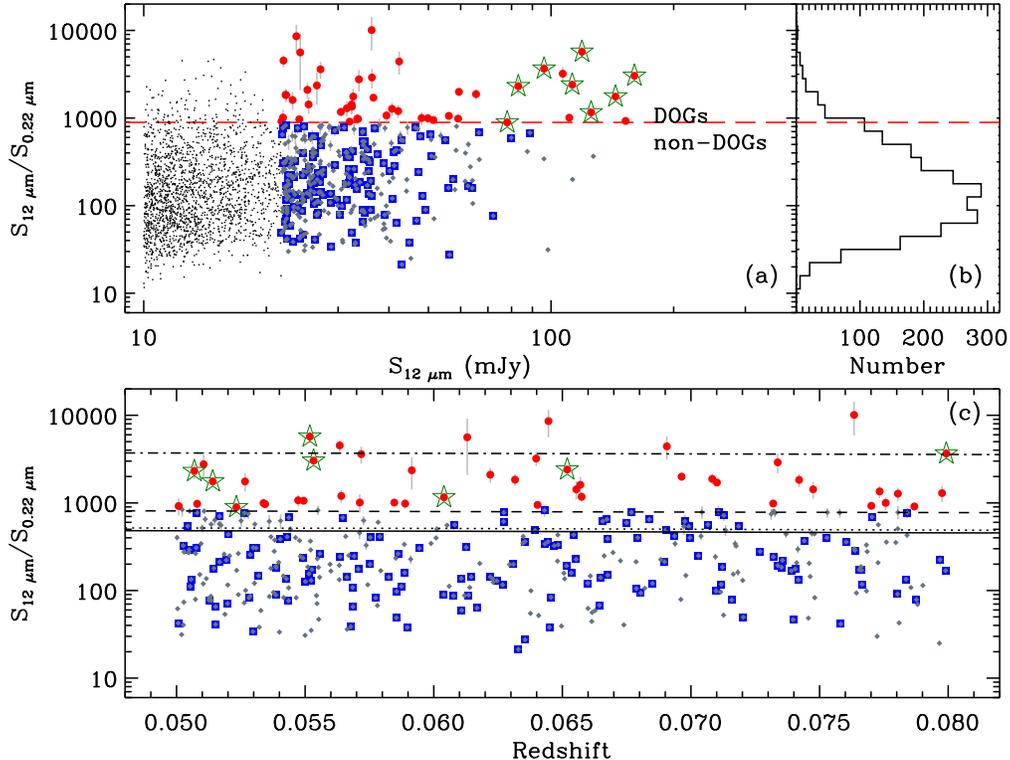}
\caption{Flux density ratio between \wise 12 $\mu$m and \galex NUV
  observations for SDSS galaxies (black dots)
  as a function of 12 $\mu$m flux density (a), and its histogram (b).
Flux density ratio as a function of redshift (c).
Green star symbols are {\it AKARI} 9 $\mu$m selected local DOGs.
Red filled circles are {\it WISE} 12 $\mu$m selected local DOGs, and
  gray diamonds are the preliminary control sample of galaxies.
Blue squares are the final control sample of galaxies
 determined in Figure \ref{fig-lirmass}.
We plot error bars only for local DOGs.
The long dashed line in (a) indicates the selection criterion 
  for these local DOGs ($S_{12\mu m}/S_{0.22 \mu m}=892$).
In the bottom panel, 
  we overplot the expected ratios from several SED templates for 
  local star-forming, or AGN-host galaxies 
  \citep[M82: solid, Arp220: dotted,
  Mrk 231: dashed, IRAS 19254-7245 South: dash dot]{pol07}.
}\label{fig-samp12um}
\end{figure*}
%%%%%%%%%%%%%%%%

%Figure 3 %%%%%%%%%%%%%%%%%%%%
\begin{figure}
\center
\includegraphics[width=80mm]{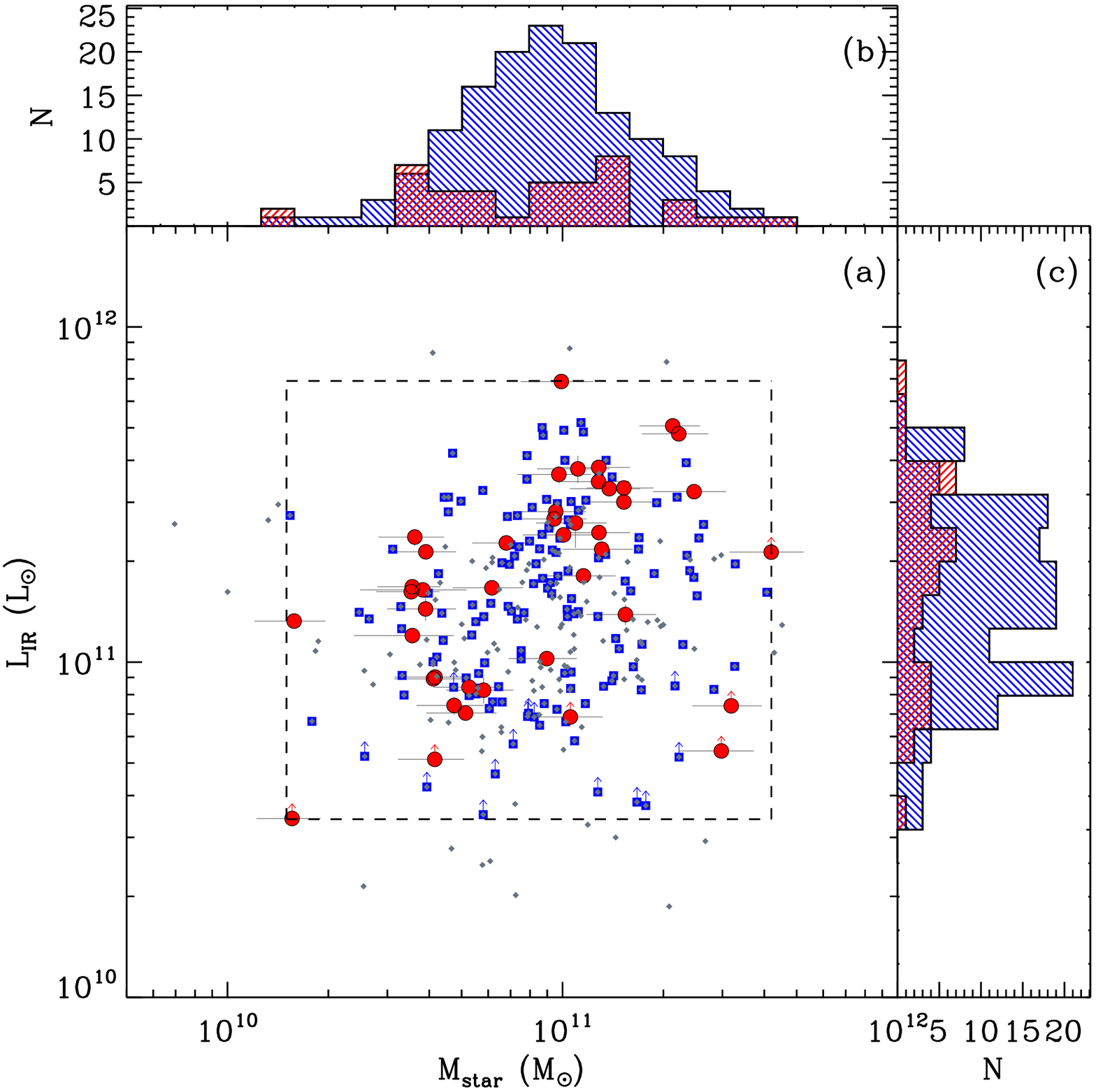}
\caption{IR luminosity ($L_{\rm IR}$) vs. stellar mass ($M_{\rm star}$)
  for {\it WISE} 12 $\mu$m selected DOGs (red filled circles),
   the preliminary control sample of galaxies (gray diamonds) and
   the final control sample of galaxies (blue squares) (a).
Dashed lines indicate the range of IR luminosity and stellar mass
  for {\it WISE} 12 $\mu$m selected DOGs. 
Arrows are lower limits to the IR luminosities (see Section \ref{sedfit}).
We plot error bars only for local DOGs.
The histograms for IR luminosity (b) and stellar mass (c).
DOGs and the final control sample
  are denoted by hatched histograms with
  orientation of 45$^\circ$ ($//$ with red color) and 
  of 315$^\circ$ ($\setminus\setminus$ with blue color) 
  relative to horizontal, respectively.
}\label{fig-lirmass}
\end{figure}
%%%%%%%%%%%%%%%%

\subsection{{\it WISE} 12 $\mu$m Selected Local DOGs}
\subsubsection{Selection Criteria}\label{12um}

To identify additional local DOGs missed by the shallow {\it AKARI} MIR all-sky survey,
  we use {\it WISE} 12 $\mu$m data instead of  {\it AKARI} 9 $\mu$m data.
We show the flux density ratio between 
  {\it WISE} 12 $\mu$m and {\it GALEX} NUV data
  as a function of the 12 $\mu$m flux density 
  in the top panel of Figure \ref{fig-samp12um}.
To determine an optimal parameter space for selecting local DOGs,
  we first plot the \akari 9 $\mu$m detected DOGs as green star symbols.
All eight objects are detected at 12 $\mu$m.
The minimum flux density ratio for these objects is 
  $S_{12\mu m}/S_{0.22 \mu m}\sim$892, and
  they are in the redshift range $0.05<z<0.08$ 
  (bottom panel\footnote{There are three {\it AKARI} 9 $\mu$m and 
  85 {\it WISE} 12 $\mu$m selected galaxies
  at $z>0.1$ that satisfy our selection criteria.
  Because of their broad redshift range (0.1$\lesssim z\lesssim$2.5), 
  the observational data probe different wavelengths.
  Therefore, we do not include them in the analysis.}).

We then select \wise 12 $\mu$m detected local DOGs
  that share the same parameter space
  as the \akari 9 $\mu$m selected DOGs 
  (i.e., $S_{12\mu m}/S_{0.22 \mu m}\geq892$ and $0.05<z<0.08$).
Among these, we consider only 47 bright 12 $\mu$m sources with 
  $S_{12 \mu m}>20$ mJy for further analysis (red filled circles).
Including the galaxies fainter than $S_{12 \mu m}=20$ mJy does not change
  our conclusions, but for comparison with high-$z$ DOGs
  we specialize to the more luminous local analogs.

\subsubsection{The Control Sample of Galaxies}\label{compsamp}

To compare the physical properties of these extremely dusty galaxies with
  other local galaxies, we construct a control sample.
We first select a preliminary control sample of galaxies 
  that are in the same range of redshift ($0.05<z<0.08$) and 
  12 $\mu$m flux density ($S_{12\mu m}\geq21.79$ 
  mJy\footnote{We select the local DOGs with $S_{12 \mu m}>20$ mJy,
  but the minimum $S_{12 \mu m}$ for the local DOGs is actually 21.79 mJy. 
  Therefore, we use $S_{12\mu m}\geq21.79$ mJy to construct 
  a control sample of galaxies.})
  as the \wise 12 $\mu$m detected local DOGs.
These control objects do not satisfy the DOG criterion 
  (i.e., $S_{12\mu m}/S_{0.22 \mu m}<892$, gray diamonds
  in Figure \ref{fig-samp12um}).

%Figure 4 %%%%%%%%%%%%%%%%%%%%
\begin{figure}
\center
\includegraphics[width=80mm]{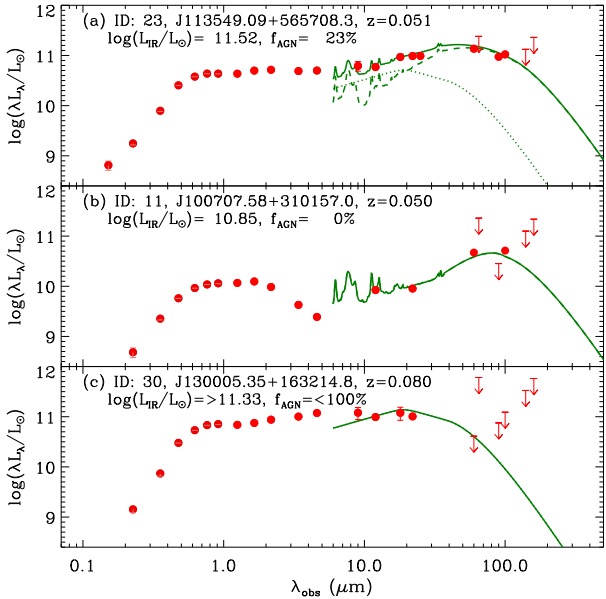}
\caption{Example SEDs of FIR detected DOGs
  with some AGN contribution (a) and
  without an AGN contribution (b), and
  of an FIR undetected DOG (c).
Filled circles are observed photometric data, and
  down arrows are upper limits.
The detection limits 
  for {\it IRAS} and {\it AKARI}
  can differ depending on the sky position.
Therefore, the down arrows shown in these panels
  indicate just averaged upper limits.
Solid, dotted, and dashed lines indicate the best-fit SEDs 
  with the DECOMPIR routine of \citet{mul11agn},
  for total, host-galaxy, and AGN components, respectively.
}\label{fig-fitsed}
\end{figure}
%%%%%%%%%%%%%%%%

%%%%%%%%%%%%%%%%%%%%%%%%%%%%%%%%%%%%%%%%%%%%%%%%%%%%%%%%%%%%%%%%%%%%%%%%%%%%%%%
%%%%%%%%%%%%%%%%%%%%%%%%         table 1       %%%%%%%%%%%%%%%%%%%%%%%%%%%%%%%%
%%%%%%%%%%%%%%%%%%%%%%%%%%%%%%%%%%%%%%%%%%%%%%%%%%%%%%%%%%%%%%%%%%%%%%%%%%%%%%%
\begin{deluxetable*}{cccccrrrc}
\tabletypesize{\tiny}
\tablewidth{0pc} %\tablenum{1}
\tablecaption{Properties of {\it WISE} 12 $\mu$m Selected DOGs
\label{tab-samp}}
\tablehead{
ID & SDSS ObjID (DR9) & R.A.$_{2000}$ & Decl.$_{2000}$ & z & M$_{\rm star}$\tablenotemark{a} & L$_{\rm IR}$ & f$_{\rm AGN}$  & CLASS\tablenotemark{b} \\
   &                  &               &                &   & ($\times10^{10}$M$_\odot$)        & ($\times10^{10}$L$_\odot$) &  (\%)          & 
}
\startdata
  1 &    1237663783124992102 & 00:12:26.83 & $-$00:48:19.45 & 0.0734 & 11.11$\pm$ 2.71 &     37.72$\pm$3.54 &   0$\pm$0 &  C \\
  2 &    1237652900763861074 & 02:25:33.60 & $-$08:25:17.80 & 0.0549 &             ... &     22.18$\pm$0.25 &   0$\pm$0 &  h \\
  3 &    1237663789024739682 & 07:25:28.47 & $+$43:43:32.39 & 0.0691 & 10.55$\pm$ 2.63 &            $>$6.86 &    $<$100 &  S \\
  4 &    1237653587407536295 & 08:03:37.33 & $+$39:26:33.13 & 0.0655 & 29.78$\pm$ 7.45 &            $>$5.43 &    $<$100 &  S \\
  5 &    1237654382512111741 & 08:42:11.60 & $+$48:18:36.31 & 0.0563 &  1.58$\pm$ 0.38 &     13.26$\pm$0.16 &   0$\pm$0 &  U \\
  6 &    1237657118948393018 & 08:42:15.29 & $+$40:25:33.29 & 0.0553 &  3.83$\pm$ 1.34 &     16.43$\pm$0.18 & 100$\pm$0 &  S \\
  7 &    1237674462024106294 & 09:04:01.02 & $+$01:27:29.12 & 0.0534 & 12.82$\pm$ 3.08 &     38.10$\pm$0.49 &   0$\pm$0 &  C \\
  8 &    1237674460413690092 & 09:07:46.91 & $+$00:34:30.55 & 0.0534 &  4.16$\pm$ 1.01 &      9.02$\pm$0.12 &   0$\pm$0 &  S \\
  9 &    1237663530802937978 & 09:38:19.17 & $+$64:37:21.26 & 0.0710 & 21.32$\pm$ 4.37 &     50.69$\pm$0.65 &   0$\pm$0 &  C \\
 10 &    1237661064416591910 & 10:01:40.47 & $+$09:54:31.54 & 0.0564 &  4.74$\pm$ 1.08 &      7.43$\pm$0.46 &  51$\pm$6 &  S \\
 11 &    1237665099004248259 & 10:07:07.58 & $+$31:01:57.00 & 0.0501 &  5.14$\pm$ 1.21 &      7.05$\pm$0.09 &   0$\pm$0 &  H \\
 12 &    1237667254540042377 & 10:09:45.63 & $+$26:11:50.54 & 0.0747 &  4.15$\pm$ 0.93 &            $>$5.13 &    $<$100 &  S \\
 13 &    1237661383848951984 & 10:11:01.09 & $+$38:15:19.74 & 0.0527 &  3.56$\pm$ 0.76 &     16.78$\pm$0.22 &   0$\pm$0 &  C \\
 14 &    1237654605860110514 & 10:17:31.29 & $+$04:36:19.04 & 0.0572 & 11.55$\pm$ 2.84 &     18.09$\pm$0.83 &   4$\pm$1 &  H \\
 15 &    1237671262281662602 & 10:17:53.90 & $+$15:56:03.73 & 0.0787 & 15.25$\pm$ 3.51 &     30.04$\pm$1.07 &  13$\pm$2 &  H \\
 16 &    1237648722831868077 & 10:33:33.15 & $+$01:06:35.15 & 0.0657 &  9.44$\pm$ 2.33 &     26.75$\pm$1.88 &   6$\pm$2 &  H \\
 17 &    1237664337173610611 & 10:44:45.46 & $+$34:15:10.58 & 0.0708 & 13.09$\pm$ 3.09 &     21.72$\pm$1.22 &  38$\pm$4 &  S \\
 18 &    1237665128536408188 & 10:56:53.37 & $+$33:19:45.68 & 0.0511 &  3.90$\pm$ 0.91 &     14.41$\pm$1.11 &   0$\pm$0 &  S \\
 19 &    1237651067886502124 & 11:02:13.01 & $+$64:59:24.86 & 0.0776 & 24.70$\pm$ 6.07 &     32.26$\pm$1.09 &  10$\pm$2 &  S \\
 20 &    1237657856606404709 & 11:08:52.62 & $+$51:02:25.71 & 0.0696 & 12.84$\pm$ 3.07 &     24.35$\pm$0.94 &  23$\pm$3 &  S \\
 21 &    1237651754005889185 & 11:19:34.25 & $+$02:31:29.19 & 0.0508 &  5.81$\pm$ 1.31 &      8.25$\pm$0.73 &  26$\pm$6 &  S \\
 22 &    1237671141477777656 & 11:29:56.35 & $-$06:24:20.48 & 0.0523 &             ... &     42.92$\pm$5.90 &   4$\pm$2 &  h \\
 23 &    1237657611801657347 & 11:35:49.09 & $+$56:57:08.27 & 0.0514 & 15.25$\pm$ 3.42 &     33.09$\pm$0.96 &  24$\pm$2 &  S \\
 24 &    1237657856610271340 & 12:04:53.90 & $+$52:23:43.39 & 0.0632 &  6.15$\pm$ 1.46 &     16.66$\pm$0.21 &   0$\pm$0 &  C \\
 25 &    1237664292075143390 & 12:12:43.86 & $+$16:11:06.09 & 0.0571 &  5.26$\pm$ 1.14 &      8.42$\pm$0.68 &  11$\pm$2 &  C \\
 26 &    1237667209992732748 & 12:21:34.35 & $+$28:49:00.12 & 0.0613 & 10.93$\pm$ 2.59 &     26.02$\pm$4.07 &   0$\pm$0 &  C \\
 27 &    1237667736660017246 & 12:56:25.47 & $+$23:20:55.05 & 0.0742 &  9.75$\pm$ 2.43 &     36.32$\pm$2.12 &   0$\pm$0 &  C \\
 28 &    1237665129084092587 & 12:56:42.72 & $+$35:07:29.92 & 0.0547 &  3.91$\pm$ 0.90 &     21.34$\pm$0.24 &   0$\pm$0 &  L \\
 29 &    1237648704581795973 & 12:56:45.02 & $+$00:11:17.61 & 0.0622 &             ... &     10.06$\pm$1.85 & 10$\pm$10 &  s \\
 30 &    1237668589190381691 & 13:00:05.35 & $+$16:32:14.80 & 0.0799 & 41.95$\pm$10.45 &           $>$21.29 &    $<$100 &  S \\
 31 &    1237667447809114245 & 13:15:17.85 & $+$24:38:08.97 & 0.0645 &  1.56$\pm$ 0.34 &            $>$3.42 &    $<$100 &  U \\
 32 &    1237665531171373168 & 13:39:26.08 & $+$25:35:19.80 & 0.0763 &  4.12$\pm$ 0.97 &      8.92$\pm$0.13 & 100$\pm$0 &  S \\
 33 &    1237665430241149030 & 13:41:02.95 & $+$29:36:42.86 & 0.0773 & 12.81$\pm$ 2.97 &     34.56$\pm$0.50 &   0$\pm$0 &  C \\
 34 &    1237665549424132213 & 13:56:03.30 & $+$25:31:12.82 & 0.0591 &  3.56$\pm$ 1.18 &     12.00$\pm$0.17 &   0$\pm$0 &  C \\
 35 &    1237665430243704865 & 14:07:00.39 & $+$28:27:14.67 & 0.0770 & 22.23$\pm$ 4.99 &     47.98$\pm$1.63 &  67$\pm$4 &  C \\
 36 &    1237661212583002226 & 14:48:19.38 & $+$44:32:32.76 & 0.0798 & 31.88$\pm$ 7.47 &            $>$7.41 &    $<$100 &  S \\
 37 &    1237662236409462990 & 14:54:27.46 & $+$06:47:19.63 & 0.0657 &  8.97$\pm$ 2.06 &     10.24$\pm$0.42 &   9$\pm$2 &  H \\
 38 &    1237665350246924437 & 14:58:26.55 & $+$24:58:15.46 & 0.0640 & 15.40$\pm$ 3.62 &     13.86$\pm$0.98 &  38$\pm$7 &  S \\
 39 &    1237648705135051235 & 15:26:37.67 & $+$00:35:33.50 & 0.0507 & 10.07$\pm$ 2.47 &     23.97$\pm$0.61 &   8$\pm$1 &  S \\
 40 &    1237662305125204020 & 15:50:01.60 & $+$27:49:00.49 & 0.0780 &  6.79$\pm$ 1.42 &     22.68$\pm$1.00 &  25$\pm$3 &  S \\
 41 &    1237662663216070833 & 15:51:53.04 & $+$27:14:33.65 & 0.0589 &  9.54$\pm$ 2.20 &     28.12$\pm$0.96 &   1$\pm$1 &  H \\
 42 &    1237662262718235201 & 16:09:48.22 & $+$04:34:52.90 & 0.0640 &  3.54$\pm$ 0.76 &     16.22$\pm$0.45 & 100$\pm$4 &  S \\
 43 &    1237662336261816691 & 16:51:21.88 & $+$21:55:26.22 & 0.0552 &  3.62$\pm$ 0.80 &     23.61$\pm$1.36 &  41$\pm$3 &  S \\
 44 &    1237661387621073252 & 16:53:37.16 & $+$30:26:09.76 & 0.0732 &             ... &     44.16$\pm$0.47 &   0$\pm$0 &  U \\
 45 &    1237668681527132368 & 17:03:30.38 & $+$45:40:47.15 & 0.0604 &             ... &     33.82$\pm$1.20 &  34$\pm$2 &  s \\
 46 &    1237656530531254308 & 17:38:01.52 & $+$56:13:25.81 & 0.0652 & 13.78$\pm$ 3.27 &     32.93$\pm$1.11 &  40$\pm$2 &  S \\
 47 &    1237663478726328339 & 22:53:32.98 & $-$00:24:42.75 & 0.0585 &  9.92$\pm$ 2.41 &     68.74$\pm$0.72 &   0$\pm$0 &  C \\
\enddata
\tablenotetext{1}{Stellar masses for the galaxies without SDSS spectra are not available.}
\tablenotetext{2}{Galaxy classification based on optical line ratios: H (SF), C (Composite), S (Seyfert), L (LINER), and U (Undetermined). The lower case letter gives the classification adopted from NED.}
\end{deluxetable*}

%%%%%%%%%%%%%%%%%%%%%%%%%%%%%%%%%%%%%%%%%%%%%%%%%%%%%%%%%%%%%%%%%%%%%%%%%%%%%%%
%%%%%%%%%%%%%%%%%%%%%%%%         table 2       %%%%%%%%%%%%%%%%%%%%%%%%%%%%%%%%
%%%%%%%%%%%%%%%%%%%%%%%%%%%%%%%%%%%%%%%%%%%%%%%%%%%%%%%%%%%%%%%%%%%%%%%%%%%%%%%
\begin{deluxetable*}{cccccccc}
\tabletypesize{\tiny}
\tablewidth{0pc}
\tablecaption{UV and Optical Photometry for {\it WISE} 12 $\mu$m Selected DOGs\tablenotemark{a}
\label{tab-uvopt}}
\tablehead{
ID & FUV   & NUV   & $u$   &  $g$  &  $r$  &  $i$  &  $z$  \\
   & (mag) & (mag) & (mag) & (mag) & (mag) & (mag) & (mag)
}
\startdata
  1 & $            ...$ & $  21.16\pm 0.27$ & $  18.61\pm 0.06$ & $  17.22\pm 0.01$ & $  16.48\pm 0.01$ & $  15.93\pm 0.01$ & $  15.73\pm 0.01$\\
  2 & $  21.60\pm 0.20$ & $  19.59\pm 0.06$ & $  17.00\pm 0.04$ & $  15.56\pm 0.02$ & $  14.80\pm 0.01$ & $  14.37\pm 0.02$ & $  14.14\pm 0.01$\\
  3 & $  21.69\pm 0.47$ & $  21.45\pm 0.32$ & $  18.70\pm 0.08$ & $  17.35\pm 0.03$ & $  16.52\pm 0.03$ & $  16.05\pm 0.02$ & $  15.75\pm 0.02$\\
  4 & $            ...$ & $  20.51\pm 0.26$ & $  17.39\pm 0.05$ & $  15.81\pm 0.02$ & $  15.02\pm 0.02$ & $  14.60\pm 0.01$ & $  14.34\pm 0.02$\\
  5 & $            ...$ & $  22.18\pm 0.17$ & $  19.07\pm 0.08$ & $  17.58\pm 0.01$ & $  16.89\pm 0.01$ & $  16.56\pm 0.01$ & $  16.35\pm 0.03$\\
  6 & $            ...$ & $  19.59\pm 0.04$ & $  18.11\pm 0.03$ & $  16.67\pm 0.01$ & $  16.14\pm 0.02$ & $  15.85\pm 0.05$ & $  15.64\pm 0.08$\\
  7 & $  20.97\pm 0.34$ & $  19.65\pm 0.14$ & $  18.08\pm 0.06$ & $  16.60\pm 0.02$ & $  15.90\pm 0.01$ & $  15.43\pm 0.01$ & $  15.15\pm 0.02$\\
  8 & $  20.05\pm 0.05$ & $  20.41\pm 0.05$ & $  18.69\pm 0.07$ & $  17.14\pm 0.01$ & $  16.48\pm 0.01$ & $  16.11\pm 0.01$ & $  15.86\pm 0.02$\\
  9 & $            ...$ & $  20.57\pm 0.09$ & $  18.07\pm 0.07$ & $  16.33\pm 0.01$ & $  15.52\pm 0.01$ & $  15.08\pm 0.01$ & $  14.76\pm 0.01$\\
 10 & $  21.27\pm 0.45$ & $  20.04\pm 0.18$ & $  18.03\pm 0.04$ & $  16.51\pm 0.02$ & $  15.79\pm 0.02$ & $  15.39\pm 0.03$ & $  15.18\pm 0.02$\\
 11 & $            ...$ & $  20.47\pm 0.24$ & $  18.30\pm 0.07$ & $  16.97\pm 0.01$ & $  16.17\pm 0.01$ & $  15.77\pm 0.01$ & $  15.52\pm 0.02$\\
 12 & $  21.54\pm 0.47$ & $  20.78\pm 0.23$ & $  19.33\pm 0.08$ & $  17.95\pm 0.01$ & $  17.18\pm 0.01$ & $  16.74\pm 0.01$ & $  16.44\pm 0.03$\\
 13 & $  21.64\pm 0.42$ & $  20.73\pm 0.25$ & $  18.49\pm 0.06$ & $  17.16\pm 0.01$ & $  16.53\pm 0.01$ & $  16.21\pm 0.01$ & $  15.99\pm 0.02$\\
 14 & $            ...$ & $  21.71\pm 0.24$ & $  20.43\pm 0.37$ & $  17.49\pm 0.01$ & $  16.55\pm 0.01$ & $  16.01\pm 0.01$ & $  15.66\pm 0.02$\\
 15 & $  21.04\pm 0.30$ & $  20.03\pm 0.11$ & $  18.24\pm 0.05$ & $  16.87\pm 0.01$ & $  16.15\pm 0.01$ & $  15.68\pm 0.01$ & $  15.54\pm 0.02$\\
 16 & $  20.77\pm 0.21$ & $  20.37\pm 0.13$ & $  19.32\pm 0.28$ & $  17.59\pm 0.39$ & $  16.76\pm 0.36$ & $  16.25\pm 0.32$ & $  15.97\pm 0.29$\\
 17 & $  21.48\pm 0.47$ & $  20.04\pm 0.15$ & $  18.09\pm 0.04$ & $  16.65\pm 0.01$ & $  15.90\pm 0.01$ & $  15.47\pm 0.01$ & $  15.22\pm 0.01$\\
 18 & $  21.54\pm 0.47$ & $  21.18\pm 0.30$ & $  18.73\pm 0.07$ & $  17.07\pm 0.01$ & $  16.30\pm 0.02$ & $  15.90\pm 0.02$ & $  15.66\pm 0.02$\\
 19 & $  20.20\pm 0.28$ & $  19.69\pm 0.18$ & $  18.55\pm 0.08$ & $  17.02\pm 0.03$ & $  16.19\pm 0.02$ & $  15.48\pm 0.02$ & $  15.37\pm 0.02$\\
 20 & $            ...$ & $  20.21\pm 0.11$ & $  18.25\pm 0.04$ & $  16.73\pm 0.02$ & $  15.98\pm 0.02$ & $  15.55\pm 0.02$ & $  15.30\pm 0.02$\\
 21 & $  21.05\pm 0.15$ & $  20.06\pm 0.14$ & $  17.87\pm 0.06$ & $  16.46\pm 0.01$ & $  15.76\pm 0.01$ & $  15.39\pm 0.01$ & $  15.18\pm 0.02$\\
 22 & $  20.43\pm 0.29$ & $  19.05\pm 0.07$ & $  17.16\pm 0.03$ & $  15.79\pm 0.01$ & $  15.04\pm 0.01$ & $  14.60\pm 0.01$ & $  14.30\pm 0.01$\\
 23 & $  20.64\pm 0.23$ & $  19.12\pm 0.07$ & $  17.01\pm 0.03$ & $  15.41\pm 0.00$ & $  14.69\pm 0.00$ & $  14.32\pm 0.00$ & $  14.13\pm 0.01$\\
 24 & $            ...$ & $  21.19\pm 0.17$ & $  18.77\pm 0.05$ & $  17.10\pm 0.01$ & $  16.41\pm 0.01$ & $  16.03\pm 0.01$ & $  15.81\pm 0.02$\\
 25 & $            ...$ & $  20.56\pm 0.25$ & $  18.72\pm 0.09$ & $  16.81\pm 0.01$ & $  16.07\pm 0.01$ & $  15.69\pm 0.01$ & $  15.46\pm 0.01$\\
 26 & $            ...$ & $  22.31\pm 0.68$ & $  19.60\pm 0.10$ & $  17.89\pm 0.01$ & $  16.95\pm 0.01$ & $  16.43\pm 0.01$ & $  16.18\pm 0.03$\\
 27 & $  21.97\pm 0.39$ & $  21.18\pm 0.18$ & $  18.98\pm 0.09$ & $  17.36\pm 0.01$ & $  16.53\pm 0.01$ & $  16.06\pm 0.01$ & $  15.77\pm 0.02$\\
 28 & $  21.28\pm 0.38$ & $  19.98\pm 0.17$ & $  18.27\pm 0.07$ & $  16.94\pm 0.01$ & $  16.33\pm 0.01$ & $  15.99\pm 0.01$ & $  15.77\pm 0.03$\\
 29 & $  22.63\pm 0.49$ & $  21.20\pm 0.21$ & $  19.14\pm 0.07$ & $  17.80\pm 0.01$ & $  17.08\pm 0.01$ & $  16.61\pm 0.01$ & $  16.40\pm 0.03$\\
 30 & $            ...$ & $  20.35\pm 0.15$ & $  18.08\pm 0.12$ & $  16.23\pm 0.01$ & $  15.31\pm 0.01$ & $  14.84\pm 0.01$ & $  14.59\pm 0.02$\\
 31 & $            ...$ & $  22.80\pm 0.37$ & $  19.82\pm 0.13$ & $  18.34\pm 0.02$ & $  17.55\pm 0.01$ & $  17.19\pm 0.01$ & $  16.97\pm 0.04$\\
 32 & $  21.55\pm 0.46$ & $  22.51\pm 0.45$ & $  19.91\pm 0.11$ & $  18.35\pm 0.01$ & $  17.60\pm 0.01$ & $  17.06\pm 0.01$ & $  16.86\pm 0.03$\\
 33 & $  21.11\pm 0.30$ & $  20.45\pm 0.18$ & $  18.42\pm 0.06$ & $  17.28\pm 0.01$ & $  16.56\pm 0.01$ & $  16.10\pm 0.01$ & $  15.94\pm 0.02$\\
 34 & $            ...$ & $  21.27\pm 0.43$ & $  19.07\pm 0.17$ & $  17.12\pm 0.01$ & $  16.40\pm 0.01$ & $  15.99\pm 0.01$ & $  15.82\pm 0.07$\\
 35 & $  20.00\pm 0.05$ & $  18.36\pm 0.04$ & $  16.64\pm 0.02$ & $  15.38\pm 0.02$ & $  14.83\pm 0.02$ & $  14.39\pm 0.02$ & $  14.27\pm 0.01$\\
 36 & $  20.52\pm 0.28$ & $  20.43\pm 0.22$ & $  17.66\pm 0.05$ & $  16.24\pm 0.01$ & $  15.29\pm 0.01$ & $  14.90\pm 0.01$ & $  14.67\pm 0.01$\\
 37 & $            ...$ & $  21.00\pm 0.25$ & $  20.43\pm 0.68$ & $  17.56\pm 0.02$ & $  16.73\pm 0.01$ & $  16.31\pm 0.01$ & $  15.98\pm 0.03$\\
 38 & $  21.06\pm 0.45$ & $  19.57\pm 0.08$ & $  17.81\pm 0.05$ & $  16.29\pm 0.01$ & $  15.55\pm 0.01$ & $  15.13\pm 0.01$ & $  14.90\pm 0.01$\\
 39 & $            ...$ & $  20.01\pm 0.20$ & $  17.62\pm 0.04$ & $  16.06\pm 0.01$ & $  15.35\pm 0.01$ & $  14.96\pm 0.01$ & $  14.73\pm 0.01$\\
 40 & $  21.19\pm 0.39$ & $  20.14\pm 0.18$ & $  18.61\pm 0.05$ & $  17.20\pm 0.02$ & $  16.49\pm 0.02$ & $  16.08\pm 0.02$ & $  15.89\pm 0.02$\\
 41 & $  20.45\pm 0.31$ & $  20.07\pm 0.13$ & $  18.55\pm 0.09$ & $  16.99\pm 0.01$ & $  16.26\pm 0.01$ & $  15.77\pm 0.01$ & $  15.48\pm 0.03$\\
 42 & $  20.90\pm 0.36$ & $  20.09\pm 0.20$ & $  18.59\pm 0.06$ & $  17.21\pm 0.03$ & $  16.56\pm 0.03$ & $  16.17\pm 0.02$ & $  15.93\pm 0.03$\\
 43 & $            ...$ & $  20.60\pm 0.17$ & $  18.74\pm 0.04$ & $  17.24\pm 0.01$ & $  16.56\pm 0.00$ & $  16.10\pm 0.00$ & $  15.89\pm 0.01$\\
 44 & $  20.62\pm 0.29$ & $  19.45\pm 0.08$ & $  17.56\pm 0.02$ & $  16.10\pm 0.01$ & $  15.36\pm 0.00$ & $  14.89\pm 0.00$ & $  14.67\pm 0.01$\\
 45 & $  19.50\pm 0.16$ & $  18.82\pm 0.08$ & $  17.60\pm 0.02$ & $  16.20\pm 0.01$ & $  15.52\pm 0.01$ & $  15.03\pm 0.01$ & $  14.88\pm 0.01$\\
 46 & $  20.80\pm 0.23$ & $  19.72\pm 0.08$ & $  17.70\pm 0.05$ & $  16.24\pm 0.02$ & $  15.56\pm 0.02$ & $  15.16\pm 0.01$ & $  14.94\pm 0.02$\\
 47 & $  20.17\pm 0.24$ & $  18.80\pm 0.08$ & $  17.41\pm 0.03$ & $  16.27\pm 0.02$ & $  15.66\pm 0.05$ & $  15.22\pm 0.07$ & $  15.04\pm 0.08$\\
\enddata
\tablenotetext{1}{All magnitudes are Galactic extinction-corrected AB magnitudes. SDSS $ugriz$ data are Petrosian magnitudes.}
\end{deluxetable*}

%%%%%%%%%%%%%%%%%%%%%%%%%%%%%%%%%%%%%%%%%%%%%%%%%%%%%%%%%%%%%%%%%%%%%%%%%%%%%%%
%%%%%%%%%%%%%%%%%%%%%%%%         table 3       %%%%%%%%%%%%%%%%%%%%%%%%%%%%%%%%
%%%%%%%%%%%%%%%%%%%%%%%%%%%%%%%%%%%%%%%%%%%%%%%%%%%%%%%%%%%%%%%%%%%%%%%%%%%%%%%
\begin{deluxetable*}{ccccrrrrrrr}
\tabletypesize{\tiny}
\tablewidth{0pc}
\tablecaption{NIR and MIR Photometry for {\it WISE} 12 $\mu$m Selected DOGs
\label{tab-nirmir}}
\tablehead{
ID & $J$\tablenotemark{a} &  $H$  & $K_s$ & 3.4 $\mu$m & 4.6 $\mu$m & 9 $\mu$m & 12 $\mu$m\tablenotemark{b} & 12 $\mu$m\tablenotemark{c}\\
   & (mag)                & (mag) & (mag) & (mJy)      & (mJy)      & (mJy)    & (mJy)                      & (mJy)
}
\startdata
  1 & $  15.37\pm 0.06$ & $  15.09\pm 0.06$ & $  14.96\pm 0.07$ & $    2.68\pm  0.06$ & $    2.34\pm  0.06$ & $              ...$ & $   36.34\pm  0.57$ & $              ...$\\
  2 & $  13.82\pm 0.03$ & $  13.50\pm 0.04$ & $  13.46\pm 0.05$ & $    9.20\pm  0.19$ & $    7.21\pm  0.14$ & $              ...$ & $   56.12\pm  0.83$ & $              ...$\\
  3 & $  15.42\pm 0.07$ & $  14.69\pm 0.06$ & $  14.20\pm 0.05$ & $   14.73\pm  0.33$ & $   21.81\pm  0.38$ & $              ...$ & $   42.38\pm  0.62$ & $              ...$\\
  4 & $  14.28\pm 0.04$ & $  14.05\pm 0.05$ & $  14.08\pm 0.06$ & $    4.87\pm  0.10$ & $    5.85\pm  0.11$ & $              ...$ & $   32.38\pm  0.48$ & $              ...$\\
  5 & $  16.24\pm 0.10$ & $  15.97\pm 0.12$ & $  16.00\pm 0.14$ & $    0.75\pm  0.02$ & $    1.08\pm  0.03$ & $              ...$ & $   22.04\pm  0.37$ & $              ...$\\
  6 & $  14.54\pm 0.04$ & $  14.10\pm 0.04$ & $  13.50\pm 0.03$ & $   31.07\pm  0.66$ & $   55.49\pm  0.97$ & $  138.18\pm  9.69$ & $  160.23\pm  2.21$ & $  162.80\pm 27.68$\\
  7 & $  14.85\pm 0.07$ & $  14.44\pm 0.06$ & $  14.33\pm 0.09$ & $    4.95\pm  0.11$ & $    4.90\pm  0.09$ & $              ...$ & $   49.83\pm  0.78$ & $              ...$\\
  8 & $  15.46\pm 0.08$ & $  15.07\pm 0.06$ & $  15.03\pm 0.09$ & $    2.77\pm  0.06$ & $    2.32\pm  0.06$ & $              ...$ & $   24.10\pm  0.40$ & $              ...$\\
  9 & $  14.46\pm 0.04$ & $  14.13\pm 0.04$ & $  14.06\pm 0.05$ & $    5.60\pm  0.11$ & $    4.86\pm  0.09$ & $              ...$ & $   36.64\pm  0.54$ & $              ...$\\
 10 & $  15.01\pm 0.12$ & $  14.46\pm 0.11$ & $  14.33\pm 0.12$ & $    5.73\pm  0.12$ & $    9.15\pm  0.19$ & $              ...$ & $   42.14\pm  0.66$ & $              ...$\\
 11 & $  15.16\pm 0.06$ & $  14.78\pm 0.07$ & $  14.76\pm 0.07$ & $    3.12\pm  0.07$ & $    2.43\pm  0.05$ & $              ...$ & $   21.79\pm  0.36$ & $              ...$\\
 12 & $  16.03\pm 0.10$ & $  15.97\pm 0.12$ & $  15.80\pm 0.12$ & $    3.56\pm  0.08$ & $    7.05\pm  0.13$ & $              ...$ & $   25.39\pm  0.42$ & $              ...$\\
 13 & $  15.66\pm 0.05$ & $  15.34\pm 0.06$ & $  15.28\pm 0.06$ & $    1.91\pm  0.04$ & $    2.13\pm  0.05$ & $              ...$ & $   32.68\pm  0.54$ & $              ...$\\
 14 & $  15.07\pm 0.06$ & $  14.64\pm 0.05$ & $  14.64\pm 0.07$ & $    3.46\pm  0.07$ & $    2.74\pm  0.06$ & $              ...$ & $   27.16\pm  0.53$ & $              ...$\\
 15 & $  15.23\pm 0.05$ & $  14.97\pm 0.06$ & $  14.84\pm 0.06$ & $    3.31\pm  0.07$ & $    2.65\pm  0.06$ & $              ...$ & $   32.03\pm  0.53$ & $              ...$\\
 16 & $  15.50\pm 0.07$ & $  15.23\pm 0.08$ & $  15.06\pm 0.09$ & $    2.95\pm  0.07$ & $    2.64\pm  0.06$ & $              ...$ & $   30.47\pm  0.48$ & $              ...$\\
 17 & $  14.78\pm 0.04$ & $  14.45\pm 0.04$ & $  14.05\pm 0.04$ & $   17.49\pm  0.37$ & $   29.63\pm  0.55$ & $              ...$ & $   65.45\pm  0.90$ & $              ...$\\
 18 & $  15.30\pm 0.05$ & $  15.12\pm 0.05$ & $  15.08\pm 0.06$ & $    2.11\pm  0.05$ & $    3.36\pm  0.07$ & $              ...$ & $   33.75\pm  0.50$ & $              ...$\\
 19 & $  14.89\pm 0.04$ & $  14.62\pm 0.05$ & $  14.47\pm 0.05$ & $    6.75\pm  0.14$ & $   11.78\pm  0.22$ & $              ...$ & $   48.12\pm  0.66$ & $              ...$\\
 20 & $  14.87\pm 0.04$ & $  14.53\pm 0.05$ & $  14.48\pm 0.05$ & $    5.11\pm  0.11$ & $    9.19\pm  0.18$ & $              ...$ & $   59.42\pm  0.82$ & $  103.40\pm 27.92$\\
 21 & $  14.95\pm 0.06$ & $  14.54\pm 0.05$ & $  14.60\pm 0.08$ & $    5.03\pm  0.13$ & $    7.69\pm  0.18$ & $              ...$ & $   33.60\pm  0.77$ & $              ...$\\
 22 & $  13.82\pm 0.04$ & $  13.48\pm 0.04$ & $  13.38\pm 0.05$ & $    9.63\pm  0.27$ & $    7.62\pm  0.20$ & $   91.51\pm 10.73$ & $   78.04\pm  1.51$ & $              ...$\\
 23 & $  13.79\pm 0.03$ & $  13.34\pm 0.03$ & $  13.00\pm 0.03$ & $   34.00\pm  0.69$ & $   47.14\pm  0.82$ & $  114.39\pm 24.69$ & $  143.99\pm  1.99$ & $  133.20\pm 30.64$\\
 24 & $  15.64\pm 0.06$ & $  15.28\pm 0.07$ & $  15.29\pm 0.07$ & $    1.87\pm  0.04$ & $    1.64\pm  0.04$ & $              ...$ & $   22.30\pm  0.35$ & $              ...$\\
 25 & $  15.20\pm 0.05$ & $  14.87\pm 0.06$ & $  14.86\pm 0.06$ & $    3.06\pm  0.07$ & $    3.81\pm  0.08$ & $              ...$ & $   21.98\pm  0.34$ & $              ...$\\
 26 & $  15.78\pm 0.07$ & $  15.31\pm 0.08$ & $  15.10\pm 0.06$ & $    2.28\pm  0.05$ & $    2.20\pm  0.05$ & $              ...$ & $   24.23\pm  0.40$ & $              ...$\\
 27 & $  15.17\pm 0.05$ & $  14.94\pm 0.06$ & $  14.72\pm 0.06$ & $    2.84\pm  0.06$ & $    2.44\pm  0.05$ & $              ...$ & $   22.36\pm  0.41$ & $              ...$\\
 28 & $  15.55\pm 0.07$ & $  15.24\pm 0.08$ & $  15.30\pm 0.08$ & $    2.04\pm  0.05$ & $    3.75\pm  0.08$ & $              ...$ & $   39.44\pm  0.62$ & $              ...$\\
 29 & $            ...$ & $            ...$ & $            ...$ & $    6.42\pm  0.12$ & $    9.32\pm  0.17$ & $              ...$ & $   25.23\pm  0.40$ & $              ...$\\
 30 & $  14.29\pm 0.04$ & $  13.89\pm 0.05$ & $  13.43\pm 0.04$ & $   27.77\pm  0.56$ & $   44.28\pm  0.86$ & $   87.88\pm 23.89$ & $   96.10\pm  1.33$ & $              ...$\\
 31 & $            ...$ & $            ...$ & $            ...$ & $    2.73\pm  0.06$ & $   11.91\pm  0.23$ & $              ...$ & $   23.70\pm  0.39$ & $              ...$\\
 32 & $            ...$ & $            ...$ & $            ...$ & $    1.65\pm  0.04$ & $    4.05\pm  0.08$ & $              ...$ & $   36.27\pm  0.57$ & $              ...$\\
 33 & $  15.45\pm 0.06$ & $  15.05\pm 0.07$ & $  15.04\pm 0.06$ & $    2.87\pm  0.06$ & $    2.77\pm  0.06$ & $              ...$ & $   32.44\pm  0.51$ & $              ...$\\
 34 & $            ...$ & $            ...$ & $            ...$ & $    3.41\pm  0.07$ & $    2.76\pm  0.06$ & $              ...$ & $   26.62\pm  0.39$ & $              ...$\\
 35 & $  14.05\pm 0.03$ & $  13.80\pm 0.04$ & $  13.44\pm 0.04$ & $   36.50\pm  0.77$ & $   51.26\pm  0.99$ & $   80.22\pm  6.16$ & $  152.45\pm  2.11$ & $              ...$\\
 36 & $  14.37\pm 0.04$ & $  13.95\pm 0.04$ & $  13.72\pm 0.04$ & $   11.14\pm  0.23$ & $   14.16\pm  0.26$ & $              ...$ & $   31.53\pm  0.46$ & $              ...$\\
 37 & $  15.42\pm 0.09$ & $  14.93\pm 0.07$ & $  14.86\pm 0.10$ & $    3.10\pm  0.06$ & $    2.70\pm  0.05$ & $              ...$ & $   23.20\pm  0.34$ & $   82.53\pm 24.76$\\
 38 & $  14.77\pm 0.06$ & $  14.39\pm 0.06$ & $  14.47\pm 0.08$ & $    4.70\pm  0.10$ & $    7.02\pm  0.14$ & $              ...$ & $   51.42\pm  0.76$ & $              ...$\\
 39 & $  14.36\pm 0.04$ & $  14.06\pm 0.05$ & $  14.11\pm 0.07$ & $    6.94\pm  0.15$ & $   14.03\pm  0.27$ & $   65.10\pm  2.42$ & $   83.09\pm  1.15$ & $              ...$\\
 40 & $  15.58\pm 0.08$ & $  15.31\pm 0.10$ & $  15.01\pm 0.07$ & $    7.42\pm  0.16$ & $   12.78\pm  0.24$ & $              ...$ & $   40.69\pm  0.60$ & $              ...$\\
 41 & $  15.02\pm 0.05$ & $  14.73\pm 0.06$ & $  14.47\pm 0.06$ & $    4.11\pm  0.08$ & $    3.45\pm  0.06$ & $              ...$ & $   33.41\pm  0.46$ & $              ...$\\
 42 & $  15.76\pm 0.08$ & $  15.41\pm 0.08$ & $  15.15\pm 0.08$ & $    8.28\pm  0.18$ & $   17.02\pm  0.30$ & $              ...$ & $  106.84\pm  1.48$ & $  158.00\pm 28.44$\\
 43 & $  15.38\pm 0.06$ & $  14.95\pm 0.06$ & $  14.20\pm 0.04$ & $   20.87\pm  0.42$ & $   40.68\pm  0.75$ & $  119.23\pm 21.73$ & $  119.11\pm  1.65$ & $  137.00\pm 21.92$\\
 44 & $  14.08\pm 0.03$ & $  13.76\pm 0.03$ & $  13.71\pm 0.03$ & $    8.04\pm  0.18$ & $    6.25\pm  0.13$ & $              ...$ & $   59.15\pm  0.82$ & $   81.69\pm 21.24$\\
 45 & $  13.88\pm 0.03$ & $  13.53\pm 0.04$ & $  13.26\pm 0.04$ & $   26.69\pm  0.54$ & $   37.00\pm  0.72$ & $  118.18\pm 15.25$ & $  125.41\pm  1.73$ & $  152.60\pm 22.89$\\
 46 & $  14.71\pm 0.05$ & $  14.33\pm 0.06$ & $  14.24\pm 0.06$ & $    9.77\pm  0.21$ & $   22.65\pm  0.42$ & $   69.90\pm  7.95$ & $  112.70\pm  1.45$ & $  114.70\pm 14.91$\\
 47 & $  14.74\pm 0.04$ & $  14.31\pm 0.05$ & $  14.40\pm 0.06$ & $    6.96\pm  0.13$ & $   11.79\pm  0.24$ & $              ...$ & $  110.85\pm  1.23$ & $              ...$\\
\enddata
\tablenotetext{1}{2MASS magnitudes are 20 mag arcsec$^{-2}$ isophotal fiducial elliptical aperture magnitudes (AB). These magnitudes represent approximately 85\% of the total flux of a galaxy (see http://www.ipac.caltech.edu/2mass/releases/allsky/doc/sec2\_3d3.html).}
\tablenotetext{2}{{\it WISE} 12 $\mu$m.}
\tablenotetext{3}{{\it IRAS} 12 $\mu$m.}
\end{deluxetable*}

%%%%%%%%%%%%%%%%%%%%%%%%%%%%%%%%%%%%%%%%%%%%%%%%%%%%%%%%%%%%%%%%%%%%%%%%%%%%%%%
%%%%%%%%%%%%%%%%%%%%%%%%         table 4       %%%%%%%%%%%%%%%%%%%%%%%%%%%%%%%%
%%%%%%%%%%%%%%%%%%%%%%%%%%%%%%%%%%%%%%%%%%%%%%%%%%%%%%%%%%%%%%%%%%%%%%%%%%%%%%%
\begin{deluxetable*}{crrrrrrrr}
\tabletypesize{\tiny}
\tablewidth{0pc}
\tablecaption{MIR and FIR Photometry for {\it WISE} 12 $\mu$m Selected DOGs
\label{tab-mirfir}}
\tablehead{
ID & 18 $\mu$m & 22 $\mu$m & 25 $\mu$m & 60 $\mu$m & 65 $\mu$m & 90 $\mu$m & 100 $\mu$m & 140 $\mu$m \\
   & (mJy)     & (mJy)     & (mJy)     & (Jy)      & (Jy)      & (Jy)      & (Jy)       & (Jy)
}
\startdata
  1 & $              ...$ & $  206.25\pm  3.61$ & $              ...$ & $    1.05\pm  0.07$ & $              ...$ & $    1.20\pm  0.04$ & $              ...$ & $              ...$\\
  2 & $              ...$ & $  111.58\pm  2.77$ & $              ...$ & $    1.75\pm  0.12$ & $              ...$ & $    1.90\pm  0.07$ & $    3.42\pm  0.24$ & $    4.26\pm  1.27$\\
  3 & $              ...$ & $   82.94\pm  2.14$ & $              ...$ & $              ...$ & $              ...$ & $              ...$ & $              ...$ & $              ...$\\
  4 & $              ...$ & $  116.52\pm  2.68$ & $              ...$ & $              ...$ & $              ...$ & $              ...$ & $              ...$ & $              ...$\\
  5 & $              ...$ & $  132.92\pm  2.69$ & $  177.30\pm 28.37$ & $    1.10\pm  0.05$ & $              ...$ & $    0.83\pm  0.02$ & $    1.17\pm  0.15$ & $              ...$\\
  6 & $  272.17\pm 15.20$ & $  342.92\pm  7.26$ & $  278.50\pm 50.13$ & $    0.36\pm  0.04$ & $              ...$ & $              ...$ & $              ...$ & $              ...$\\
  7 & $              ...$ & $  161.28\pm  2.97$ & $  236.90\pm 40.27$ & $    3.06\pm  0.15$ & $    3.31\pm  0.41$ & $    3.53\pm  0.11$ & $    4.42\pm  0.40$ & $    4.00\pm  0.77$\\
  8 & $              ...$ & $   49.43\pm  1.50$ & $              ...$ & $    0.76\pm  0.05$ & $              ...$ & $    0.80\pm  0.07$ & $    1.04\pm  0.16$ & $              ...$\\
  9 & $              ...$ & $  110.76\pm  2.86$ & $  167.80\pm 26.85$ & $    1.72\pm  0.10$ & $              ...$ & $    1.98\pm  0.10$ & $    3.40\pm  0.31$ & $              ...$\\
 10 & $              ...$ & $  119.12\pm  2.52$ & $  202.10\pm 40.42$ & $    0.21\pm  0.04$ & $              ...$ & $              ...$ & $              ...$ & $              ...$\\
 11 & $              ...$ & $   42.74\pm  1.42$ & $              ...$ & $    0.61\pm  0.04$ & $              ...$ & $              ...$ & $    1.11\pm  0.14$ & $              ...$\\
 12 & $              ...$ & $   54.35\pm  1.95$ & $              ...$ & $              ...$ & $              ...$ & $              ...$ & $              ...$ & $              ...$\\
 13 & $              ...$ & $  203.42\pm  4.50$ & $  288.00\pm 40.32$ & $    1.48\pm  0.09$ & $              ...$ & $    1.07\pm  0.11$ & $    1.55\pm  0.14$ & $              ...$\\
 14 & $              ...$ & $   74.13\pm  2.39$ & $              ...$ & $    0.83\pm  0.07$ & $              ...$ & $    0.96\pm  0.03$ & $    1.50\pm  0.13$ & $              ...$\\
 15 & $              ...$ & $   81.51\pm  2.85$ & $              ...$ & $    0.81\pm  0.06$ & $              ...$ & $    0.94\pm  0.03$ & $              ...$ & $              ...$\\
 16 & $              ...$ & $  101.39\pm  2.52$ & $              ...$ & $              ...$ & $              ...$ & $    1.28\pm  0.06$ & $              ...$ & $              ...$\\
 17 & $              ...$ & $  138.16\pm  2.80$ & $              ...$ & $    0.54\pm  0.04$ & $              ...$ & $    0.75\pm  0.23$ & $    0.65\pm  0.13$ & $              ...$\\
 18 & $              ...$ & $  143.74\pm  3.05$ & $  193.40\pm 30.94$ & $    0.81\pm  0.05$ & $              ...$ & $    0.51\pm  0.09$ & $    0.80\pm  0.13$ & $              ...$\\
 19 & $              ...$ & $  184.33\pm  3.73$ & $  216.80\pm 19.51$ & $    0.54\pm  0.04$ & $              ...$ & $              ...$ & $    0.95\pm  0.18$ & $              ...$\\
 20 & $              ...$ & $  200.81\pm  4.07$ & $  203.50\pm 28.49$ & $    0.49\pm  0.05$ & $              ...$ & $              ...$ & $    0.83\pm  0.15$ & $              ...$\\
 21 & $              ...$ & $   83.94\pm  4.56$ & $              ...$ & $    0.42\pm  0.06$ & $              ...$ & $    0.68\pm  0.14$ & $    0.83\pm  0.22$ & $              ...$\\
 22 & $              ...$ & $  181.80\pm  7.70$ & $  290.80\pm 52.34$ & $    2.60\pm  0.18$ & $    2.90\pm  0.20$ & $    3.67\pm  0.04$ & $    4.19\pm  0.38$ & $    4.45\pm  1.05$\\
 23 & $  343.40\pm 21.14$ & $  438.11\pm 10.09$ & $  500.80\pm 35.06$ & $    1.67\pm  0.10$ & $              ...$ & $    1.73\pm  0.08$ & $    2.15\pm  0.17$ & $              ...$\\
 24 & $              ...$ & $  121.11\pm  2.34$ & $  142.60\pm 21.39$ & $    0.82\pm  0.06$ & $              ...$ & $    0.64\pm  0.08$ & $    0.86\pm  0.18$ & $              ...$\\
 25 & $              ...$ & $   75.37\pm  2.08$ & $              ...$ & $    0.38\pm  0.06$ & $              ...$ & $              ...$ & $              ...$ & $              ...$\\
 26 & $              ...$ & $  101.11\pm  2.51$ & $              ...$ & $              ...$ & $              ...$ & $    1.93\pm  0.11$ & $              ...$ & $    2.85\pm  0.37$\\
 27 & $              ...$ & $   82.87\pm  2.14$ & $  143.10\pm 35.77$ & $    1.60\pm  0.11$ & $              ...$ & $    1.62\pm  0.06$ & $    2.55\pm  0.25$ & $              ...$\\
 28 & $              ...$ & $  197.69\pm  3.82$ & $  303.60\pm 48.58$ & $    2.66\pm  0.16$ & $              ...$ & $    2.00\pm  0.04$ & $    2.52\pm  0.28$ & $              ...$\\
 29 & $              ...$ & $   50.53\pm  1.77$ & $              ...$ & $    0.27\pm  0.06$ & $              ...$ & $              ...$ & $    0.59\pm  0.14$ & $              ...$\\
 30 & $  175.05\pm 50.81$ & $  181.30\pm  4.68$ & $              ...$ & $              ...$ & $              ...$ & $              ...$ & $              ...$ & $              ...$\\
 31 & $              ...$ & $   49.70\pm  1.60$ & $              ...$ & $              ...$ & $              ...$ & $              ...$ & $              ...$ & $              ...$\\
 32 & $   94.93\pm  8.66$ & $  167.80\pm  3.55$ & $  204.30\pm 57.20$ & $    0.22\pm  0.04$ & $              ...$ & $              ...$ & $              ...$ & $              ...$\\
 33 & $              ...$ & $  126.94\pm  2.57$ & $  205.30\pm 45.17$ & $    1.48\pm  0.12$ & $              ...$ & $    1.45\pm  0.08$ & $    1.94\pm  0.27$ & $              ...$\\
 34 & $              ...$ & $   48.30\pm  1.51$ & $              ...$ & $    0.71\pm  0.06$ & $              ...$ & $    0.82\pm  0.05$ & $    1.22\pm  0.15$ & $              ...$\\
 35 & $  305.41\pm 30.17$ & $  398.46\pm  7.71$ & $  399.40\pm 39.94$ & $    0.73\pm  0.06$ & $              ...$ & $              ...$ & $    0.94\pm  0.15$ & $              ...$\\
 36 & $              ...$ & $   73.11\pm  1.68$ & $              ...$ & $              ...$ & $              ...$ & $              ...$ & $              ...$ & $              ...$\\
 37 & $              ...$ & $   34.64\pm  1.12$ & $              ...$ & $    0.52\pm  0.05$ & $              ...$ & $    0.75\pm  0.04$ & $    1.18\pm  0.14$ & $              ...$\\
 38 & $  112.20\pm 12.27$ & $  221.61\pm  3.47$ & $  161.70\pm 21.02$ & $    0.36\pm  0.04$ & $              ...$ & $              ...$ & $              ...$ & $              ...$\\
 39 & $  219.55\pm 24.55$ & $  348.01\pm  5.13$ & $  359.20\pm 39.51$ & $    0.92\pm  0.06$ & $              ...$ & $              ...$ & $    1.33\pm  0.21$ & $              ...$\\
 40 & $              ...$ & $  123.93\pm  2.28$ & $  127.10\pm 20.34$ & $    0.49\pm  0.04$ & $              ...$ & $              ...$ & $    0.69\pm  0.17$ & $              ...$\\
 41 & $              ...$ & $   90.45\pm  1.92$ & $  132.20\pm 25.12$ & $    1.31\pm  0.07$ & $              ...$ & $    1.52\pm  0.10$ & $    2.52\pm  0.23$ & $              ...$\\
 42 & $  247.55\pm 35.80$ & $  281.31\pm  4.40$ & $  317.80\pm 31.78$ & $    0.20\pm  0.04$ & $              ...$ & $              ...$ & $              ...$ & $              ...$\\
 43 & $              ...$ & $  324.78\pm  5.68$ & $  410.20\pm 32.82$ & $    1.01\pm  0.07$ & $              ...$ & $              ...$ & $    0.92\pm  0.21$ & $              ...$\\
 44 & $              ...$ & $  127.29\pm  2.46$ & $  144.50\pm 18.78$ & $    1.48\pm  0.10$ & $              ...$ & $    1.85\pm  0.08$ & $    3.47\pm  0.21$ & $              ...$\\
 45 & $  235.96\pm 23.60$ & $  361.07\pm  5.99$ & $  415.90\pm 29.11$ & $    1.16\pm  0.07$ & $              ...$ & $    0.88\pm  0.05$ & $    1.46\pm  0.18$ & $              ...$\\
 46 & $  241.33\pm 13.94$ & $  312.17\pm  4.60$ & $  383.90\pm 15.36$ & $    0.92\pm  0.05$ & $              ...$ & $    0.82\pm  0.04$ & $    1.22\pm  0.17$ & $              ...$\\
 47 & $  310.73\pm 37.67$ & $  618.28\pm 10.25$ & $  716.30\pm 57.30$ & $    5.14\pm  0.36$ & $    3.77\pm  0.32$ & $    5.04\pm  0.18$ & $    5.03\pm  0.55$ & $    5.97\pm  0.94$\\
\enddata
\end{deluxetable*}

To have an unbiased control sample,
  the redshift, IR luminosity, and stellar mass distributions
  of the control sample should match those of local DOGs.
We first examine the redshift distribution using 
  the Kolmogorov-Smirnov (K-S) test to determine whether 
  the DOGs and the control sample
  are drawn from the same distribution.
The K-S test cannot reject
  the hypothesis that the redshift distributions of the two samples
  are extracted from the same parent population.
We then compare the IR luminosities ($L_{\rm IR}$) and 
  stellar masses ($M_{\rm star}$)
  of the two samples in Figure \ref{fig-lirmass} 
  (DOGs: red circles, control: gray diamonds).
We compute the IR luminosities of DOGs and the control sample
  from the SED fit to the IR photometric data.
We explain the details of this fitting in the next section.
Stellar mass estimates are adopted
  from the MPA/JHU DR7 value-added galaxy 
  catalog\footnote{http://www.mpa-garching.mpg.de/SDSS/DR7/Data/stellarmass.html}
  (VAGC).
These estimates are based on the fit of SDSS five-band photometry 
  to the models of \citet{bc03} (see also \citealt{kau03}).
We convert the stellar masses in the MPA/JHU DR7 VAGC
  that are based on the Kroupa IMF \citep{kro01}
  to those with a Salpeter IMF \citep{sal55} 
  by dividing them by a factor of 0.7 \citep{elb07}.

For the IR luminosity distribution, the K-S test rejects
  the hypothesis that the distributions of the two samples
  are extracted from the same parent population
  with a confidence level of 99.9\%.
Therefore, we revise the construction of the control sample
  by randomly selecting galaxies
  among the preliminary control sample of galaxies (i.e., gray diamonds)
  to have the same IR luminosity distribution as the DOGs.
We restrict this process to galaxies in the same range
  of IR luminosity and stellar mass as the DOGs (i.e., those within the dashed box).
We set the number of control sample objects to be three times the number of DOGs
  (i.e., 141 galaxies in the control sample).
The resulting distributions of redshift, IR luminosity, and stellar mass
  for the control sample
  are similar to those for the 47 DOGs, according to K-S tests.
We show these 141 control sample of galaxies with blue squares 
  in Figures \ref{fig-samp12um} and \ref{fig-lirmass}.

\section{Analysis of Spectral Energy Distributions of Local DOGs}\label{sed}
\subsection{Spectral Energy Distribution Modeling}\label{sedfit}

We compute the IR luminosities of DOGs and the control sample
  using the SED templates and fitting routine of \citet{mul11agn}, 
  DECOMPIR\footnote{http://sites.google.com/site/decompir}.
This routine requires observational data 
  at $\lambda>6$ $\mu$m for the fit;
  we use measured flux densities
  of \wise 12 and 22 $\mu$m, \akari 9, 18, 65, 90, 140 and 160 $\mu$m,
  \iras 12, 25, 60 and 100 $\mu$m.
We use only reliable flux density measurements;
  flux quality flags are either `high' or `moderate' for the {\it IRAS} data and
  `high' for the {\it AKARI} data, and
  we require signal-to-noise ratios (S/Ns) $\geq$ 3 for \wise data.
Because some bands partially overlap
  (e.g., \wise 12 $\mu$m and \iras 12 $\mu$m,
  {\it IRAS} 60 $\mu$m and {\it AKARI} 65 $\mu$m,
  and {\it IRAS} 100 $\mu$m and {\it AKARI} 90 $\mu$m),
  we use only one measurement
  (i.e., \wise 12 $\mu$m, {\it IRAS} 60 $\mu$m and 100 $\mu$m)
  for the fit to avoid over-weighting
  when both flux densities are measured.
 
For a given galaxy,
  this routine decomposes the observed SED into 
  two components (i.e., a host-galaxy and an AGN).
We can thus measure the contribution of (buried) AGN 
  to the total infrared energy budget of a galaxy.
The SED templates consist of one AGN SED and 
  five groups of host-galaxy SEDs, referred to as `SB1' through `SB5'.
These templates are constructed 
  from {\it Spitzer} infrared spectrograph (IRS) spectra and 
  {\it IRAS} photometric data of AGN-host and 
  starburst galaxies \citep{mul11agn}.

We apply this routine to the DOGs and the control sample,
  and choose the best-fit solution with the lowest $\chi^2$ value
  for each galaxy.
We compute the uncertainties in the IR luminosity and in the AGN contribution
  by randomly selecting flux densities
  at each band within the associated error distribution 
  (assumed to be Gaussian) and then refitting.
Figure \ref{fig-fitsed} shows example SEDs of DOGs
  with the best-fit AGN (dotted line) and host-galaxy (dashed line) templates.
For the DOG in the top panel (a),
  the estimated AGN contribution to the total IR luminosity
  is 23\%; the optical spectral classification also indicates that
  this DOG is a Seyfert galaxy. 
The DOG in the middle panel (b)
  does not contain any AGN component,
  consistent with its optical spectral classification (i.e., star-forming).

Among 47 DOGs and 141 control sample of galaxies,
  6 DOGs and 14 control sample of galaxies are not detected
  in the FIR bands (i.e., neither at \iras 60 $\mu$m nor at \akari 90 $\mu$m).
These FIR undetected galaxies have photometric data 
  only up to the \wise 22 $\mu$m band.
These non-detections do not result from lack of coverage by the 
  \iras or \akari all-sky surveys;
  they occur because the FIR flux densities for these objects 
  are below the detection limits of \iras or {\it AKARI}.
Non-detection in the FIR bands 
  means that the IR SEDs are mainly dominated by
  a hot AGN dust component that is peaked 
  in the MIR bands rather than in the FIR bands.
For these galaxies, we fit the SEDs with only an AGN template
  to estimate their IR luminosities.
The measured IR luminosities then indicate 
  lower limits to the total IR luminosities.
Similarly, we set the AGN contribution to the total IR luminosity
  in these FIR undetected galaxies to be 100\%.
The bottom panel (c) in Figure \ref{fig-fitsed} shows
 an SED representative of this example.

%Figure 5 %%%%%%%%%%%%%%%%%%%%
\begin{figure*}
\center
\includegraphics[width=175mm]{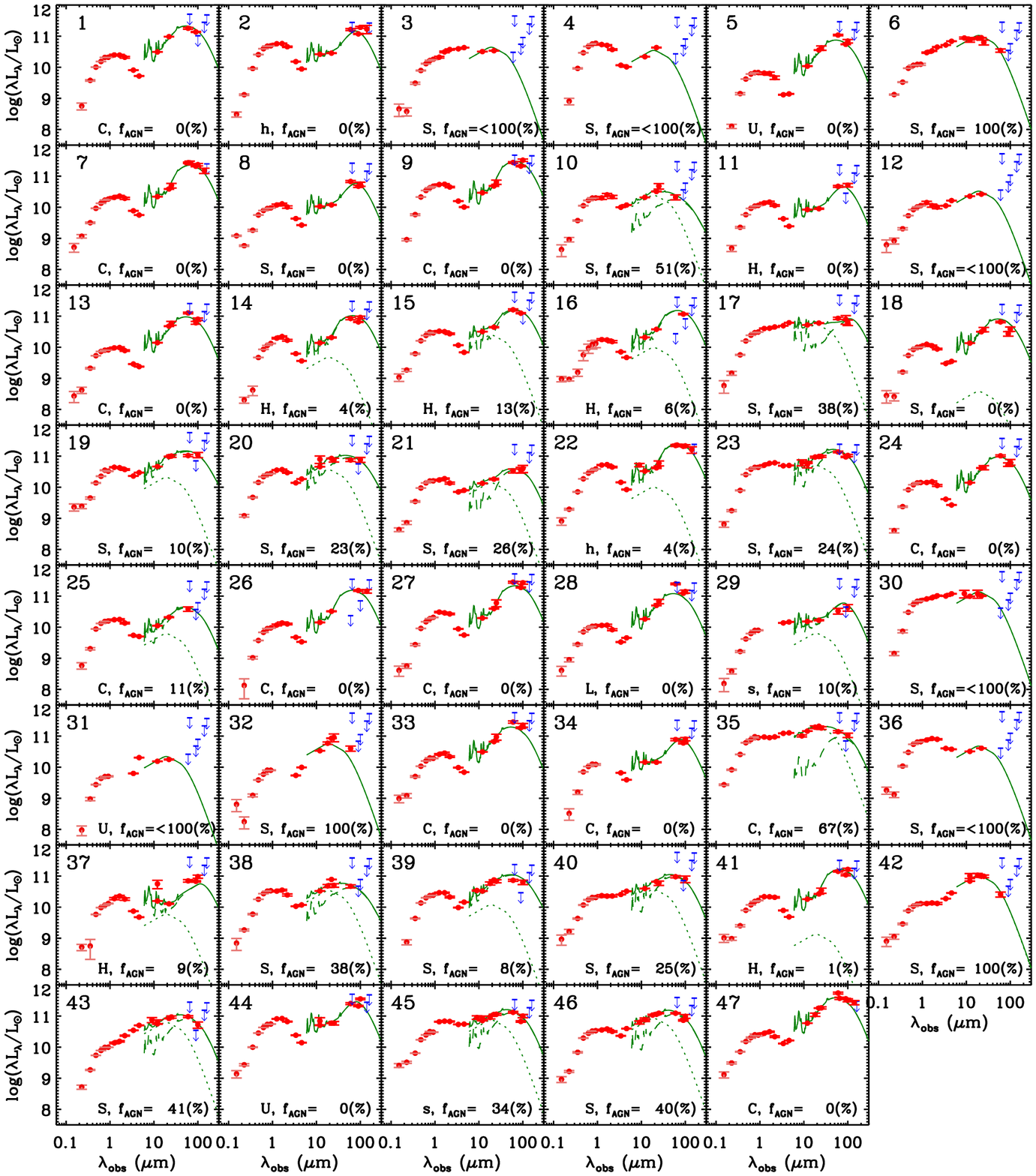}
\caption{SEDs of 47 DOGs.
Red filled circles are observed photometric data, and
  blue down arrows are upper limits.
Solid, dotted, and dashed lines indicate the best-fit SEDs 
  with the DECOMPIR routine of \citet{mul11agn},
  for total, host-galaxy, and AGN components, respectively.
Number in the upper left corner of each panel is the identification in Table \ref{tab-samp}.
Galaxy classification based on optical line ratios
  (H: SF, C: Composite, S: Seyfert, L: LINER, U: Undetermined)
  and the AGN contribution to the total IR luminosity
  are shown in the bottom of each panel.
The lower case letter gives the classification adopted from NED.
}\label{fig-sed}
\end{figure*}
%%%%%%%%%%%%%%%%

%Figure 6 %%%%%%%%%%%%%%%%%%%%
\begin{figure*}
\center
\includegraphics[width=175mm]{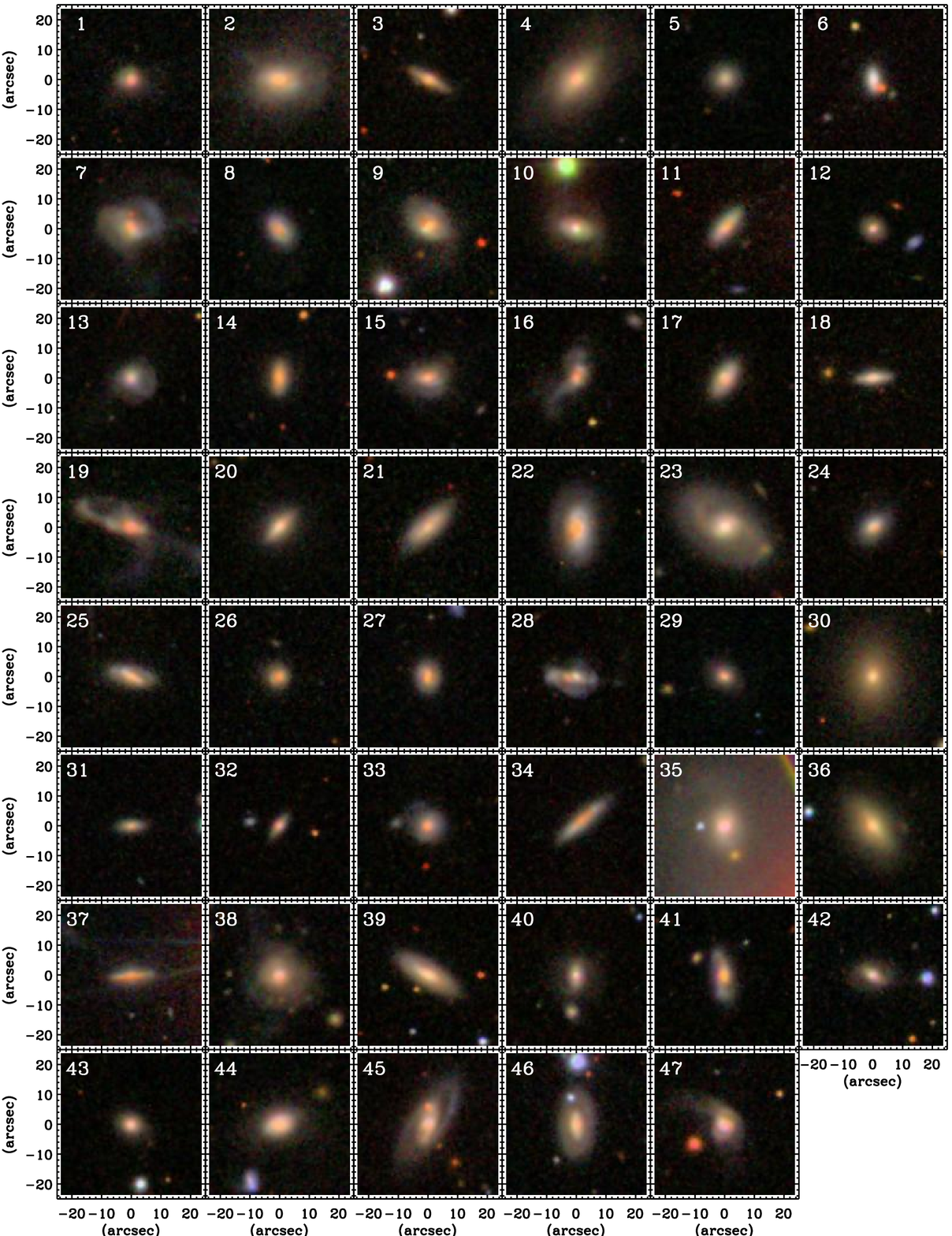}
\caption{Optical color images ($50\arcsec \times 50 \arcsec$) of 47 DOGs
  (RGB color composites from $irg$ bands).
Number in the upper left corner of each panel is 
  the identification in Table \ref{tab-samp}.
}\label{fig-chart}
\end{figure*}
%%%%%%%%%%%%%%%%

We list the 47 local analogs of DOGs in Table \ref{tab-samp}
  along with their SDSS identification, Right Ascension, Declination, 
  spectroscopic redshift, stellar mass, IR luminosity,
  AGN contribution to the total IR luminosity, and emission line classification. 
We also list the relevant UV/optical photometric data in Table \ref{tab-uvopt},
  NIR/MIR photometric data in Table \ref{tab-nirmir}, and 
  MIR/FIR photometric data in Table \ref{tab-mirfir}.
With these photometric data, 
  we plot the observed SEDs of the 47 DOGs in Figure \ref{fig-sed};
  this plot includes many examples of the three SED types in Figure \ref{fig-fitsed}.

We also show SDSS color images of all the DOGs in Figure \ref{fig-chart}.
The figure shows a variety of optical morphologies among DOGs
  from those with merging features (e.g., ID: 7, 13, 16, 19, 47)
  to highly-inclined disk galaxies (e.g., ID: 3, 11, 21, 34, 37).
Interestingly, there are some galaxies 
  with large bulge/spheroids (e.g., ID: 4, 30, 36).
There are other early-type galaxies with IR activity in the local universe 
  (e.g., \citealt{kna89,jhlee10beg, smith12,hwa12a2199, ko13}).
All of these early-type DOGs have Seyfert
  optical spectra; they are not detected in the FIR bands,
  suggesting AGN-dominated SEDs (see Figure \ref{fig-sed}).

%Figure 7 %%%%%%%%%%%%%%%%%%%%
\begin{figure}
\center
\includegraphics[width=85mm]{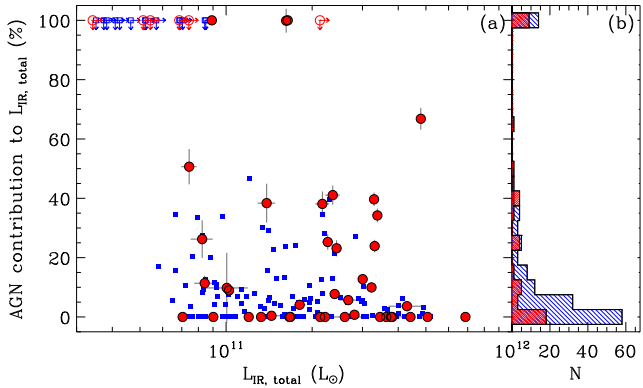}
\caption{AGN contribution to the total IR luminosity
  as a function of total IR luminosity
  for DOGs (red circles) and the control sample (blue squares) (a), and
  its histogram (b).
Filled and open symbols indicate
  FIR (i.e., \iras 60 $\mu$m or \akari 90 $\mu$m) 
  detected and undetected galaxies, respectively.
Arrows indicate lower and upper limits
  for the total IR luminosity and AGN contribution, respectively.
We plot error bars only for local DOGs.
DOGs and the control sample are denoted by hatched histograms with
  orientation of 45$^\circ$ ($//$ with red color) and 
  of 315$^\circ$ ($\setminus\setminus$ with blue color) 
  relative to horizontal, respectively.
}\label{fig-agncont}
\end{figure}
%%%%%%%%%%%%%%%%

%Figure 8 %%%%%%%%%%%%%%%%%%%%
\begin{figure*}
\center
\includegraphics[width=150mm]{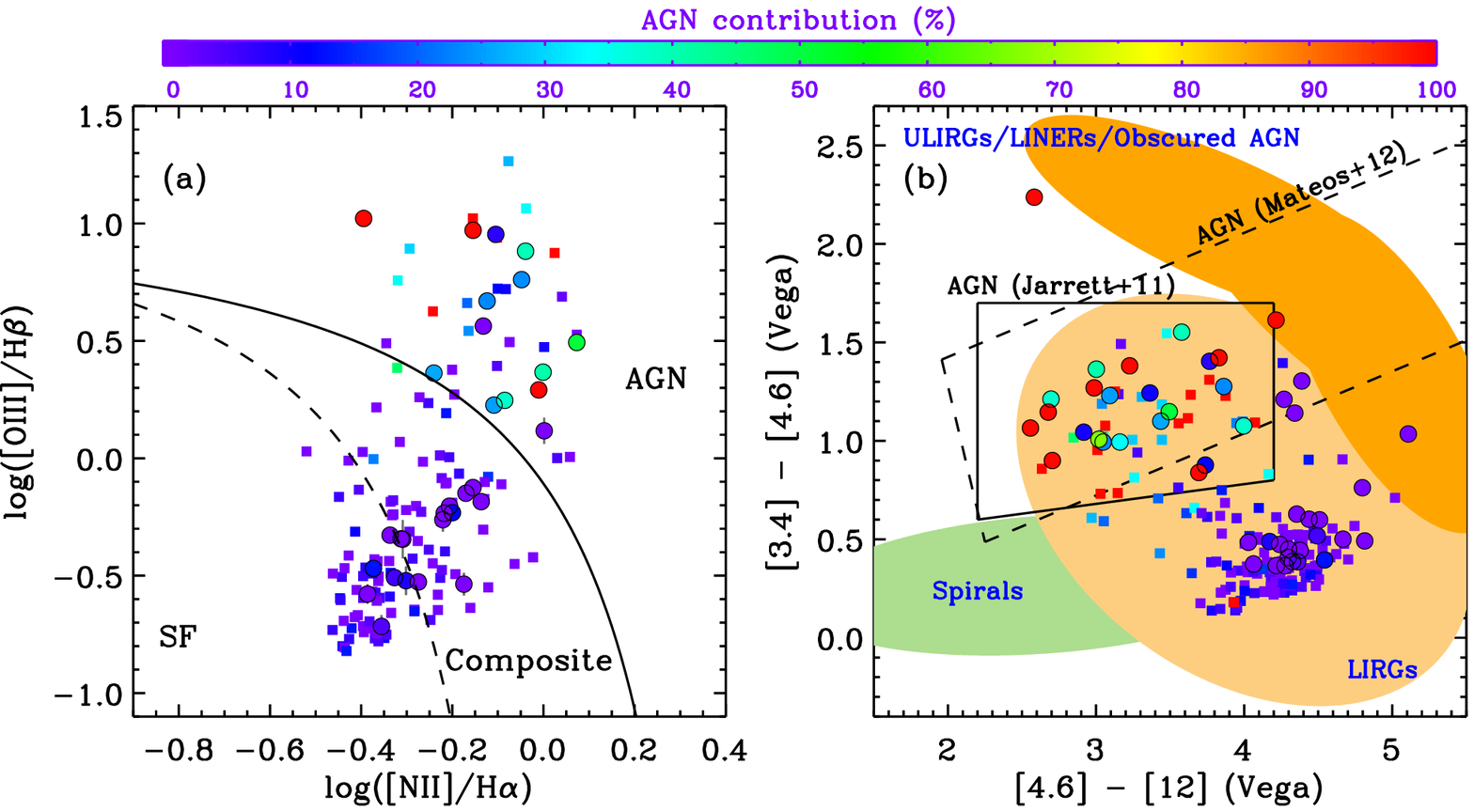}
\caption{AGN diagnostic diagrams for DOGs
   based on optical [O III]/H$\beta$ vs. [N II]/H$\alpha$ line ratios (a)
   and on {\it WISE} colors (b).
Filled circles and squares indicate 
  {\it WISE} 12 $\mu$m selected DOGs and the control sample, respectively.
Different colored symbols represent different 
  AGN contributions measured from the SED decomposition 
  (color coded as shown by the color bar to the top). 
We plot error bars only for local DOGs.
Solid and dashed lines in (a) indicate the extreme starburst \citep{kew01}
  and pure SF limits \citep{kau03agn}, respectively.
Solid and dashed lines in (b) are the MIR AGN selection criteria proposed by 
  \citet{jar11} and \citet{mat12}, respectively.
We mark several regions occupied by different classes of objects
 (spirals, LIRGs and ULIRGs; see Figure 12 in \citealt{wri10}).
}\label{fig-agnfit}
\end{figure*}
%%%%%%%%%%%%%%%%

In Figure \ref{fig-agncont},
  we show the AGN contribution to the total IR luminosities
  of DOGs (red circles) and the control sample (blue squares)
  as a function of total IR luminosity.
We plot FIR detected (filled symbols) and undetected (open symbols)
  galaxies separately.
We show open symbols with arrows to represent the upper/lower limits
  for AGN contribution and total IR luminosities.
There is no correlation between the AGN contribution
  and the total IR luminosity for the DOGs or for the control sample;
  in contrast, ULIRGs do show a correlation \citep{vei09,leejc12akari}.

\subsection{Comparison of AGN Diagnostics}

To compare the measured AGN contribution with other AGN selection methods,
  we plot the optical line ratios of DOGs (circles) and 
  the control sample (squares)
  in the left panel of Figure \ref{fig-agnfit}.
Different colored symbols represent different 
  AGN contributions measured from the SED decomposition 
  (color coded as shown by the color bar to the top).
As expected,
  most galaxies with a small AGN contribution (purple symbols)
  lie in the regions of star-forming and composite galaxies;
  those with a large AGN contribution (greenish and reddish symbols)
  lie mostly in the AGN region.

We use these line ratio diagrams
  to determine the optical spectral classification of the DOGs
  listed in Table \ref{tab-samp} and in Figure \ref{fig-sed}.
We adopt the criteria of \citet{kew06} 
  based on the Baldwin-Phillips-Terlevich (BPT) 
  emission-line ratio diagrams \citep{bpt81,vo87}. 
For galaxies with S/Ns$\geq$3
   in the strong emission-lines H$\beta$, [OIII] $\lambda$5007,
   H$\alpha$, [NII] $\lambda$6584, and [SII]$\lambda\lambda$6717,6731,
   we base the classification on their positions 
   in the line ratio diagrams
   with [OIII]/H$\beta$ plotted against
   [NII]/H$\alpha$, [SII]/H$\alpha$, and [OI]/H$\alpha$.
These classes are star-forming galaxies, Seyferts, 
    low-ionization nuclear emission-line regions (LINERs), 
    composite galaxies, and ambiguous galaxies 
    (see \citealt{kew06} for more details).
Composite galaxies host a mixture of star formation and AGN,
  and lie between the extreme starburst line \citep{kew01} and the
  pure star formation line \citep{kau03agn}
  in the [OIII]/H$\beta$ vs. [NII]/H$\alpha$ line ratio diagram 
  (see the left panel in Figure \ref{fig-agnfit}).
Ambiguous galaxies are those classified as one
  type in one or two diagrams, but as another type
  in the other diagrams. % (see \citealt{kew06} for more details).
For four galaxies with `ambiguous' type (ID: 17, 21, 36 and 40),
  we replace the galaxy classification with the one
  based only on the [OIII]/H$\beta$ vs. [NII]/H$\alpha$ line ratio diagram.
We assign `undetermined' type to those that do not satisfy the S/N criteria.
For seven galaxies that we cannot classify because of a low S/N spectrum
  or the absence of an SDSS spectrum,
  we adopt the classification from NED if available.

In the right panel of Figure \ref{fig-agnfit}, 
  we plot the DOGs and the control sample
  in the \wise color-color diagram.
We mark several regions occupied by different types of objects
  (spirals, luminous infrared galaxies (LIRGs), 
  and ULIRGs; see Figure 12 in \citealt{wri10}).
We also overplot two AGN selection criteria proposed by
  \citet[solid lines]{jar11} and \citet[dashed lines]{mat12}.
Most of the DOGs and the control sample
  are distributed in the region of LIRGs,
  confirming that they are IR luminous objects.
%  confirming that they are indeed dusty objects.
The two AGN selection criteria
  are really efficient in selecting AGN-host galaxies;
  most of the galaxies with a small AGN contribution (purple symbols)
  are outside the AGN selection boxes
  (see also \citealt{stern12,kir13} for other AGN criteria based on \wise colors).

\section{Comparison of Local DOGs with the control Sample}\label{comp}

To study how special the local analogs of DOGs are
  among MIR selected galaxies,
  we compare the physical properties of these DOGs
  with those of the control galaxy sample.
We first compare UV and infrared properties in Section \ref{uvir}.
We then discuss photometric/spectroscopic parameters and 
  environments of DOGs in Sections \ref{parameter} and \ref{environ}, respectively.

%\subsection{UV and Infrared Properties of DOGs and the control Sample}\label{uvir}
\subsection{Dust Obscuration in DOGs}\label{uvir}
%\subsubsection{UV and Infrared Flux Densities of DOGs}\label{ratio}

\begin{figure*}
\center
\includegraphics[width=150mm]{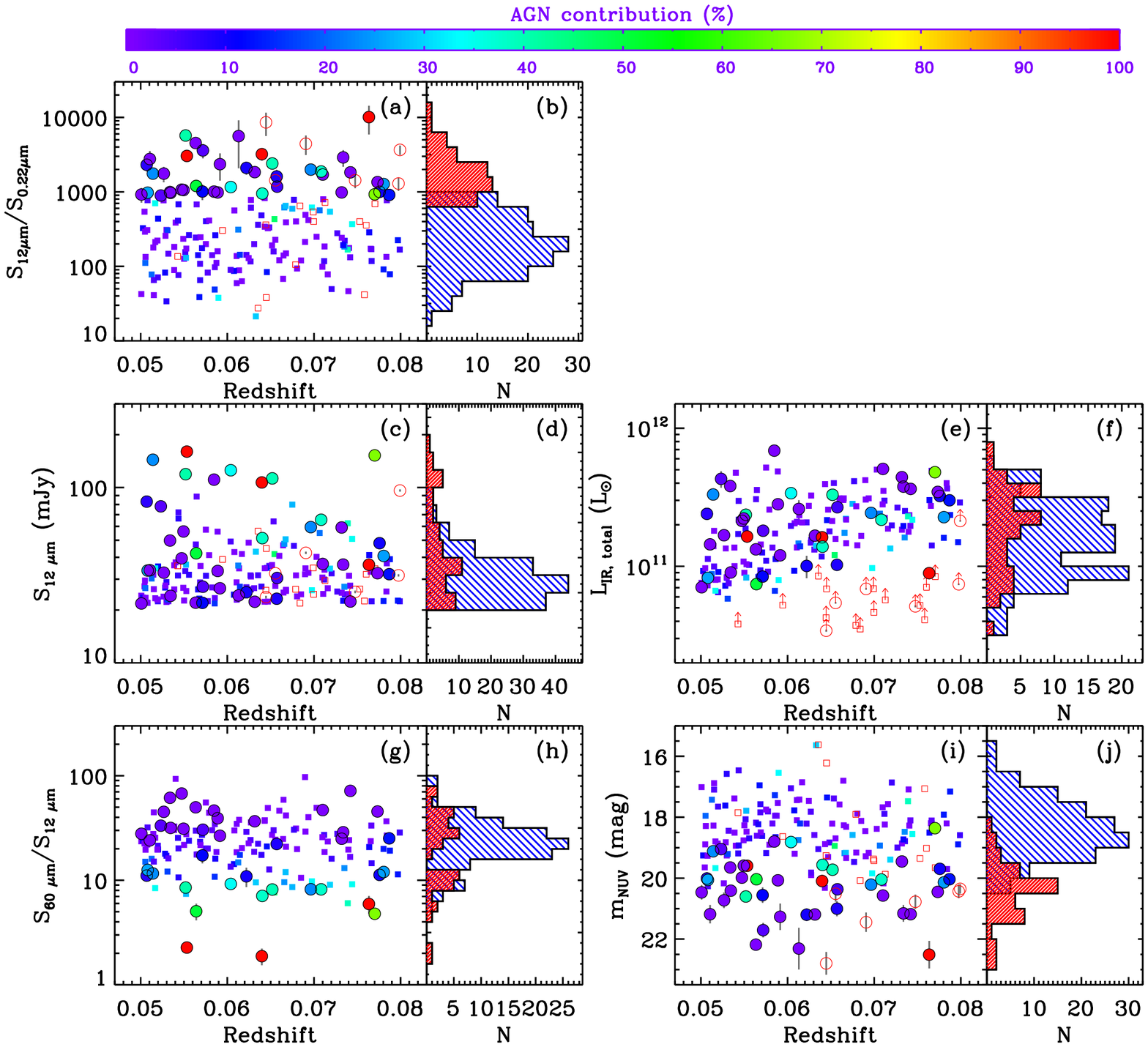}
\caption{Flux density ratios between \wise 12 $\mu$m and \galex NUV
    for local DOGs (circles) and for the control sample (squares)
    as a function of redshift (a), and their histograms (b).
Filled and open symbols indicate
  FIR (i.e., \iras 60 $\mu$m or \akari 90 $\mu$m) 
  detected and undetected galaxies, respectively.
Different colored symbols represent different 
  AGN contributions measured from the SED decomposition 
  (color coded as shown by the color bar to the top).
We plot error bars only for local DOGs.
DOGs and the control sample are denoted by hatched histograms with
  orientation of 45$^\circ$ ($//$ with red color) and 
  of 315$^\circ$ ($\setminus\setminus$ with blue color) 
  relative to horizontal, respectively.
Same as (a-b), but for {\it WISE} 12 $\mu$m flux density (c-d),
  for total IR luminosity (e-f), 
  for flux density ratios between \iras 60 $\mu$m and \wise 12 $\mu$m (g-h), and
  for NUV magnitude (i-j).
Arrows in (e) indicate lower limits to the total IR luminosities.
}\label{fig-flux}
\end{figure*}

To examine observational bias in our sample selection,
  we first show the flux density ratio between 
  {\it WISE} 12 $\mu$m and {\it GALEX} NUV
  as a function of redshift in the top left panel of Figure \ref{fig-flux}.
This flux density ratio is the one 
  we use to select DOGs (see Figure \ref{fig-samp12um}).
DOGs are indicated by circles, and the control sample 
  is indicated by squares.
We plot FIR detected and undetected galaxies
  with filled and open symbols, respectively.  
The figure indicates that
  the flux density ratio ($S_{12 \mu m}/S_{0.22 \mu m}$) shows no dependence 
  either on the measured AGN contribution or on the FIR detection.
However, the majority of FIR undetected galaxies (open symbols)
  have $z>0.06$.
This effect occurs simply because the FIR detection limit increases with redshift
  (see middle right panel).
However, this redshift dependence does not introduce any bias in our results because
  both the DOGs and the control sample are affected in the same way.
  
The middle left panel shows the \wise 12 $\mu$m flux density
  as a function of redshift.
The median flux densities of DOGs and the control sample
  are similar ($36\pm5$ vs. $30\pm2$ mJy), but
  the dispersion is much larger for DOGs than for the control sample.
There are some bright DOGs with $S_{12 \mu m}>100$ mJy.
However, there are no such bright galaxies among the control sample
  even though there are initially some non-DOGs 
  with $S_{12 \mu m}>100$ mJy in the top panel of Figure \ref{fig-samp12um}.
Most relatively faint DOGs (e.g., $S_{12 \mu m}<40$ mJy)
  have a low AGN contribution (bluish symbols),
  but many bright DOGs contain a
  significant AGN contribution (green or red symbols).
This result is similar to the one for high-$z$ DOGs;
  bright 24 $\mu$m high-$z$ DOGs
  are more AGN-dominated \citep{dey08}.

We note that high 12 $\mu$m flux densities do not necessarily imply
  high IR luminosities because IR luminosity estimates
  are more sensitive to FIR data
  and depend on detailed SED modeling.
Thus we do not see any dependence of IR luminosity
  on AGN contribution for FIR detected galaxies (filled symbols)
  in the middle right panel.
The middle right panel also shows that
  FIR undetected galaxies (open symbols)
  have lower IR luminosities than
  FIR detected galaxies (filled symbols), 
  mainly as a result of the FIR detection limits.

The different AGN contribution among DOGs
  is more evident in the plot of 
  flux density ratio between \iras 60 $\mu$m and \wise 12 $\mu$m
  (bottom left panel).
The control sample 
  shows a scatter around the peak at a ratio of 20 (see blue histogram in (h)).
However, the histogram of DOGs shows a clear bimodal 
  distribution (red histogram in (h)).
Most DOGs with a small AGN contribution (i.e., $f_{\rm AGN}<20\%$)
  have large $S_{60 \mu m}/S_{12 \mu m}$ (i.e., $>$15);
  they are distributed in a range similar to the control sample.
However, DOGs with a large AGN contribution
  have small $S_{60 \mu m}/S_{12 \mu m}$ (i.e., $<$15).
This bifurcation is also apparent for 
  the flux density ratios between \iras 100 $\mu$m 
  and \wise 12 $\mu$m (not shown here).
The presence of two types of DOGs in the local universe
  (i.e., DOGs with small $S_{60 \mu m}/S_{12 \mu m}$ and
  large AGN contribution
  vs. DOGs with large $S_{60 \mu m}/S_{12 \mu m}$ and 
  small AGN contribution; 
  see also middle left panel for $S_{12 \mu m}$ in Figure \ref{fig-flux})
  is similar to the situation for high-$z$ DOGs;
  there are power-law (AGN-dominated and 24 $\mu$m bright) and 
  bump (SF-dominated and 24 $\mu$m faint) DOGs \citep{dey08}.

The bottom right panel indicates
  that NUV magnitudes of DOGs are systematically fainter 
  than for the control sample of galaxies.
This difference suggests that 
  the systematic UV faintness of DOGs is mainly responsible for 
  the extreme ratios between 
  {\it WISE} 12 $\mu$m and {\it GALEX} NUV flux densities.
The NUV magnitude distribution for DOGs with a large AGN contribution 
  (e.g., $f_{\rm AGN}\geq20\%$) is
  the same as for those with a small AGN contribution 
  (e.g., $f_{\rm AGN}<20\%$).
A K-S test confirms this impression.

If one of the bands between \galex NUV and \wise 12 $\mu$m
  is more important than the other band
  in explaining the extreme $S_{12 \mu m}/S_{0.22 \mu m}$, % for DOGs,
  the distribution of DOGs should be significantly different from
  the control sample only in one panel.
The middle left panel in Figure \ref{fig-flux}
  shows that the flux density 
  $S_{12 \mu m}$ for DOGs is on average
  larger than for the control sample.
The ratio of the medians of the two samples is 1.2.
The K-S test also rejects the hypothesis that 
  the $S_{12 \mu m}$ distributions of the two samples
  are extracted from the same parent population
  with a confidence level of 99.8\%.
The bottom right panel shows that the flux density
  $S_{2.17 \mu m}$ for the DOGs is again
  different from the control sample.
The flux density ratio of the medians of the two samples is 5.8,
  much larger than for $S_{12 \mu m}$.

This result indicates that
  both $S_{12 \mu m}$ and $S_{0.22 \mu m}$ are responsible for 
  the extreme $S_{12 \mu m}/S_{0.22 \mu m}$ flux density ratio for DOGs.
However, the larger difference in $S_{0.22 \mu m}$
  than for $S_{12 \mu m}$ (i.e., factors of 5.8 vs. 1.2)
  strongly suggests that
  the UV faintness of DOGs is the main factor leading to the 
  extreme $S_{12 \mu m}/S_{0.22 \mu m}$.

\subsection{Optical Structure and Star Formation Activity of DOGs}\label{parameter}

\begin{figure*}
\center
\includegraphics[width=150mm]{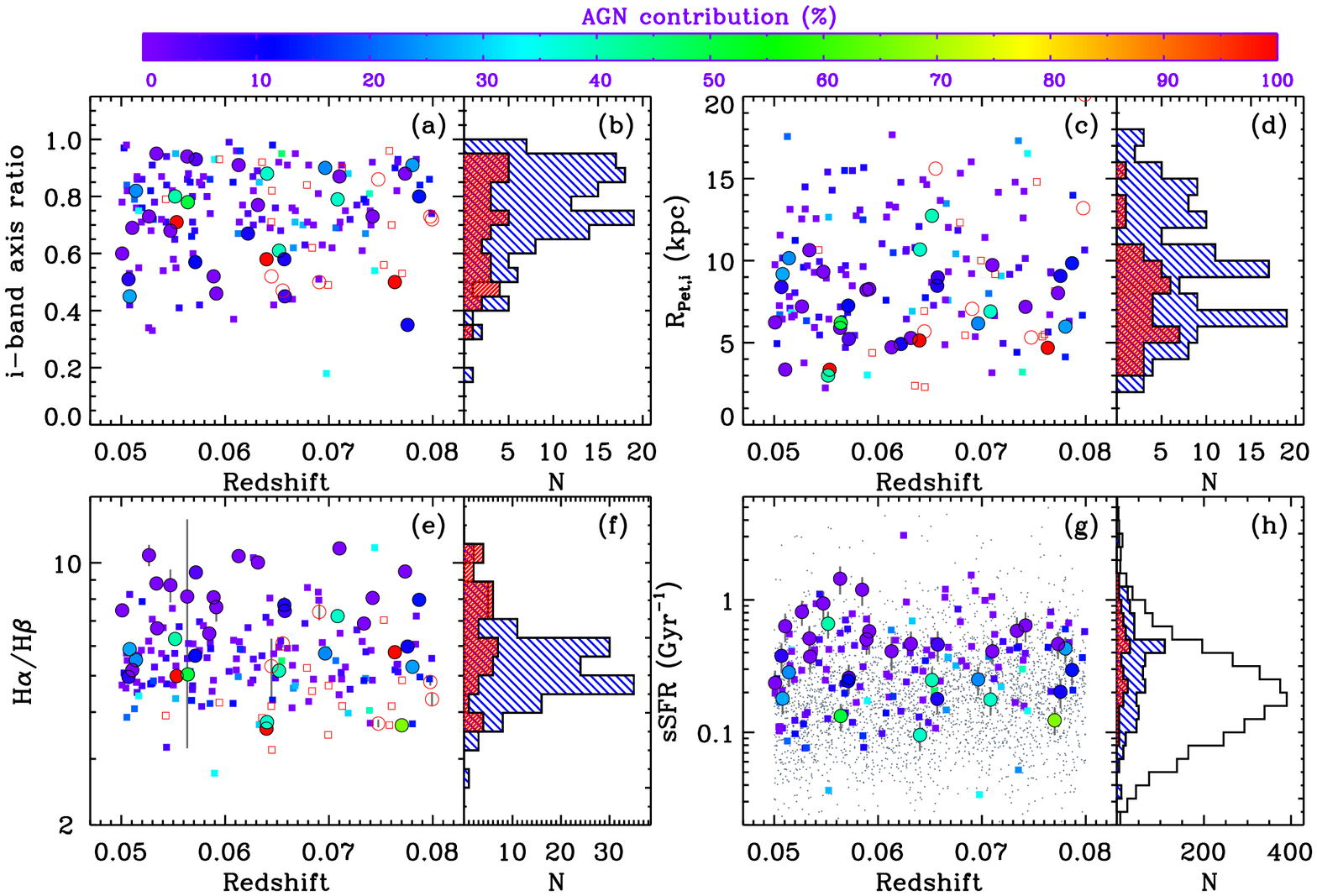}
\caption{Same as Figure \ref{fig-flux}, but for
  $i$-band axis ratio (a-b),
  $i$-band Petrosian radius (c-d), 
   Balmer decrement (H$\alpha$/H$\beta$) (e-f), and
  sSFRs (g-h).
We plot error bars only for local DOGs in (e) and (g).
In (g-h), FIR detected galaxies
  are shown as gray dots and as an open histogram. 
To better show the histograms in (h),
  we multiply the histograms of DOGs and the control sample of galaxies by five.
}\label{fig-opt}
\end{figure*}

We plot several photometric and spectroscopic parameters
  for the DOGs and for the control sample in Figure \ref{fig-opt}.
  
The top left panel shows
  $i$-band axis ratios of DOGs and the control sample.
The ratio is the seeing-corrected $i$-band isophotal axis ratio
  adopted from the Korea Institute for Advanced Study (KIAS) 
  VAGC\footnote{http://astro.kias.re.kr/vagc/dr7/} \citep{choi10}.
A K-S test for the axis ratio distributions of DOGs and the control sample
  rejects the hypothesis that the axis ratio distributions of the two samples
  are extracted from the same parent population
  with only a confidence level of 88\%. 
However, a Wilcoxon Rank-Sum test
  rejects the hypothesis that the two samples
  have the same mean with a confidence level of 96\%.
One interesting feature in this figure is that
  the fraction of DOGs with small axis ratios (e.g., $\leq0.6$) %i=53 degree
  among DOGs is larger ($36\pm7$\%) than
  that among the control sample of galaxies ($17\pm3$\%).
The large dust obscuration of some DOGs 
  may simply result from the high inclination of disk galaxies
  (see also the color images in Figure \ref{fig-chart}).

%Figure 9 %%%%%%%%%%%%%%%%%%%%
%%%%%%%%%%%%%%%%

We also plot $i$-band Petrosian radii of the DOGs (i.e., galaxy-size indicator)
  as a function of redshift in the top right panel. 
The Petrosian radius is adopted from the KIAS VAGC,
  calculated using elliptical annuli.
This estimate is typically larger than the value
  based on circular annuli in the SDSS 
  photometric database \citep{choi07}.
The majority of DOGs are smaller than $\sim$10 kpc;
  the control sample has a wider range of sizes.
A K-S test rejects
  the hypothesis that the Petrosian radius
  distributions of the two samples are extracted
  from the same parent population with a confidence level of 99\%.
On the other hand, the K-S test for the distributions of Petrosian radii
  between DOGs with large and small AGN contribution
  (i.e., $f_{\rm AGN}\geq20\%$ vs. $f_{\rm AGN}<20\%$)
  indicates that the distributions of the two samples
  are not different.

The bottom left panel shows
  the flux ratios between H$\alpha$ and H$\beta$ (i.e., Balmer decrement)
  as a function of redshift.
As expected, 
  most DOGs and the control sample have H$\alpha/$H$\beta$ values
  larger than an intrinsic H$\alpha/$H$\beta$ ratio of 3.1
  for AGN-dominated galaxies and 
  H$\alpha/$H$\beta=$2.86 for SF-dominated galaxies
  (in the nominal case B recombination
  for $T=10,000$ K and $n_e\approx 10$ cm$^{-3}$ 
  with no dust, \citealt{ost06}).
Not surprisingly,
  the DOGs have on average larger values than the control sample.
A K-S test supports this with a confidence level of 99.9\%.
The majority of DOGs with a small AGN contribution (purple symbols) 
  have larger H$\alpha/$H$\beta$ values than the DOGs 
  with a large AGN contribution (green and red symbols),
  suggesting that the DOGs with a small AGN contribution
  are more heavily obscured.

%Figure 10 %%%%%%%%%%%%%%%%%%%%
%%%%%%%%%%%%%%%%

In bottom right panel, 
  we compare sSFRs of DOGs with those of the control sample.  
We convert the IR luminosity (after removing the AGN contribution) into a SFR
  based on the relation in \citet{ken98} with the assumption of a Salpeter IMF:
  SFR ($M_\odot$ yr$^{-1}$) $= 1.72\times10^{-10}L_{\rm IR} (L_\odot)$. 
Note that we also use the Salpeter IMF for stellar mass estimates 
  (see Section \ref{compsamp}).
The K-S test for DOGs and the control sample rejects
  the hypothesis that the sSFRs of the two samples are extracted
  from the same parent population with only a confidence level of 65\%,
  indicating no significant difference between the two samples.
The sSFRs of DOGs with a large AGN contribution 
  (e.g, $f_{\rm AGN}\geq20\%$) tend to
  be smaller than those of DOGs with a small AGN contribution 
  (e.g, $f_{\rm AGN}<20\%$).
This conclusion is supported by the K-S test with a confidence level of 99\%.

To compare the sSFRs of DOGs with 
  the general trend for IR bright galaxies,
  we also plot FIR detected SDSS galaxies (gray dots) %at the same redshift range
  regardless of their 12 $\mu$m flux densities.
The DOGs %and the control sample of galaxies
  tend to have larger sSFRs than the FIR detected galaxies.
This result is also confirmed by the K-S test that rejects
 the hypothesis that sSFRs of the two samples are extracted
 from the same parent population with a confidence level of 99.9\%.
However, the median sSFRs of DOGs and the control sample are larger
  than those of FIR detected galaxies by only a factor $\lesssim2$,
  suggesting that the star formation modes of DOGs and 
  the control sample
  are not significantly different from typical IR bright galaxies.
In other words, the small difference in sSFRs between
  DOGs and FIR detected galaxies indicates that most DOGs 
  are not starburst systems with much larger 
  (i.e., factors $>2$) sSFRs than typical IR bright galaxies \citep{elb11}.

%Figure 11 %%%%%%%%%%%%%%%%%%%%
\begin{figure*}
\center
\includegraphics[width=150mm]{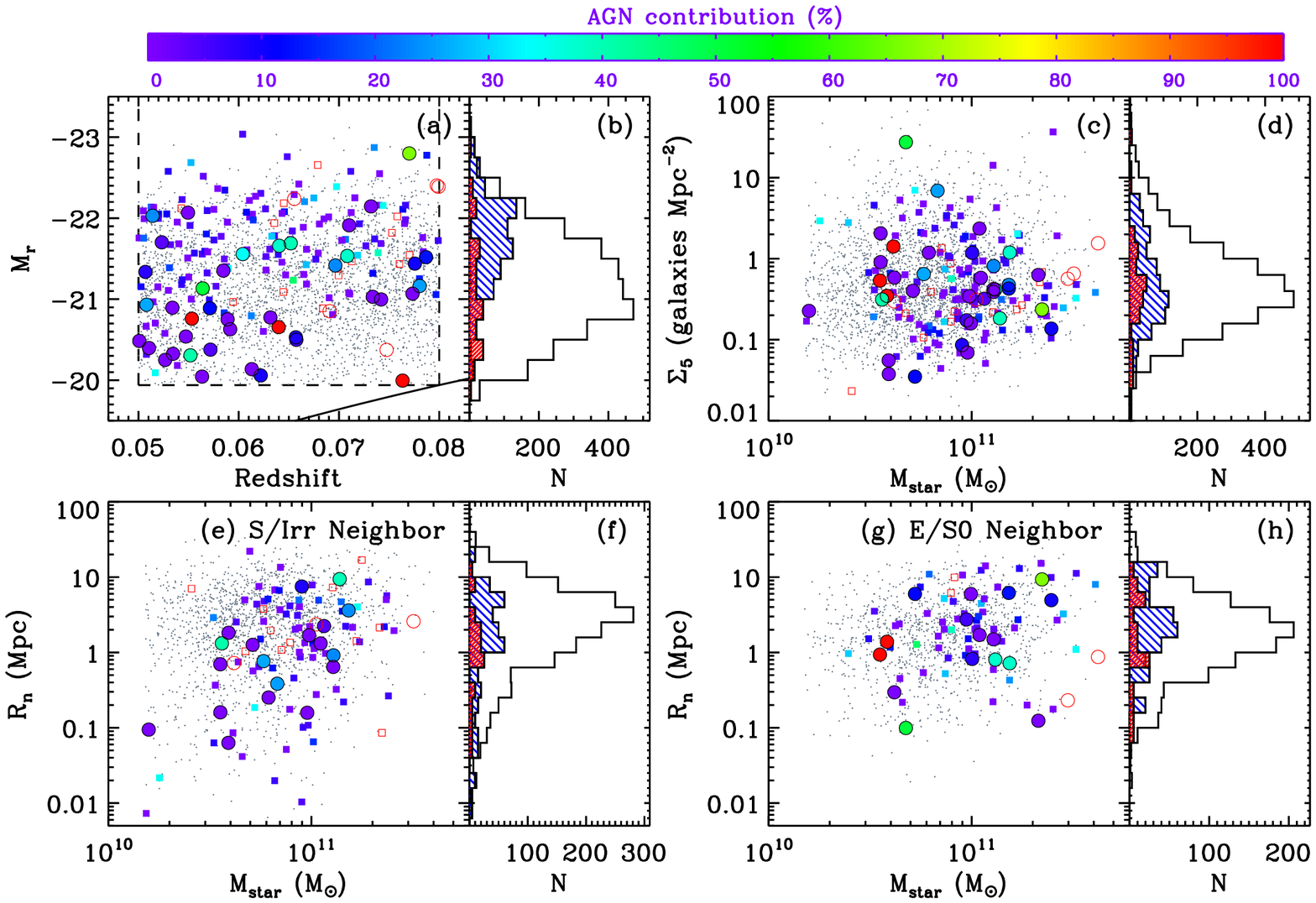}
\caption{Same as Figure \ref{fig-flux}, but for
  $r$-band absolute magnitude (a-b),
  surface galaxy number density ($\Sigma_5$) (c-d),
  distance to the nearest neighbor galaxy for
  late-type neighbor case (e-f)
  and for early-type neighbor case (g-h).  
Environmental parameters in panels (c, e and g)
  are plotted as a function of stellar mass.
FIR detected galaxies
  are shown as gray dots and as an open histogram. 
Dashed lines in (a) define the volume-limited sample.
The bottom solid curve corresponds to the apparent magnitude limit ($m_r=17.77$)
  for the main galaxy sample in SDSS \citep{choi07}.
To better show the histograms,
  we multiply the histograms of DOGs and the control sample of galaxies by five.
}\label{fig-env}
\end{figure*}
%%%%%%%%%%%%%%%%

\subsection{Large- and Small-scale Environments of DOGs}\label{environ}

Because star formation or nuclear activity of galaxies
  is strongly related to their environment
  (e.g., \citealt{pc09, bm09, hwa12agn}),
  we next compare the environments of DOGs with those
  of the control sample.
Here we consider two environmental indicators:
  a surface galaxy number density estimated from 
    the five nearest neighbor galaxies ($\Sigma_5$)
    as a large-scale environmental parameter, and
  a distance to the nearest neighbor galaxy ($R_{\rm n}$)
    as a small-scale environmental parameter.
 
The background density, $\Sigma_5$, 
  is defined by $\Sigma_5=5(\pi D^2_{p,5})^{-1}$,
  where $D_{p,5}$ is the projected distance to the 5th-nearest neighbor.
The fifth-nearest neighbor to each target galaxy is identified 
  in a volume-limited sample ($0.05<z<0.08$ and $M_r\leq-19.94$,
  dashed line in the top left panel of Figure \ref{fig-env})
  among the neighbor galaxies 
  that have relative velocities
  $\Delta v=|v_{\rm neighbors}- v_{\rm target}|<1500$ km s$^{-1}$
  to exclude foreground and background galaxies.
  
To define the small-scale environmental parameter attributed to the nearest neighbor,
  we first search for the nearest neighbor of a target galaxy
  that is the closest to the target galaxy projected on the sky
  and that satisfies conditions on magnitude and relative velocity.
We search for the nearest neighbor galaxy among galaxies
  with magnitudes brighter than $M_r=M_{r,\rm target}+0.5$ and 
    with relative velocities less than 
    $\Delta v=600$ km s$^{-1}$
    for early-type target galaxies and less than $\Delta \upsilon=400$ km s$^{-1}$ 
    for late-type target galaxies.
These velocity limits cover most close neighbors including 
  dusty star-forming galaxies in the local universe
  (see Figure 2 of \citealt{bar00} and Figure 1 of \citealt{park08}).
The magnitude criterion selects galaxies mostly in major interacting pairs;
  these objects should be the most effective 
  in triggering SFA \citep{woods07,cox08,hwa10lirg}.
To have a fair sample of neighbor galaxies in our sample,
   we select target galaxies
   among those with $M_{r,\rm target}=-19.94-0.5$
   where $M_r\leq-19.94$ is the magnitude limit of the volume-limited sample
   (see top left panel in Figure \ref{fig-env}).

The top left panel of Figure \ref{fig-env}
  shows the $r$-band absolute magnitudes of DOGs (circles) and 
  the control sample (squares).
For comparison, we also plot the FIR detected SDSS galaxies (gray dots).
The distribution of DOGs 
  does not differ from FIR detected galaxies.
However, the DOGs tend be fainter than the control sample,
  confirmed by a K-S test with a confidence level of 99.9\%.
The distributions of redshift and stellar mass 
  for DOGs and the control sample are similar (see Figure \ref{fig-lirmass}).
Thus, the different $r$-band absolute magnitude distributions
  simply reflect the difference in the amount of dust extinction;
  DOGs are more dust obscured by definition.
Thus, they are optically fainter than the control sample.

The top right panel
  shows $\Sigma_5$ distributions of DOGs, the control sample,
  and FIR detected SDSS galaxies as a function of stellar mass.
The $\Sigma_5$ distribution of DOGs does not differ
  from the control sample or from the FIR detected galaxies,
  as confirmed by a K-S test.
FIR detected galaxies are usually absent in high-density regions 
  in the local universe because galaxies in these regions 
  have lost (or consumed) the gas/dust, 
  necessary for their IR activity \citep{hwa10lirg}.
DOGs and the control sample, similar to FIR detected galaxies,
  are also mostly located in low-density regions.
  
The $\Sigma_5$ distribution of DOGs with a large AGN contribution 
  (e.g., $f_{\rm AGN}\geq20\%$)
  is similar to those with a small AGN contribution 
  (e.g., $f_{\rm AGN}<20\%$).
Although there are two DOGs in relatively high-density regions 
  (e.g., $\Sigma_5>6$ Mpc$^{-2}$) and
  both of them have a large AGN contribution,
  the K-S test cannot reject
  the hypothesis that the $\Sigma_5$ distributions of the two samples
  are extracted from the same parent population
  (but see \citealt{bro08} for stronger clustering for bright 
  (probably more AGN-dominated) DOGs at $z\sim2$).

In the bottom panels,
  we plot the projected distance
  to the nearest neighbor galaxy of DOGs
  as a function of their stellar mass.
Because galaxy properties are strongly correlated with 
  the morphological type of the nearest neighbor galaxy \citep{pc09,hwa10lirg},
  we plot the distribution of neighbor separation separately
  according to their neighbor's morphology:
  late-type neighbor case (bottom left panel) and 
  early-type neighbor case (bottom right panel).
We adopt galaxy morphology data from the KIAS VAGC.

The bottom panels show that 
  there are more close pairs (e.g., $R_n<0.1$ Mpc)
  for the late-type neighbor case (bottom left panel)
  than for the early-type neighbor case (bottom right panel).
This is true for all the samples (DOGs, control sample and
  FIR detected galaxies),
  confirming the dependence of IR activity of dusty galaxies on the
  morphological type (i.e., cold gas fraction) of 
  the nearest neighbor galaxy \citep{hwa10lirg}.
Moreover,
  both panels show that the majority of DOGs
  have neighbor separation $R_n\gtrsim0.3$ Mpc,
  indicating that the majority of DOGs are not currently interacting galaxies. 
However, it does not necessarily imply
  that galaxies with large pair separation 
  are currently quiescent systems
  (see Section \ref{discuss} for details).
The neighbor separation distribution of DOGs
  does not differ from the control sample or
  from the FIR detected galaxies.

%Figure 12 %%%%%%%%%%%%%%%%%%%%
\begin{figure*}
\center
\includegraphics[width=160mm]{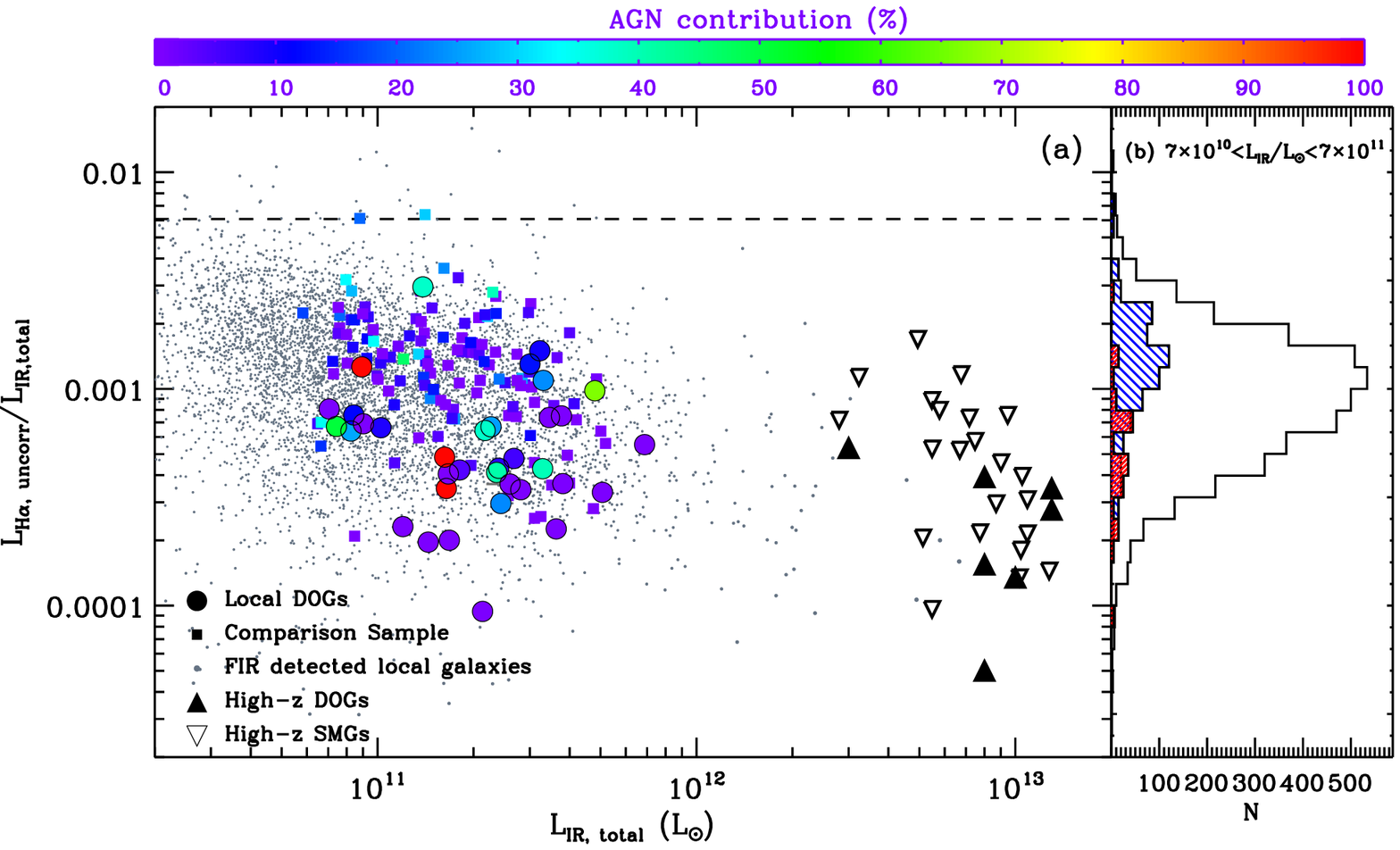}
\caption{L$_{\rm H\alpha, uncorr}$/L$_{\rm IR, total}$ vs. L$_{\rm IR, total}$
  for local DOGs (circles) and the control sample (squares) (a),
  and their histograms (b).
$L_{\rm H\alpha, uncorr}$ is observed H$\alpha$ luminosity
  after aperture correction, but without dust-extinction correction.
Different colored symbols represent different 
  AGN contributions measured from the SED decomposition 
  (color coded as shown by the color bar to the top).
We plot error bars only for local DOGs (not visible because of large symbols).
For comparison, we also plot the FIR detected SDSS galaxies
  at $z>0.05$ (gray dots),
  high-$z$ DOGs (filled triangles, \citealt{brand07}) and 
  SMGs (open upside down triangles, \citealt{swi04}).
Dashed line indicates when H$\alpha$ and L$_{\rm IR, total}$ 
  SFR indicators would agree.
To have a fair comparison in (b),
  we use only galaxies 
  at $7\times10^{10}<L_{\rm IR, total}/L_{\odot}<7\times10^{11}$.
Local DOGs and the control sample are 
  denoted by hatched histograms with
  orientation of 45$^\circ$ ($//$ with red color) and 
  of 315$^\circ$ ($\setminus\setminus$ with blue color) 
  relative to horizontal, respectively.
The open histogram refers to FIR detected SDSS galaxies.
To better show the histograms,
  we multiply the histograms of 
  local DOGs and the control sample by five.
}\label{fig-extin}
\end{figure*}
%%%%%%%%%%%%%%%%

\section{Comparison of Local and High-z DOGs}\label{highz}

Because of the limited observational data for SDSS galaxies,
  the selection criteria for the local DOGs are not exactly identical
  to those of high-$z$ DOGs.
However, both local and high-$z$ DOGs have 
  extreme flux density ratios between rest-frame MIR and UV.
These local DOGs are reasonable local analogs of high-$z$ DOGs.
In this section,
  we compare physical properties of local and high-$z$ DOGs.

Before we compare local DOGs with high-$z$ ones,
  we emphasize that
  the IR luminosities of current samples of local and high-$z$ DOGs differ.
The high-$z$ DOGs we discuss
  are mainly from \citet{dey08} who identified $\sim$2600 DOGs
  with $S_{24 \mu m}\geq300$ $\mu$Jy
  in the NOAO Deep Wide-Field Survey Bo\"{o}tes field,
  from \citet{saj12} who identified 26 DOGs
  with $S_{24 \mu m}\geq890$ $\mu$Jy in the 
  {\it Spitzer} Extragalactic First Look Survey 
  field\footnote{http://ssc.spitzer.caltech.edu/fls/}, and
  from \citet{pen12} who identified 
  $\sim$60 DOGs with $S_{24 \mu m}\geq53$ $\mu$Jy in the 
  Great Observatories Origins Deep Survey (GOODS).
Although the stellar masses of local and high-$z$ DOGs are similar,
  the IR luminosities of most high-$z$ DOGs 
  exceed $10^{12}$ $L_{\odot}$ (i.e., ULIRGs)
  simply because of the detection limits \citep{buss12,mel12,saj12,pen12}.
However, the IR luminosities of local DOGs
  are an order of magnitude lower than for the high-$z$ DOGs 
  (see Figure \ref{fig-lirmass}). 
We keep this luminosity difference between the two samples
   in mind for the following analysis.

It should also be noted that
  the universe at $z\sim2$ differs from that at $z\sim0$.
The main relevant difference is the gas (or dust) fraction 
  in galaxies \citep{dad10,tac10,mag12,sar13}.
Because the gas fraction in galaxies at $z\sim2$ is greater
  than for at $z\sim0$ (also dust),
  galaxies at $z\sim2$ can more easily be IR luminous 
  than those at $z\sim0$
  without invoking galaxy-galaxy interactions or mergers \citep{dad10sflaw,elb11}.

%Figure 13 %%%%%%%%%%%%%%%%%%%%
\begin{figure*}
\center
\includegraphics[width=160mm]{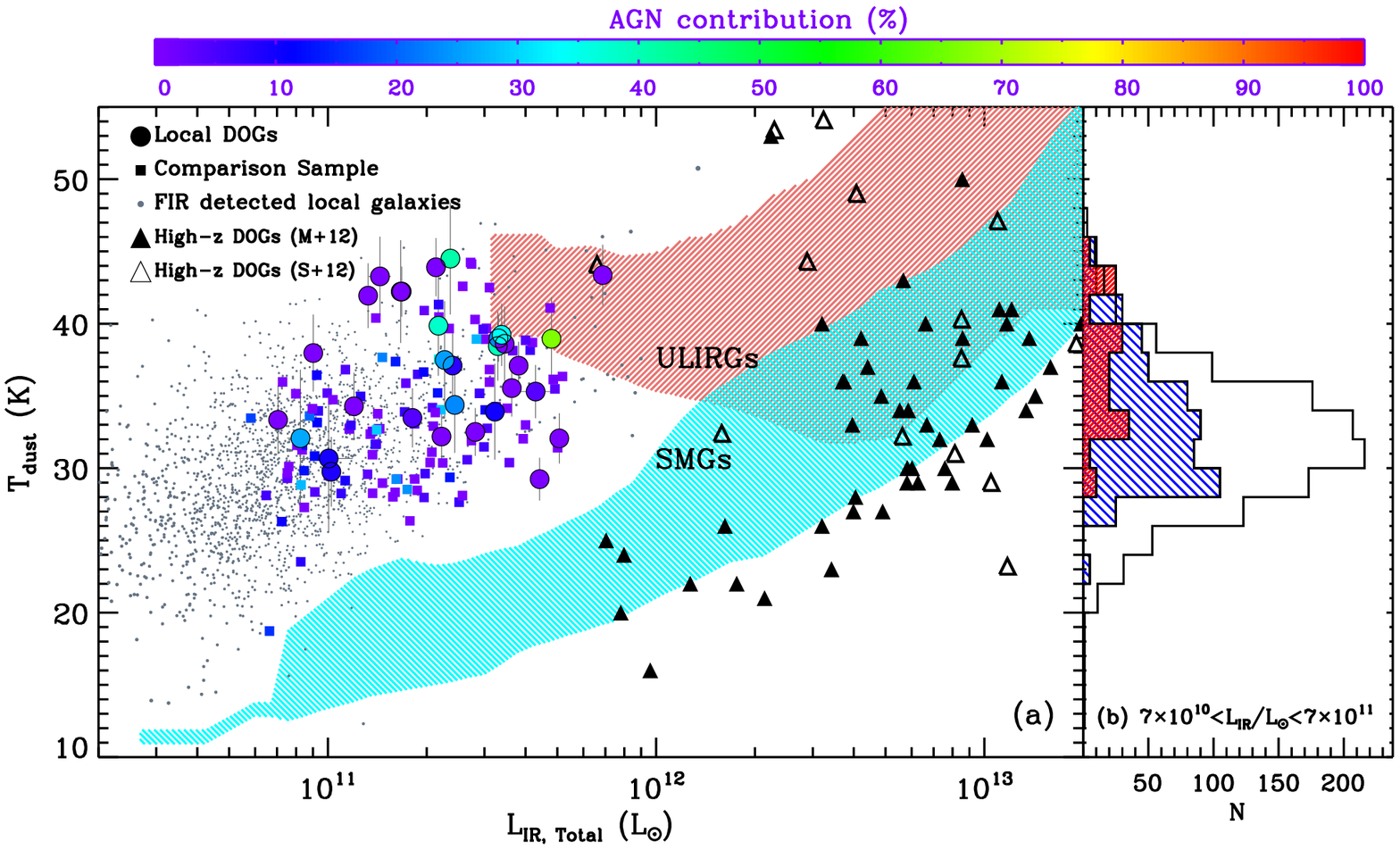}
\caption{Dust temperature vs. total IR luminosity
  for local DOGs (circles) and for the control sample (squares) (a), 
  and their histograms (b).
Different colored symbols represent different 
  AGN contributions measured from the SED decomposition 
  (color coded as shown by the color bar to the top).
We plot error bars only for local DOGs.
For comparison, we also plot FIR detected SDSS galaxies
  at the same redshift range as local DOGs (i.e., $0.05<z<0.08$, gray dots),
  and high-$z$ DOGs (filled triangles from \citealt{mel12} and open triangles
  from \citealt{saj12}).
The loci of ULIRGs at $0.1\lesssim z\lesssim3$ 
  \citep{yang07,you09td,gmag10bumpy} 
  and high-$z$ SMGs \citep{chap05,kov06,bmag12}
  are shown as regions filled by pink and blue color, respectively.
To have a fair comparison in (b),
  we use only galaxies 
  at $7\times10^{10}<L_{\rm IR, total}/L_{\odot}<7\times10^{11}$.
Local DOGs and the control sample are denoted by hatched histograms with
  orientation of 45$^\circ$ ($//$ with red color) and 
  of 315$^\circ$ ($\setminus\setminus$ with blue color) 
  relative to horizontal, respectively.
The open histogram refers to FIR detected local galaxies.
To better show the histograms,
  we multiply the histograms of 
  local DOGs and the control sample by five.
}\label{fig-lirtdust}
\end{figure*}
%%%%%%%%%%%%%%%%

\subsection{The AGN Contribution in DOGs}

Among many similarities between local and high-$z$ DOGs,
  the most striking is
  that both have two subclasses: AGN- and SF-dominated DOGs.
For example,
  high-$z$ DOGs are usually grouped into two classes
  based on their rest-frame NIR/MIR SEDs;
  there are bump and power-law DOGs \citep{dey08}.
Power-law DOGs are relatively bright at 24 $\mu$m,
  and contain an AGN dust component in contrast with bump DOGs.
Similarly, local DOGs can be divided into two classes.
The local DOGs with a large AGN contribution are bright at 12 $\mu$m.
They also contain an AGN dust component
  similar to power-law high-$z$ DOGs 
  (see middle left and bottom left panels in Figure \ref{fig-flux}).

Secondly, the sSFRs of local DOGs with a small AGN contribution
  are larger than those with a large AGN contribution 
  (see bottom right panel in Figure \ref{fig-opt}).
\citet{mel12} found a similar result for high-$z$ DOGs.
They use IR8 value ($\equiv L_{\rm IR}/\nu L_\nu$(8 $\mu$m)),
  a good proxy for sSFRs of IR bright galaxies 
  in a given redshift \citep{elb11}, as the basis for their conclusion.

Another interesting comparison is that
  galaxy size might depend on the AGN contribution.
For high-$z$ DOGs,
  some studies suggest that
  power-law objects tend to be smaller and more concentrated
  than bump objects \citep{mel09,don10,buss09hstp,buss11hstb}.
However, the top right panel in Figure \ref{fig-opt} 
  suggests no size difference between 
  local DOGs with small and large AGN contribution.
It is interesting to note that
  \citet{buss11hstb} found similar sizes 
  for high-$z$ bump and power-law DOGs
  when they considered only the objects with $H<22.5$ (mag).
Recent deep, near-infrared surveys using {\it HST} Wide Field Camera 3 (WFC3)
  including the Cosmic Assembly Near-Infrared Dark Energy Legacy Survey 
  (CANDELS; \citealt{gro11})
  will be useful for resolving this issue 
  by increasing the statistics for high-$z$ DOGs.

\subsection{Dust Obscuration of Local and High-$z$ DOGs}
 
Both local and high-$z$ DOGs are, by definition,
  galaxies with large dust obscuration that
  also affects their optical spectra.
The Balmer decrement of local DOGs 
  is in the range H$\alpha$/H$\beta\approx$ 3.6$-$11 
  (see bottom left panel in Figure \ref{fig-opt}),
  equivalent to $A$(H$\alpha$)$\approx0.7-3.8$
  if we assume the \citet{cal00} extinction curve
  and the intrinsic ratio of H$\alpha/$H$\beta\approx$2.86 
  (case B recombination for $T=10,000$ K and $n_e\approx 10$ cm$^{-3}$ 
  with no dust; \citealt{ost06}).
These amounts of extinction are slightly smaller than those
  of high-$z$ DOGs (e.g., $A$(H$\alpha$)$\approx2.4-4.6$ 
  for three high-$z$ DOGs, see \citealt{brand07}).
However, the amount of dust extinction can depend on IR luminosity.
Therefore, it is not clear whether
  this difference in the amount of dust extinction
  between local and high-$z$ DOGs
  results from evolution in dust properties or
  from different IR luminosities.
  
To distinguish these two effects,
  we plot the ratio between H$\alpha$ and IR luminosities
  of DOGs ($L_{\rm H\alpha, uncorr}/L_{\rm IR,total}$)
  as a function of IR luminosity in Figure \ref{fig-extin}.
Because the Balmer decrement is not available for many high-$z$ DOGs,
  we use $L_{\rm H\alpha, uncorr}/L_{\rm IR,total}$ 
  as a proxy for dust extinction.
Here $L_{\rm H\alpha, uncorr}$ is the observed H$\alpha$ luminosity
  after aperture correction, but without any dust-extinction correction.
To compare the behavior of DOGs with
  the general trend for IR bright galaxies,
  we also plot the FIR detected SDSS galaxies (gray dots).

One interesting feature in Figure \ref{fig-extin} is that
  $L_{\rm H\alpha, uncorr}/L_{\rm IR,total}$ of galaxies
  decreases (i.e., more extincted) with increasing IR luminosity.
As expected,
  $L_{\rm H\alpha, uncorr}/L_{\rm IR,total}$ of local DOGs (filled circles)
  is smaller (i.e., more extincted)
  than for the control sample (filled squares) or
  for the FIR detected galaxies (gray dots).
On the other hand,
  $L_{\rm H\alpha, uncorr}/L_{\rm IR,total}$ of local DOGs (filled circles) is larger
  (i.e., less extincted) than for high-$z$ DOGs (filled triangles).
Because there are no local DOGs 
  in the same IR luminosity range as high-$z$ DOGs,
  it is difficult to tell whether the difference 
  in $L_{\rm H\alpha, uncorr}/L_{\rm IR,total}$
  between local and high-$z$ DOGs
  results from evolution in dust properties or
  from different IR luminosities.
However, FIR detected galaxies at $L_{\rm IR,total}>10^{12}$ ($L_\odot$)
  seem to be smoothly connected to high-$z$ DOGs (filled triangles),
  suggesting that the difference in $L_{\rm H\alpha, uncorr}/L_{\rm IR,total}$
  between local and high-$z$ DOGs 
  results mainly from different IR luminosities rather than from
  evolution in dust properties.
Interestingly,
  the $L_{\rm H\alpha, uncorr}/L_{\rm IR,total}$ distribution of local DOGs 
  (filled circles) is similar to that for submillimeter galaxies at $z\sim2$ 
  (open upside down triangles, \citealt{swi04}),
  even though their IR luminosities differ
  by more than an order of magnitude.

\subsection{Dust Temperature of Local and High-$z$ DOGs}
  
To further compare the dust properties of local and high-$z$ DOGs,
  we plot their dust temperature as a function of IR luminosity
  in Figure \ref{fig-lirtdust}.
We estimate the dust temperature of our sample galaxies
  including local DOGs (filled circles), the control sample (filled squares),
  and FIR detected SDSS galaxies (gray dots)
  by converting the flux density ratio of \iras 60 $\mu$m and 100 $\mu$m
  into a dust temperature.
We use the following conversion equation, originally derived in \citet{hwa11inter}:
\begin{equation}
T_{\rm dust}~({\rm K})=(43.0\pm 0.3) + (37.0\pm 1.5){\rm log}(S_{60 \mu m}/S_{100 \mu m}).
\label{eq-tdust}
\end{equation}

Hwang et al. derived this conversion equation
  by comparing the \iras 60/100 $\mu$m flux density ratio
  with the dust temperature directly measured
  from the SED fit of \akari 140 or 160 $\mu$m detected SDSS galaxies.
They fit the observational data
  with a modified black body model and
  fix the emissivity parameter to $\beta=1.5$\ 
  (see \citealt{hwa10tdust,elb10} for more details).

As expected,
  Figure \ref{fig-lirtdust} shows that the dust temperature of local galaxies
  including DOGs, control sample, and FIR detected galaxies
  increases with increasing IR luminosity \citep{chap03,hwa10tdust,amb10}.
These galaxies are smoothly connected to
  the region occupied by ULIRGs at $0.1\lesssim z\lesssim 3.0$ (pink region).
In a given IR luminosity range 
  (e.g., $7\times10^{11}<L_{\rm IR, total}/L_\odot<7\times10^{12}$
  where local DOGs exist),
  the dust temperature of local DOGs (see red histogram in the right panel) 
  is on average higher than
  for the control sample and for the FIR detected galaxies.

Most high-$z$ DOGs in \citet[filled triangles]{mel12}
  and \citet[open triangles]{saj12}
  have dust temperatures similar to or lower than local DOGs,
  even though the IR luminosities of high-$z$ DOGs 
  are much higher than those of local DOGs.
Moreover, the majority of high-$z$ DOGs
  follow the trend of high-$z$ submillimeter galaxies 
  (SMGs, blue region),
  but have much lower dust temperature
  than ULIRGs at $L_{\rm IR, total}/L_\odot\sim10^{12}$.
  
These data suggest that high-$z$ DOGs
  have dust temperature similar to or lower than other IR bright galaxies 
  at similar redshift. 
However, the dust temperature of high-$z$ DOGs
  is measured only for those detected 
  both at 250 $\mu$m and 350 $\mu$m \citep{mel12,saj12}.
DOGs not detected in these bands 
  (i.e., many DOGs with high dust temperature)
  are unlikely to appear in this plot.
Moreover, blending issue in the photometry of 
  these {\it Herschel} 250$-$500 $\mu$m bands
  can also introduce a systematic bias toward lower
  dust temperature estimates 
  (\citealt{hwa10tdust}; R. Leiton et al. 2013, in preparation).

The lack of high-$z$ DOGs with high dust temperature
  at $L_{\rm IR, total}/L_\odot\sim10^{12}$
  could result simply from selection effects 
  (see also discussion in \citealt{mel12} and their Figure 10).
This issue, especially for high-$z$ IR bright galaxies,
  is examined well in \citet{gmag10bumpy}, \citet{chapin11} and \citet{sym13}.
The detection of some high-$z$ DOGs with high dust temperature 
  (e.g., $T_{\rm dust}\gtrsim 60$ K with $L_{\rm IR}/L_\odot>10^{13}$)
  among \wise selected galaxies 
  demonstrates that they do exist 
  (\citealt{wu12}; there are also some in \citealt{saj12}).

\section{Discussion: Diverse Nature of Local and High-$z$ DOGs}\label{discuss}

There are apparently two types of local DOGs
  with small and large AGN contribution to their IR luminosities;
  high-$z$ DOGs show the same two populations.
The bottom left panel in Figure \ref{fig-flux} shows that
  the flux density ratios between \iras 60 $\mu$m
  and \wise 12 $\mu$m are systematically smaller 
  for local DOGs with a large AGN contribution
  than those with a small AGN contribution.
This trend is also apparent for high-$z$ DOGs 
  (i.e., power-law DOGs vs. bump DOGs),
  even though the observed bands are not identical
  (e.g., see Figure 7 in \citealt{mel12} for the flux density ratio
  of $S_{\rm 250 \mu m}/S_{\rm 24 \mu m}$ for $z\sim2$ DOGs).
Because rest-frame 60$-$100 $\mu$m data are closely
  related to the total IR luminosities of dusty galaxies and 12 $\mu$m data 
  are sensitive to a hot AGN dust component,
  a smaller flux density ratio between FIR and MIR bands
  for a smaller AGN contribution
  is consistent with expectation \citep{vei09,mul11agn,kir12}.

The main reason for the extreme $S_{12 \mu m}/S_{0.22 \mu m}$ 
  flux density ratios in local DOGs
  is the abnormal faintness in the NUV
  rather than the extreme brightness at 12 $\mu$m
  (see middle left ($S_{12 \mu m}$) and 
  bottom right ($S_{0.22 \mu m}$) panels in Figure \ref{fig-flux}).
This result is also valid for AGN-dominated DOGs.
This result is supported by the larger
  H$\alpha$/H$\beta$ flux ratios (i.e., Balmer decrement) of
  local DOGs relative to the control sample.

The abnormal faintness in the NUV
  we find for local DOGs is consistent with the conclusion
  for high-$z$ DOGs \citep{pen12}.
Penner et al. further suggest that
  the large dust obscuration in DOGs could result from the 
  spatial coincidence between dust and massive stars
  or from a large dust content.
In both cases, merging process
  between galaxies that can significantly change the dust geometry
  or galaxy inclination can play an important role.
Simulations also suggest that bright DOGs are recent merger products,
  but fainter DOGs could have a diverse origin 
  including the large inclination of disk galaxies \citep{nar10}.  
The optical color images of local DOGs in Figure \ref{fig-chart}
  indeed show that there are some galaxies in the process of merging
  (e.g., ID: 7, 13, 16, 19, 47; see also \citealt{mel09,don10,buss11hstb} 
  for high-$z$ DOGs with merging features).

The larger fraction of galaxies with small axis ratios ($<0.6$) 
  among local DOGs than that among the control sample of galaxies
  (i.e., $36\pm7$\% for DOGs, but $17\pm3$\% for control sample;
  see top left panel in Figure \ref{fig-opt} and 
  color images in Figure \ref{fig-chart}) and
  the large pair separation for 
  local DOGs (see bottom panels in Figure \ref{fig-env})
  suggest that the large dust obscuration of some DOGs
  simply results from their large inclination.
The optical color images of our local DOGs in Figure \ref{fig-chart}
  confirm that some DOGs are indeed highly-inclined disk galaxies
  (e.g., ID: 3, 11, 21, 34, 37).
However, it should be noted that 
  galaxies with large pair separation do not necessarily imply
  that they are currently quiescent systems.
When two galaxies have just finished merging, 
  the new nearest neighbor galaxy
  of the merger product may be far away.
Therefore, some DOGs with large pair separation (e.g., $R_n\gtrsim 0.3$ Mpc)
  could be recent merger remnants (or late stage mergers).
Actually, \citet{park08} showed
  that, at fixed background density, postmerger features including
  large displacement of the galaxy nucleus from the center, turmoil
  features, and/or very close double cores, are more frequently
  seen in the isolated galaxies than less isolated ones.
The optical color images in Figure \ref{fig-chart} show 
  some of these cases for local DOGs,
  but the fraction of these systems is not large. 
  
High resolution optical images for high-$z$ DOGs also suggest
  that many DOGs are undisturbed disk galaxies
  rather than merging objects \citep{kar12,sch12}.
Simulations by \citet{nar10} also show that
  isolated disk galaxies can satisfy DOG selection criteria 
  when they are edge-on.
It is also interesting that
  half or more of local DOGs have 
  neither small axis ratios (i.e., edge-on systems)
  nor obvious signs of galaxy-galaxy interactions and mergers
  (e.g., ID: 5, 27, 30, 36, 43).

These results suggest
  both local and high-$z$ DOGs are diverse populations,
  ranging from merging galaxies to edge-on disk galaxies.
We do not find clear evidence of evolution between subclasses 
  (e.g., from SF-dominated DOGs to AGN-dominated DOGs).
There are also some non-DOGs 
  with a large AGN contribution (see Figure \ref{fig-agncont}), 
  suggesting that the amount of dust obscuration
  is not closely related to the presence of AGN. 
These results imply that DOGs are not in 
  a unique phase of galaxy evolutionary sequence.
Instead they seem to be the high-end tail of the dust obscuration distribution 
  originating from various physical mechanisms.

\section{Conclusions}\label{sum}

By combining \galex UV and \akari/\wise MIR all-sky survey data,
  we identify 47 local analogs of DOGs
  with $S_{12\mu m}/S_{0.22 \mu m}\geq$892
  and $S_{12\mu m}>20$ mJy at $0.05<z<0.08$ in the SDSS DR7.
We compare their physical properties with
  other 12 $\mu$m selected galaxies that do not satisfy DOG criteria
  to study how special the DOGs are among MIR selected galaxies.
  %among dusty galaxies.
We also compare the physical properties of local DOGs with
  high-$z$ DOGs.
Our primary results are:

\begin{enumerate}

\item We fit the IR photometric data for local DOGs and the control sample
  with AGN/starburst SED templates
  to estimate their IR luminosities and the AGN contribution 
  to their IR luminosities.
Their total IR luminosities are in the range
 $3.4\times10^{10}  ({\rm L}_\odot) \lesssim L_{\rm IR} \lesssim 7.0\times10^{11}  ({\rm L}_\odot)$
 with a median $L_{\rm IR}$ of $2.1\times10^{11}$ (${\rm L}_\odot$).
Among 47 DOGs,
  20 ($43\%$) have a relatively large AGN contribution
  (i.e., $f_{\rm AGN}\geq 20\%$),
  and 27 ($57\%$) DOGs with a small AGN contribution (i.e., $f_{\rm AGN}<20\%$).

\item Comparison of several flux density ratios
  between DOGs and the control sample
  indicates that the extreme flux density ratios between
  MIR and UV bands for DOGs
  result mainly from abnormal faintness in the UV
  rather than from extreme brightness in the MIR.

\item The $i$-band axis ratio distributions of 
  DOGs and the control sample
  show that the fraction of galaxies with small axis ratios (i.e., $<0.6$)
  among DOGs ($36\pm7$\%) is larger than
  that among the control sample ($17\pm3$\%).
There is no obvious sign of interaction in the majority of DOGs.
%The majority of DOGs are not currently interacting galaxies.  
These results suggest, and optical images show, that some DOGs
  have large dust obscuration simply because of their high inclination.

\item The sSFRs of DOGs are similar to those of the control sample.
However, these sSFRs are slightly larger than 
  for typical IR bright galaxies at similar redshift
  by a factor of $\lesssim2$.
The DOGs with a large AGN contribution  (e.g, $f_{\rm AGN}\geq20\%$) 
  appear to have smaller sSFRs than those with a small AGN contribution
  (i.e., $f_{\rm AGN}<20\%$).

\item The large- and small-scale environments of DOGs
  are similar to those of the control sample
  and of other IR bright galaxies at similar redshift.

\end{enumerate}

From the comparison between local and high-$z$ DOGs, we find:

\begin{enumerate}

\item Many physical properties of local DOGs are similar to 
  those of high-$z$ DOGs,
  even though the IR luminosities of local DOGs
  are an order of magnitude lower than for the high-$z$ counterparts.
These properties include
  the presence of two classes (AGN- and SF-dominated) of DOGs,
  different properties between the two classes,
  abnormal faintness in the UV rather than extreme brightness in the MIR, and
  diverse optical morphology.

\item There are some differences between local and high-$z$ DOGs.
These differences include
  larger L$_{\rm H\alpha}$/L$_{\rm IR, total}$ for local DOGs
  than for high-$z$ DOGs,
  few local DOGs with dust temperatures as low as high-$z$ DOGs, and
  similar size distributions between AGN- and SF-dominated DOGs
  at low redshift in contrast with different distributions at high redshift.
However, these differences may result from different IR luminosities and/or
  different selection effects.

\end{enumerate}

Our results suggest that local DOGs
  indeed share a common underlying physical origin with high-$z$ DOGs.
Both local and high-$z$ DOGs are diverse in nature.
They seem to be in the high-end tail of the dust obscuration distribution.
Their dusty nature results from a range of physical mechanisms
  rather than from a unique phase in the galaxy evolutionary sequence.

Some differences between local and high-$z$ DOGs
  including the amount of dust obscuration and the dust temperature
  may result mainly from different IR luminosities and/or
  complex selection effects.
To resolve this issue,
  it is necessary to have galaxy samples with similar IR luminosities
  by exploring larger local volumes with appropriate bands 
  to access the rare luminous low-$z$ DOGs.
Deeper data sets could identify less luminous high-$z$ DOGs
  to overlap the low-$z$ sample.
FIR and submillimeter data
  for larger number of DOGs in both low and high redshifts
  would mitigate the selection effects and
  would provide direct measures of dust mass and temperature.
 
\acknowledgments

We thank the anonymous referee for many insightful comments
that improved the manuscript.
HSH acknowledges the Smithsonian Institution for the support of his post-doctoral fellowship.
The Smithsonian Institution also supports the research of MJG.
This publication makes use of data products from the Wide-field Infrared Survey Explorer, 
which is a joint project of the University of California, Los Angeles, 
and the Jet Propulsion Laboratory/California Institute of Technology, 
funded by the National Aeronautics and Space Administration.
This research is based on observations with AKARI, a JAXA project with the participation of ESA.
Funding for the SDSS and SDSS-II has been provided by the Alfred P. Sloan 
Foundation, the Participating Institutions, the National Science 
Foundation, the U.S. Department of Energy, the National Aeronautics and 
Space Administration, the Japanese Monbukagakusho, the Max Planck 
Society, and the Higher Education Funding Council for England. 
The SDSS Web Site is http://www.sdss.org/.
The SDSS is managed by the Astrophysical Research Consortium for the 
Participating Institutions. The Participating Institutions are the 
American Museum of Natural History, Astrophysical Institute Potsdam, 
University of Basel, Cambridge University, Case Western Reserve University, 
University of Chicago, Drexel University, Fermilab, the Institute for 
Advanced Study, the Japan Participation Group, Johns Hopkins University, 
the Joint Institute for Nuclear Astrophysics, the Kavli Institute for 
Particle Astrophysics and Cosmology, the Korean Scientist Group, the 
Chinese Academy of Sciences (LAMOST), Los Alamos National Laboratory, 
the Max-Planck-Institute for Astronomy (MPIA), the Max-Planck-Institute 
for Astrophysics (MPA), New Mexico State University, Ohio State University, 
University of Pittsburgh, University of Portsmouth, Princeton University,
the United States Naval Observatory, and the University of Washington. 
This research has made use of the NASA/IPAC Extragalactic Database (NED) 
which is operated by the Jet Propulsion Laboratory, California Institute of Technology, 
under contract with the National Aeronautics and Space Administration.

\bibliographystyle{apj} % style aa.bst
%\nocite{*}
\bibliography{ref_hshwang} % your references Yourfile.bib

\begin{thebibliography}{106}
\expandafter\ifx\csname natexlab\endcsname\relax\def\natexlab#1{#1}\fi

\bibitem[{{Ahn} {et~al.}(2012){Ahn}, {Alexandroff}, {Allende Prieto},
  {Anderson}, {Anderton}, {Andrews}, {Aubourg}, {Bailey}, {Balbinot}, {Barnes},
  \& et~al.}]{ahn12}
{Ahn}, C.~P., {Alexandroff}, R., {Allende Prieto}, C., {et~al.} 2012, \apjs,
  203, 21

\bibitem[{{Amblard} {et~al.}(2010){Amblard}, {Cooray}, {Serra}, {Temi},
  {Barton}, {Negrello}, {Auld}, {Baes}, {Baldry}, {Bamford}, {Blain}, {Bock},
  {Bonfield}, {Burgarella}, {Buttiglione}, {Cameron}, {Cava}, {Clements},
  {Croom}, {Dariush}, {de Zotti}, {Driver}, {Dunlop}, {Dunne}, {Dye}, {Eales},
  {Frayer}, {Fritz}, {Gardner}, {Gonzalez-Nuevo}, {Herranz}, {Hill}, {Hopkins},
  {Hughes}, {Ibar}, {Ivison}, {Jarvis}, {Jones}, {Kelvin}, {Lagache}, {Leeuw},
  {Liske}, {Lopez-Caniego}, {Loveday}, {Maddox}, {Micha{\l}owski}, {Norberg},
  {Parkinson}, {Peacock}, {Pearson}, {Pascale}, {Pohlen}, {Popescu},
  {Prescott}, {Robotham}, {Rigby}, {Rodighiero}, {Samui}, {Sansom}, {Scott},
  {Serjeant}, {Sharp}, {Sibthorpe}, {Smith}, {Thompson}, {Tuffs}, {Valtchanov},
  {van Kampen}, {van der Werf}, {Verma}, {Vieira}, \& {Vlahakis}}]{amb10}
{Amblard}, A., {Cooray}, A., {Serra}, P., {et~al.} 2010, \aap, 518, L9

\bibitem[{{Baldwin} {et~al.}(1981){Baldwin}, {Phillips}, \&
  {Terlevich}}]{bpt81}
{Baldwin}, J.~A., {Phillips}, M.~M., \& {Terlevich}, R. 1981, \pasp, 93, 5

\bibitem[{{Barton} {et~al.}(2000){Barton}, {Geller}, \& {Kenyon}}]{bar00}
{Barton}, E.~J., {Geller}, M.~J., \& {Kenyon}, S.~J. 2000, \apj, 530, 660

\bibitem[{{Behroozi} {et~al.}(2012){Behroozi}, {Wechsler}, \& {Conroy}}]{beh12}
{Behroozi}, P.~S., {Wechsler}, R.~H., \& {Conroy}, C. 2012, ApJ, submitted
  (arXiv:1207.6105)

\bibitem[{{Berta} {et~al.}(2003){Berta}, {Fritz}, {Franceschini}, {Bressan}, \&
  {Pernechele}}]{berta03}
{Berta}, S., {Fritz}, J., {Franceschini}, A., {Bressan}, A., \& {Pernechele},
  C. 2003, \aap, 403, 119

\bibitem[{{Blanton} \& {Moustakas}(2009)}]{bm09}
{Blanton}, M.~R., \& {Moustakas}, J. 2009, \araa, 47, 159

\bibitem[{{Blanton} \& {Roweis}(2007)}]{bla07kcorr}
{Blanton}, M.~R., \& {Roweis}, S. 2007, \aj, 133, 734

\bibitem[{{Brand} {et~al.}(2007){Brand}, {Dey}, {Desai}, {Soifer}, {Bian},
  {Armus}, {Brown}, {Le Floc'h}, {Higdon}, {Houck}, {Jannuzi}, \&
  {Weedman}}]{brand07}
{Brand}, K., {Dey}, A., {Desai}, V., {et~al.} 2007, \apj, 663, 204

\bibitem[{{Brodwin} {et~al.}(2008){Brodwin}, {Dey}, {Brown}, {Pope}, {Armus},
  {Bussmann}, {Desai}, {Jannuzi}, \& {Le Floc'h}}]{bro08}
{Brodwin}, M., {Dey}, A., {Brown}, M.~J.~I., {et~al.} 2008, \apjl, 687, L65

\bibitem[{{Bruzual} \& {Charlot}(2003)}]{bc03}
{Bruzual}, G., \& {Charlot}, S. 2003, \mnras, 344, 1000

\bibitem[{{Bussmann} {et~al.}(2009{\natexlab{a}}){Bussmann}, {Dey}, {Lotz},
  {Armus}, {Brand}, {Brown}, {Desai}, {Eisenhardt}, {Higdon}, {Higdon},
  {Jannuzi}, {Le Floc'h}, {Melbourne}, {Soifer}, \& {Weedman}}]{buss09hstp}
{Bussmann}, R.~S., {Dey}, A., {Lotz}, J., {et~al.} 2009{\natexlab{a}}, \apj,
  693, 750

\bibitem[{{Bussmann} {et~al.}(2009{\natexlab{b}}){Bussmann}, {Dey}, {Borys},
  {Desai}, {Jannuzi}, {Le Floc'h}, {Melbourne}, {Sheth}, \& {Soifer}}]{buss09}
{Bussmann}, R.~S., {Dey}, A., {Borys}, C., {et~al.} 2009{\natexlab{b}}, \apj,
  705, 184

\bibitem[{{Bussmann} {et~al.}(2011){Bussmann}, {Dey}, {Lotz}, {Armus}, {Brown},
  {Desai}, {Eisenhardt}, {Higdon}, {Higdon}, {Jannuzi}, {Le Floc'h},
  {Melbourne}, {Soifer}, \& {Weedman}}]{buss11hstb}
{Bussmann}, R.~S., {Dey}, A., {Lotz}, J., {et~al.} 2011, \apj, 733, 21

\bibitem[{{Bussmann} {et~al.}(2012){Bussmann}, {Dey}, {Armus}, {Brown},
  {Desai}, {Gonzalez}, {Jannuzi}, {Melbourne}, \& {Soifer}}]{buss12}
{Bussmann}, R.~S., {Dey}, A., {Armus}, L., {et~al.} 2012, \apj, 744, 150

\bibitem[{{Calzetti} {et~al.}(2000){Calzetti}, {Armus}, {Bohlin}, {Kinney},
  {Koornneef}, \& {Storchi-Bergmann}}]{cal00}
{Calzetti}, D., {Armus}, L., {Bohlin}, R.~C., {et~al.} 2000, \apj, 533, 682

\bibitem[{{Chapin} {et~al.}(2011){Chapin}, {Chapman}, {Coppin}, {Devlin},
  {Dunlop}, {Greve}, {Halpern}, {Hasselfield}, {Hughes}, {Ivison}, {Marsden},
  {Moncelsi}, {Netterfield}, {Pascale}, {Scott}, {Smail}, {Viero}, {Walter},
  {Weiss}, \& {van der Werf}}]{chapin11}
{Chapin}, E.~L., {Chapman}, S.~C., {Coppin}, K.~E., {et~al.} 2011, \mnras, 411,
  505

\bibitem[{{Chapman} {et~al.}(2005){Chapman}, {Blain}, {Smail}, \&
  {Ivison}}]{chap05}
{Chapman}, S.~C., {Blain}, A.~W., {Smail}, I., \& {Ivison}, R.~J. 2005, \apj,
  622, 772

\bibitem[{{Chapman} {et~al.}(2003){Chapman}, {Helou}, {Lewis}, \&
  {Dale}}]{chap03}
{Chapman}, S.~C., {Helou}, G., {Lewis}, G.~F., \& {Dale}, D.~A. 2003, \apj,
  588, 186

\bibitem[{{Choi} {et~al.}(2010){Choi}, {Han}, \& {Kim}}]{choi10}
{Choi}, Y., {Han}, D., \& {Kim}, S.~S. 2010, J. Korean Astron. Soc., 43, 191

\bibitem[{{Choi} {et~al.}(2007){Choi}, {Park}, \& {Vogeley}}]{choi07}
{Choi}, Y., {Park}, C., \& {Vogeley}, M.~S. 2007, \apj, 658, 884

\bibitem[{{Cox} {et~al.}(2008){Cox}, {Jonsson}, {Somerville}, {Primack}, \&
  {Dekel}}]{cox08}
{Cox}, T.~J., {Jonsson}, P., {Somerville}, R.~S., {Primack}, J.~R., \& {Dekel},
  A. 2008, \mnras, 384, 386

\bibitem[{{Daddi} {et~al.}(2010{\natexlab{a}}){Daddi}, {Elbaz}, {Walter},
  {Bournaud}, {Salmi}, {Carilli}, {Dannerbauer}, {Dickinson}, {Monaco}, \&
  {Riechers}}]{dad10sflaw}
{Daddi}, E., {Elbaz}, D., {Walter}, F., {et~al.} 2010{\natexlab{a}}, \apjl,
  714, L118

\bibitem[{{Daddi} {et~al.}(2010{\natexlab{b}}){Daddi}, {Bournaud}, {Walter},
  {Dannerbauer}, {Carilli}, {Dickinson}, {Elbaz}, {Morrison}, {Riechers},
  {Onodera}, {Salmi}, {Krips}, \& {Stern}}]{dad10}
{Daddi}, E., {Bournaud}, F., {Walter}, F., {et~al.} 2010{\natexlab{b}}, \apj,
  713, 686

\bibitem[{{Desai} {et~al.}(2009){Desai}, {Soifer}, {Dey}, {Le Floc'h}, {Armus},
  {Brand}, {Brown}, {Brodwin}, {Jannuzi}, {Houck}, {Weedman}, {Ashby},
  {Gonzalez}, {Huang}, {Smith}, {Teplitz}, {Willner}, \& {Melbourne}}]{desai09}
{Desai}, V., {Soifer}, B.~T., {Dey}, A., {et~al.} 2009, \apj, 700, 1190

\bibitem[{{Dey} {et~al.}(2008){Dey}, {Soifer}, {Desai}, {Brand}, {Le Floc'h},
  {Brown}, {Jannuzi}, {Armus}, {Bussmann}, {Brodwin}, {Bian}, {Eisenhardt},
  {Higdon}, {Weedman}, \& {Willner}}]{dey08}
{Dey}, A., {Soifer}, B.~T., {Desai}, V., {et~al.} 2008, \apj, 677, 943

\bibitem[{{Dickinson} {et~al.}(2003){Dickinson}, {Giavalisco}, \& {GOODS
  Team}}]{dic03}
{Dickinson}, M., {Giavalisco}, M., \& {GOODS Team}. 2003, in The Mass of
  Galaxies at Low and High Redshift, ed. {R.~Bender \& A.~Renzini}, 324--+

\bibitem[{{Donley} {et~al.}(2010){Donley}, {Rieke}, {Alexander}, {Egami}, \&
  {P{\'e}rez-Gonz{\'a}lez}}]{don10}
{Donley}, J.~L., {Rieke}, G.~H., {Alexander}, D.~M., {Egami}, E., \&
  {P{\'e}rez-Gonz{\'a}lez}, P.~G. 2010, \apj, 719, 1393

\bibitem[{{Donoso} {et~al.}(2012){Donoso}, {Yan}, {Tsai}, {Eisenhardt},
  {Stern}, {Assef}, {Leisawitz}, {Jarrett}, \& {Stanford}}]{don12}
{Donoso}, E., {Yan}, L., {Tsai}, C., {et~al.} 2012, \apj, 748, 80

\bibitem[{{Elbaz} {et~al.}(2007){Elbaz}, {Daddi}, {Le Borgne}, {Dickinson},
  {Alexander}, {Chary}, {Starck}, {Brandt}, {Kitzbichler}, {MacDonald},
  {Nonino}, {Popesso}, {Stern}, \& {Vanzella}}]{elb07}
{Elbaz}, D., {Daddi}, E., {Le Borgne}, D., {et~al.} 2007, \aap, 468, 33

\bibitem[{{Elbaz} {et~al.}(2010){Elbaz}, {Hwang}, {Magnelli}, {Daddi},
  {Aussel}, {Altieri}, {Amblard}, {Andreani}, {Arumugam}, {Auld}, {Babbedge},
  {Berta}, {Blain}, {Bock}, {Bongiovanni}, {Boselli}, {Buat}, {Burgarella},
  {Castro-Rodriguez}, {Cava}, {Cepa}, {Chanial}, {Chary}, {Cimatti},
  {Clements}, {Conley}, {Conversi}, {Cooray}, {Dickinson}, {Dominguez},
  {Dowell}, {Dunlop}, {Dwek}, {Eales}, {Farrah}, {F{\"o}rster Schreiber},
  {Fox}, {Franceschini}, {Gear}, {Genzel}, {Glenn}, {Griffin}, {Gruppioni},
  {Halpern}, {Hatziminaoglou}, {Ibar}, {Isaak}, {Ivison}, {Lagache}, {Le
  Borgne}, {Le Floc'h}, {Levenson}, {Lu}, {Lutz}, {Madden}, {Maffei}, {Magdis},
  {Mainetti}, {Maiolino}, {Marchetti}, {Mortier}, {Nguyen}, {Nordon},
  {O'Halloran}, {Okumura}, {Oliver}, {Omont}, {Page}, {Panuzzo},
  {Papageorgiou}, {Pearson}, {Perez Fournon}, {P{\'e}rez Garc{\'{\i}}a},
  {Poglitsch}, {Pohlen}, {Popesso}, {Pozzi}, {Rawlings}, {Rigopoulou},
  {Riguccini}, {Rizzo}, {Rodighiero}, {Roseboom}, {Rowan-Robinson},
  {Saintonge}, {Sanchez Portal}, {Santini}, {Sauvage}, {Schulz}, {Scott},
  {Seymour}, {Shao}, {Shupe}, {Smith}, {Stevens}, {Sturm}, {Symeonidis},
  {Tacconi}, {Trichas}, {Tugwell}, {Vaccari}, {Valtchanov}, {Vieira},
  {Vigroux}, {Wang}, {Ward}, {Wright}, {Xu}, \& {Zemcov}}]{elb10}
{Elbaz}, D., {Hwang}, H.~S., {Magnelli}, B., {et~al.} 2010, \aap, 518, L29

\bibitem[{{Elbaz} {et~al.}(2011){Elbaz}, {Dickinson}, {Hwang},
  {D{\'{\i}}az-Santos}, {Magdis}, {Magnelli}, {Le Borgne}, {Galliano},
  {Pannella}, {Chanial}, {Armus}, {Charmandaris}, {Daddi}, {Aussel}, {Popesso},
  {Kartaltepe}, {Altieri}, {Valtchanov}, {Coia}, {Dannerbauer}, {Dasyra},
  {Leiton}, {Mazzarella}, {Alexander}, {Buat}, {Burgarella}, {Chary}, {Gilli},
  {Ivison}, {Juneau}, {Le Floc'h}, {Lutz}, {Morrison}, {Mullaney}, {Murphy},
  {Pope}, {Scott}, {Brodwin}, {Calzetti}, {Cesarsky}, {Charlot}, {Dole},
  {Eisenhardt}, {Ferguson}, {F{\"o}rster Schreiber}, {Frayer}, {Giavalisco},
  {Huynh}, {Koekemoer}, {Papovich}, {Reddy}, {Surace}, {Teplitz}, {Yun}, \&
  {Wilson}}]{elb11}
{Elbaz}, D., {Dickinson}, M., {Hwang}, H.~S., {et~al.} 2011, \aap, 533, 119

\bibitem[{{Fiore} {et~al.}(2008){Fiore}, {Grazian}, {Santini}, {Puccetti},
  {Brusa}, {Feruglio}, {Fontana}, {Giallongo}, {Comastri}, {Gruppioni},
  {Pozzi}, {Zamorani}, \& {Vignali}}]{fio08}
{Fiore}, F., {Grazian}, A., {Santini}, P., {et~al.} 2008, \apj, 672, 94

\bibitem[{{Grogin} {et~al.}(2011){Grogin}, {Kocevski}, {Faber}, {Ferguson},
  {Koekemoer}, {Riess}, {Acquaviva}, {Alexander}, {Almaini}, {Ashby}, {Barden},
  {Bell}, {Bournaud}, {Brown}, {Caputi}, {Casertano}, {Cassata}, {Castellano},
  {Challis}, {Chary}, {Cheung}, {Cirasuolo}, {Conselice}, {Roshan Cooray},
  {Croton}, {Daddi}, {Dahlen}, {Dav{\'e}}, {de Mello}, {Dekel}, {Dickinson},
  {Dolch}, {Donley}, {Dunlop}, {Dutton}, {Elbaz}, {Fazio}, {Filippenko},
  {Finkelstein}, {Fontana}, {Gardner}, {Garnavich}, {Gawiser}, {Giavalisco},
  {Grazian}, {Guo}, {Hathi}, {H{\"a}ussler}, {Hopkins}, {Huang}, {Huang},
  {Jha}, {Kartaltepe}, {Kirshner}, {Koo}, {Lai}, {Lee}, {Li}, {Lotz}, {Lucas},
  {Madau}, {McCarthy}, {McGrath}, {McIntosh}, {McLure}, {Mobasher},
  {Moustakas}, {Mozena}, {Nandra}, {Newman}, {Niemi}, {Noeske}, {Papovich},
  {Pentericci}, {Pope}, {Primack}, {Rajan}, {Ravindranath}, {Reddy}, {Renzini},
  {Rix}, {Robaina}, {Rodney}, {Rosario}, {Rosati}, {Salimbeni}, {Scarlata},
  {Siana}, {Simard}, {Smidt}, {Somerville}, {Spinrad}, {Straughn}, {Strolger},
  {Telford}, {Teplitz}, {Trump}, {van der Wel}, {Villforth}, {Wechsler},
  {Weiner}, {Wiklind}, {Wild}, {Wilson}, {Wuyts}, {Yan}, \& {Yun}}]{gro11}
{Grogin}, N.~A., {Kocevski}, D.~D., {Faber}, S.~M., {et~al.} 2011, \apjs, 197,
  35

\bibitem[{{Houck} {et~al.}(2005){Houck}, {Soifer}, {Weedman}, {Higdon},
  {Higdon}, {Herter}, {Brown}, {Dey}, {Jannuzi}, {Le Floc'h}, {Rieke}, {Armus},
  {Charmandaris}, {Brandl}, \& {Teplitz}}]{hou05}
{Houck}, J.~R., {Soifer}, B.~T., {Weedman}, D., {et~al.} 2005, \apjl, 622, L105

\bibitem[{{Hwang} {et~al.}(2010{\natexlab{a}}){Hwang}, {Elbaz}, {Lee}, {Jeong},
  {Park}, {Lee}, \& {Lee}}]{hwa10lirg}
{Hwang}, H.~S., {Elbaz}, D., {Lee}, J.~C., {et~al.} 2010{\natexlab{a}}, \aap,
  522, 33

\bibitem[{{Hwang} {et~al.}(2012{\natexlab{a}}){Hwang}, {Geller}, {Diaferio}, \&
  {Rines}}]{hwa12a2199}
{Hwang}, H.~S., {Geller}, M.~J., {Diaferio}, A., \& {Rines}, K.~J.
  2012{\natexlab{a}}, \apj, 752, 64

\bibitem[{{Hwang} {et~al.}(2012{\natexlab{b}}){Hwang}, {Geller}, {Kurtz},
  {Dell'Antonio}, \& {Fabricant}}]{hwa12shels}
{Hwang}, H.~S., {Geller}, M.~J., {Kurtz}, M.~J., {Dell'Antonio}, I.~P., \&
  {Fabricant}, D.~G. 2012{\natexlab{b}}, \apj, 758, 25

\bibitem[{{Hwang} {et~al.}(2012{\natexlab{c}}){Hwang}, {Park}, {Elbaz}, \&
  {Choi}}]{hwa12agn}
{Hwang}, H.~S., {Park}, C., {Elbaz}, D., \& {Choi}, Y.-Y. 2012{\natexlab{c}},
  \aap, 538, 15

\bibitem[{{Hwang} {et~al.}(2010{\natexlab{b}}){Hwang}, {Elbaz}, {Magdis},
  {Daddi}, {Symeonidis}, {Altieri}, {Amblard}, {Andreani}, {Arumugam}, {Auld},
  {Aussel}, {Babbedge}, {Berta}, {Blain}, {Bock}, {Bongiovanni}, {Boselli},
  {Buat}, {Burgarella}, {Castro-Rodr{\'{\i}}guez}, {Cava}, {Cepa}, {Chanial},
  {Chapin}, {Chary}, {Cimatti}, {Clements}, {Conley}, {Conversi}, {Cooray},
  {Dannerbauer}, {Dickinson}, {Dominguez}, {Dowell}, {Dunlop}, {Dwek}, {Eales},
  {Farrah}, {Schreiber}, {Fox}, {Franceschini}, {Gear}, {Genzel}, {Glenn},
  {Griffin}, {Gruppioni}, {Halpern}, {Hatziminaoglou}, {Ibar}, {Isaak},
  {Ivison}, {Jeong}, {Lagache}, {Le Borgne}, {Le Floc'h}, {Lee}, {Lee}, {Lee},
  {Levenson}, {Lu}, {Lutz}, {Madden}, {Maffei}, {Magnelli}, {Mainetti},
  {Maiolino}, {Marchetti}, {Mortier}, {Nguyen}, {Nordon}, {O'Halloran},
  {Okumura}, {Oliver}, {Omont}, {Page}, {Panuzzo}, {Papageorgiou}, {Pearson},
  {P{\'e}rez-Fournon}, {Garc{\'{\i}}a}, {Poglitsch}, {Pohlen}, {Popesso},
  {Pozzi}, {Rawlings}, {Rigopoulou}, {Riguccini}, {Rizzo}, {Rodighiero},
  {Roseboom}, {Rowan-Robinson}, {Saintonge}, {Portal}, {Santini}, {Sauvage},
  {Schulz}, {Scott}, {Seymour}, {Shao}, {Shupe}, {Smith}, {Stevens}, {Sturm},
  {Tacconi}, {Trichas}, {Tugwell}, {Vaccari}, {Valtchanov}, {Vieira},
  {Vigroux}, {Wang}, {Ward}, {Wright}, {Xu}, \& {Zemcov}}]{hwa10tdust}
{Hwang}, H.~S., {Elbaz}, D., {Magdis}, G., {et~al.} 2010{\natexlab{b}}, \mnras,
  409, 75

\bibitem[{{Hwang} {et~al.}(2011){Hwang}, {Elbaz}, {Dickinson}, {Charmandaris},
  {Daddi}, {Le Borgne}, {Buat}, {Magdis}, {Altieri}, {Aussel}, {Coia},
  {Dannerbauer}, {Dasyra}, {Kartaltepe}, {Leiton}, {Magnelli}, {Popesso}, \&
  {Valtchanov}}]{hwa11inter}
{Hwang}, H.~S., {Elbaz}, D., {Dickinson}, M., {et~al.} 2011, \aap, 535, 60

\bibitem[{{Ishihara} {et~al.}(2010){Ishihara}, {Onaka}, {Kataza}, {Salama},
  {Alfageme}, {Cassatella}, {Cox}, {Garc{\'{\i}}a-Lario}, {Stephenson},
  {Cohen}, {Fujishiro}, {Fujiwara}, {Hasegawa}, {Ita}, {Kim}, {Matsuhara},
  {Murakami}, {M{\"u}ller}, {Nakagawa}, {Ohyama}, {Oyabu}, {Pyo}, {Sakon},
  {Shibai}, {Takita}, {Tanab{\'e}}, {Uemizu}, {Ueno}, {Usui}, {Wada},
  {Watarai}, {Yamamura}, \& {Yamauchi}}]{ish10}
{Ishihara}, D., {Onaka}, T., {Kataza}, H., {et~al.} 2010, \aap, 514, 1

\bibitem[{{Jarrett} {et~al.}(2000){Jarrett}, {Chester}, {Cutri}, {Schneider},
  {Skrutskie}, \& {Huchra}}]{jar00}
{Jarrett}, T.~H., {Chester}, T., {Cutri}, R., {et~al.} 2000, \aj, 119, 2498

\bibitem[{{Jarrett} {et~al.}(2011){Jarrett}, {Cohen}, {Masci}, {Wright},
  {Stern}, {Benford}, {Blain}, {Carey}, {Cutri}, {Eisenhardt}, {Lonsdale},
  {Mainzer}, {Marsh}, {Padgett}, {Petty}, {Ressler}, {Skrutskie}, {Stanford},
  {Surace}, {Tsai}, {Wheelock}, \& {Yan}}]{jar11}
{Jarrett}, T.~H., {Cohen}, M., {Masci}, F., {et~al.} 2011, \apj, 735, 112

\bibitem[{{Kartaltepe} {et~al.}(2012){Kartaltepe}, {Dickinson}, {Alexander},
  {Bell}, {Dahlen}, {Elbaz}, {Faber}, {Lotz}, {McIntosh}, {Wiklind}, {Altieri},
  {Aussel}, {Bethermin}, {Bournaud}, {Charmandaris}, {Conselice}, {Cooray},
  {Dannerbauer}, {Dav{\'e}}, {Dunlop}, {Dekel}, {Ferguson}, {Grogin}, {Hwang},
  {Ivison}, {Kocevski}, {Koekemoer}, {Koo}, {Lai}, {Leiton}, {Lucas}, {Lutz},
  {Magdis}, {Magnelli}, {Morrison}, {Mozena}, {Mullaney}, {Newman}, {Pope},
  {Popesso}, {van der Wel}, {Weiner}, \& {Wuyts}}]{kar12}
{Kartaltepe}, J.~S., {Dickinson}, M., {Alexander}, D.~M., {et~al.} 2012, \apj,
  757, 23

\bibitem[{{Kauffmann} {et~al.}(2003{\natexlab{a}}){Kauffmann}, {Heckman},
  {White}, {Charlot}, {Tremonti}, {Brinchmann}, {Bruzual}, {Peng}, {Seibert},
  {Bernardi}, {Blanton}, {Brinkmann}, {Castander}, {Cs{\'a}bai}, {Fukugita},
  {Ivezic}, {Munn}, {Nichol}, {Padmanabhan}, {Thakar}, {Weinberg}, \&
  {York}}]{kau03}
{Kauffmann}, G., {Heckman}, T.~M., {White}, S.~D.~M., {et~al.}
  2003{\natexlab{a}}, \mnras, 341, 33

\bibitem[{{Kauffmann} {et~al.}(2003{\natexlab{b}}){Kauffmann}, {Heckman},
  {Tremonti}, {Brinchmann}, {Charlot}, {White}, {Ridgway}, {Brinkmann},
  {Fukugita}, {Hall}, {Ivezi{\'c}}, {Richards}, \& {Schneider}}]{kau03agn}
{Kauffmann}, G., {Heckman}, T.~M., {Tremonti}, C., {et~al.} 2003{\natexlab{b}},
  \mnras, 346, 1055

\bibitem[{{Kawada} {et~al.}(2007){Kawada}, {Baba}, {Barthel}, {Clements},
  {Cohen}, {Doi}, {Figueredo}, {Fujiwara}, {Goto}, {Hasegawa}, {Hibi}, {Hirao},
  {Hiromoto}, {Jeong}, {Kaneda}, {Kawai}, {Kawamura}, {Kester}, {Kii},
  {Kobayashi}, {Kwon}, {Lee}, {Makiuti}, {Matsuo}, {Matsuura}, {M{\"u}ller},
  {Murakami}, {Nagata}, {Nakagawa}, {Narita}, {Noda}, {Oh}, {Okada}, {Okuda},
  {Oliver}, {Ootsubo}, {Pak}, {Park}, {Pearson}, {Rowan-Robinson}, {Saito},
  {Salama}, {Sato}, {Savage}, {Serjeant}, {Shibai}, {Shirahata}, {Sohn},
  {Suzuki}, {Takagi}, {Takahashi}, {Thomson}, {Usui}, {Verdugo}, {Watabe},
  {White}, {Wang}, {Yamamura}, {Yamauchi}, \& {Yasuda}}]{kaw07}
{Kawada}, M., {Baba}, H., {Barthel}, P.~D., {et~al.} 2007, \pasj, 59, 389

\bibitem[{{Kennicutt}(1998)}]{ken98}
{Kennicutt}, Jr., R.~C. 1998, \araa, 36, 189

\bibitem[{{Kewley} {et~al.}(2001){Kewley}, {Dopita}, {Sutherland}, {Heisler},
  \& {Trevena}}]{kew01}
{Kewley}, L.~J., {Dopita}, M.~A., {Sutherland}, R.~S., {Heisler}, C.~A., \&
  {Trevena}, J. 2001, \apj, 556, 121

\bibitem[{{Kewley} {et~al.}(2006){Kewley}, {Groves}, {Kauffmann}, \&
  {Heckman}}]{kew06}
{Kewley}, L.~J., {Groves}, B., {Kauffmann}, G., \& {Heckman}, T. 2006, \mnras,
  372, 961

\bibitem[{{Kewley} {et~al.}(2005){Kewley}, {Jansen}, \& {Geller}}]{kew05}
{Kewley}, L.~J., {Jansen}, R.~A., \& {Geller}, M.~J. 2005, \pasp, 117, 227

\bibitem[{{Kirkpatrick} {et~al.}(2012){Kirkpatrick}, {Pope}, {Alexander},
  {Charmandaris}, {Daddi}, {Dickinson}, {Elbaz}, {Gabor}, {Hwang}, {Ivison},
  {Mullaney}, {Pannella}, {Scott}, {Altieri}, {Aussel}, {Bournaud}, {Buat},
  {Coia}, {Dannerbauer}, {Dasyra}, {Kartaltepe}, {Leiton}, {Lin}, {Magdis},
  {Magnelli}, {Morrison}, {Popesso}, \& {Valtchanov}}]{kir12}
{Kirkpatrick}, A., {Pope}, A., {Alexander}, D.~M., {et~al.} 2012, \apj, 759,
  139

\bibitem[{{Kirkpatrick} {et~al.}(2013){Kirkpatrick}, {Pope}, {Charmandaris},
  {Daddi}, {Elbaz}, {Hwang}, {Pannella}, {Scott}, {Altieri}, {Aussel}, {Coia},
  {Dannerbauer}, {Dasyra}, {Dickinson}, {Kartaltepe}, {Leiton}, {Magdis},
  {Magnelli}, {Popesso}, \& {Valtchanov}}]{kir13}
{Kirkpatrick}, A., {Pope}, A., {Charmandaris}, V., {et~al.} 2013, \apj, 763,
  123

\bibitem[{{Knapp} {et~al.}(1989){Knapp}, {Guhathakurta}, {Kim}, \&
  {Jura}}]{kna89}
{Knapp}, G.~R., {Guhathakurta}, P., {Kim}, D.-W., \& {Jura}, M.~A. 1989, \apjs,
  70, 329

\bibitem[{{Ko} {et~al.}(2013){Ko}, {Hwang}, {Lee}, \& {Sohn}}]{ko13}
{Ko}, J., {Hwang}, H.~S., {Lee}, J.~C., \& {Sohn}, Y.-J. 2013, \apj, 767, 90

\bibitem[{{Kov{\'a}cs} {et~al.}(2006){Kov{\'a}cs}, {Chapman}, {Dowell},
  {Blain}, {Ivison}, {Smail}, \& {Phillips}}]{kov06}
{Kov{\'a}cs}, A., {Chapman}, S.~C., {Dowell}, C.~D., {et~al.} 2006, \apj, 650,
  592

\bibitem[{{Kroupa}(2001)}]{kro01}
{Kroupa}, P. 2001, \mnras, 322, 231

\bibitem[{{Lee} {et~al.}(2012){Lee}, {Hwang}, {Lee}, {Kim}, \&
  {Lee}}]{leejc12akari}
{Lee}, J.~C., {Hwang}, H.~S., {Lee}, M.~G., {Kim}, M., \& {Lee}, J.~H. 2012,
  \apj, 756, 95

\bibitem[{{Lee} {et~al.}(2010){Lee}, {Hwang}, {Lee}, {Lee}, \&
  {Matsuhara}}]{jhlee10beg}
{Lee}, J.~H., {Hwang}, H.~S., {Lee}, M.~G., {Lee}, J.~C., \& {Matsuhara}, H.
  2010, \apj, 719, 1946

\bibitem[{{Lonsdale} {et~al.}(2009){Lonsdale}, {Polletta}, {Omont}, {Shupe},
  {Berta}, {Zylka}, {Siana}, {Lutz}, {Farrah}, {Smith}, {Lagache}, {DeBreuck},
  {Owen}, {Beelen}, {Weedman}, {Franceschini}, {Clements}, {Tacconi},
  {Afonso-Luis}, {P{\'e}rez-Fournon}, {Cox}, \& {Bertoldi}}]{lon09}
{Lonsdale}, C.~J., {Polletta}, M.~d.~C., {Omont}, A., {et~al.} 2009, \apj, 692,
  422

\bibitem[{{Magdis} {et~al.}(2010){Magdis}, {Elbaz}, {Hwang}, {Amblard},
  {Arumugam}, {Aussel}, {Blain}, {Bock}, {Boselli}, {Buat},
  {Castro-Rodr{\'{\i}}guez}, {Cava}, {Chanial}, {Clements}, {Conley},
  {Conversi}, {Cooray}, {Dowell}, {Dwek}, {Eales}, {Farrah}, {Franceschini},
  {Glenn}, {Griffin}, {Halpern}, {Hatziminaoglou}, {Huang}, {Ibar}, {Isaak},
  {Le Floc'h}, {Lagache}, {Levenson}, {Lonsdale}, {Lu}, {Madden}, {Maffei},
  {Mainetti}, {Marchetti}, {Morrison}, {Nguyen}, {O'Halloran}, {Oliver},
  {Omont}, {Owen}, {Page}, {Pannella}, {Panuzzo}, {Papageorgiou}, {Pearson},
  {P{\'e}rez-Fournon}, {Pohlen}, {Rigopoulou}, {Rizzo}, {Roseboom},
  {Rowan-Robinson}, {Schulz}, {Scott}, {Seymour}, {Shupe}, {Smith}, {Stevens},
  {Strazzullo}, {Symeonidis}, {Trichas}, {Tugwell}, {Vaccari}, {Valtchanov},
  {Vigroux}, {Wang}, {Wright}, {Xu}, \& {Zemcov}}]{gmag10bumpy}
{Magdis}, G.~E., {Elbaz}, D., {Hwang}, H.~S., {et~al.} 2010, \mnras, 409, 22

\bibitem[{{Magdis} {et~al.}(2012){Magdis}, {Daddi}, {B{\'e}thermin}, {Sargent},
  {Elbaz}, {Pannella}, {Dickinson}, {Dannerbauer}, {da Cunha}, {Walter},
  {Rigopoulou}, {Charmandaris}, {Hwang}, \& {Kartaltepe}}]{mag12}
{Magdis}, G.~E., {Daddi}, E., {B{\'e}thermin}, M., {et~al.} 2012, \apj, 760, 6

\bibitem[{{Magnelli} {et~al.}(2012){Magnelli}, {Lutz}, {Santini}, {Saintonge},
  {Berta}, {Albrecht}, {Altieri}, {Andreani}, {Aussel}, {Bertoldi},
  {B{\'e}thermin}, {Bongiovanni}, {Capak}, {Chapman}, {Cepa}, {Cimatti},
  {Cooray}, {Daddi}, {Danielson}, {Dannerbauer}, {Dunlop}, {Elbaz}, {Farrah},
  {F{\"o}rster Schreiber}, {Genzel}, {Hwang}, {Ibar}, {Ivison}, {Le Floc'h},
  {Magdis}, {Maiolino}, {Nordon}, {Oliver}, {P{\'e}rez Garc{\'{\i}}a},
  {Poglitsch}, {Popesso}, {Pozzi}, {Riguccini}, {Rodighiero}, {Rosario},
  {Roseboom}, {Salvato}, {Sanchez-Portal}, {Scott}, {Smail}, {Sturm},
  {Swinbank}, {Tacconi}, {Valtchanov}, {Wang}, \& {Wuyts}}]{bmag12}
{Magnelli}, B., {Lutz}, D., {Santini}, P., {et~al.} 2012, \aap, 539, A155

\bibitem[{{Martin} {et~al.}(2005){Martin}, {Fanson}, {Schiminovich},
  {Morrissey}, {Friedman}, {Barlow}, {Conrow}, {Grange}, {Jelinsky},
  {Milliard}, {Siegmund}, {Bianchi}, {Byun}, {Donas}, {Forster}, {Heckman},
  {Lee}, {Madore}, {Malina}, {Neff}, {Rich}, {Small}, {Surber}, {Szalay},
  {Welsh}, \& {Wyder}}]{mar05}
{Martin}, D.~C., {Fanson}, J., {Schiminovich}, D., {et~al.} 2005, \apjl, 619,
  L1

\bibitem[{{Mateos} {et~al.}(2012){Mateos}, {Alonso-Herrero}, {Carrera},
  {Blain}, {Watson}, {Barcons}, {Braito}, {Severgnini}, {Donley}, \&
  {Stern}}]{mat12}
{Mateos}, S., {Alonso-Herrero}, A., {Carrera}, F.~J., {et~al.} 2012, \mnras,
  426, 3271

\bibitem[{{Melbourne} {et~al.}(2009){Melbourne}, {Bussman}, {Brand}, {Desai},
  {Armus}, {Dey}, {Jannuzi}, {Houck}, {Matthews}, \& {Soifer}}]{mel09}
{Melbourne}, J., {Bussman}, R.~S., {Brand}, K., {et~al.} 2009, \aj, 137, 4854

\bibitem[{{Melbourne} {et~al.}(2011){Melbourne}, {Peng}, {Soifer}, {Urrutia},
  {Desai}, {Armus}, {Bussmann}, {Dey}, \& {Matthews}}]{mel11}
{Melbourne}, J., {Peng}, C.~Y., {Soifer}, B.~T., {et~al.} 2011, \aj, 141, 141

\bibitem[{{Melbourne} {et~al.}(2012){Melbourne}, {Soifer}, {Desai}, {Pope},
  {Armus}, {Dey}, {Bussmann}, {Jannuzi}, \& {Alberts}}]{mel12}
{Melbourne}, J., {Soifer}, B.~T., {Desai}, V., {et~al.} 2012, \aj, 143, 125

\bibitem[{{Moshir} {et~al.}(1992){Moshir}, {Kopman}, \& {Conrow}}]{mos92}
{Moshir}, M., {Kopman}, G., \& {Conrow}, T.~A.~O. 1992, {IRAS Faint Source
  Survey, Explanatory supplement version 2}

\bibitem[{{Mullaney} {et~al.}(2011){Mullaney}, {Alexander}, {Goulding}, \&
  {Hickox}}]{mul11agn}
{Mullaney}, J.~R., {Alexander}, D.~M., {Goulding}, A.~D., \& {Hickox}, R.~C.
  2011, \mnras, 414, 1082

\bibitem[{{Murakami} {et~al.}(2007){Murakami}, {Baba}, {Barthel}, {Clements},
  {Cohen}, {Doi}, {Enya}, {Figueredo}, {Fujishiro}, {Fujiwara}, {Fujiwara},
  {Garcia-Lario}, {Goto}, {Hasegawa}, {Hibi}, {Hirao}, {Hiromoto}, {Hong},
  {Imai}, {Ishigaki}, {Ishiguro}, {Ishihara}, {Ita}, {Jeong}, {Jeong},
  {Kaneda}, {Kataza}, {Kawada}, {Kawai}, {Kawamura}, {Kessler}, {Kester},
  {Kii}, {Kim}, {Kim}, {Kobayashi}, {Koo}, {Kwon}, {Lee}, {Lorente}, {Makiuti},
  {Matsuhara}, {Matsumoto}, {Matsuo}, {Matsuura}, {M{\"u}ller}, {Murakami},
  {Nagata}, {Nakagawa}, {Naoi}, {Narita}, {Noda}, {Oh}, {Ohnishi}, {Ohyama},
  {Okada}, {Okuda}, {Oliver}, {Onaka}, {Ootsubo}, {Oyabu}, {Pak}, {Park},
  {Pearson}, {Rowan-Robinson}, {Saito}, {Sakon}, {Salama}, {Sato}, {Savage},
  {Serjeant}, {Shibai}, {Shirahata}, {Sohn}, {Suzuki}, {Takagi}, {Takahashi},
  {Tanab{\'e}}, {Takeuchi}, {Takita}, {Thomson}, {Uemizu}, {Ueno}, {Usui},
  {Verdugo}, {Wada}, {Wang}, {Watabe}, {Watarai}, {White}, {Yamamura},
  {Yamauchi}, \& {Yasuda}}]{mur07}
{Murakami}, H., {Baba}, H., {Barthel}, P., {et~al.} 2007, \pasj, 59, 369

\bibitem[{{Narayanan} {et~al.}(2010){Narayanan}, {Dey}, {Hayward}, {Cox},
  {Bussmann}, {Brodwin}, {Jonsson}, {Hopkins}, {Groves}, {Younger}, \&
  {Hernquist}}]{nar10}
{Narayanan}, D., {Dey}, A., {Hayward}, C.~C., {et~al.} 2010, \mnras, 407, 1701

\bibitem[{{Neugebauer} {et~al.}(1984){Neugebauer}, {Habing}, {van Duinen},
  {Aumann}, {Baud}, {Beichman}, {Beintema}, {Boggess}, {Clegg}, {de Jong},
  {Emerson}, {Gautier}, {Gillett}, {Harris}, {Hauser}, {Houck}, {Jennings},
  {Low}, {Marsden}, {Miley}, {Olnon}, {Pottasch}, {Raimond}, {Rowan-Robinson},
  {Soifer}, {Walker}, {Wesselius}, \& {Young}}]{neu84}
{Neugebauer}, G., {Habing}, H.~J., {van Duinen}, R., {et~al.} 1984, \apjl, 278,
  L1

\bibitem[{{Obri{\'c}} {et~al.}(2006){Obri{\'c}}, {Ivezi{\'c}}, {Best},
  {Lupton}, {Tremonti}, {Brinchmann}, {Ag{\"u}eros}, {Knapp}, {Gunn},
  {Rockosi}, {Schlegel}, {Finkbeiner}, {Ga{\'c}e{\v s}a}, {Smol{\v c}i{\'c}},
  {Anderson}, {Voges}, {Juri{\'c}}, {Siverd}, {Steinhardt}, {Jagoda},
  {Blanton}, \& {Schneider}}]{obr06}
{Obri{\'c}}, M., {Ivezi{\'c}}, {\v Z}., {Best}, P.~N., {et~al.} 2006, \mnras,
  370, 1677

\bibitem[{{Osterbrock} \& {Ferland}(2006)}]{ost06}
{Osterbrock}, D.~E., \& {Ferland}, G.~J. 2006, {Astrophysics of Gaseous Nebulae
  and Active Galactic Nuclei}, ed. {Osterbrock, D.~E.~\& Ferland, G.~J.}

\bibitem[{{Park} \& {Choi}(2009)}]{pc09}
{Park}, C., \& {Choi}, Y. 2009, \apj, 691, 1828

\bibitem[{{Park} {et~al.}(2008){Park}, {Gott}, \& {Choi}}]{park08}
{Park}, C., {Gott}, J.~R.~I., \& {Choi}, Y. 2008, \apj, 674, 784

\bibitem[{{Penner} {et~al.}(2012){Penner}, {Dickinson}, {Pope}, {Dey},
  {Magnelli}, {Pannella}, {Altieri}, {Aussel}, {Buat}, {Bussmann},
  {Charmandaris}, {Coia}, {Daddi}, {Dannerbauer}, {Elbaz}, {Hwang},
  {Kartaltepe}, {Lin}, {Magdis}, {Morrison}, {Popesso}, {Scott}, \&
  {Valtchanov}}]{pen12}
{Penner}, K., {Dickinson}, M., {Pope}, A., {et~al.} 2012, \apj, 759, 28

\bibitem[{{Polletta} {et~al.}(2007){Polletta}, {Tajer}, {Maraschi},
  {Trinchieri}, {Lonsdale}, {Chiappetti}, {Andreon}, {Pierre}, {Le F{\`e}vre},
  {Zamorani}, {Maccagni}, {Garcet}, {Surdej}, {Franceschini}, {Alloin},
  {Shupe}, {Surace}, {Fang}, {Rowan-Robinson}, {Smith}, \& {Tresse}}]{pol07}
{Polletta}, M., {Tajer}, M., {Maraschi}, L., {et~al.} 2007, \apj, 663, 81

\bibitem[{{Pope} {et~al.}(2008){Pope}, {Bussmann}, {Dey}, {Meger}, {Alexander},
  {Brodwin}, {Chary}, {Dickinson}, {Frayer}, {Greve}, {Huynh}, {Lin},
  {Morrison}, {Scott}, \& {Yan}}]{pope08}
{Pope}, A., {Bussmann}, R.~S., {Dey}, A., {et~al.} 2008, \apj, 689, 127

\bibitem[{{Reddy} {et~al.}(2012){Reddy}, {Dickinson}, {Elbaz}, {Morrison},
  {Giavalisco}, {Ivison}, {Papovich}, {Scott}, {Buat}, {Burgarella},
  {Charmandaris}, {Daddi}, {Magdis}, {Murphy}, {Altieri}, {Aussel},
  {Dannerbauer}, {Dasyra}, {Hwang}, {Kartaltepe}, {Leiton}, {Magnelli}, \&
  {Popesso}}]{red12}
{Reddy}, N., {Dickinson}, M., {Elbaz}, D., {et~al.} 2012, \apj, 744, 154

\bibitem[{{Reddy} {et~al.}(2008){Reddy}, {Steidel}, {Pettini}, {Adelberger},
  {Shapley}, {Erb}, \& {Dickinson}}]{red08}
{Reddy}, N.~A., {Steidel}, C.~C., {Pettini}, M., {et~al.} 2008, \apjs, 175, 48

\bibitem[{{Ree} {et~al.}(2007){Ree}, {Lee}, {Yi}, {Yoon}, {Rich}, {Deharveng},
  {Sohn}, {Kaviraj}, {Rhee}, {Sheen}, {Schawinski}, {Rey}, {Boselli}, {Rhee},
  {Donas}, {Seibert}, {Wyder}, {Barlow}, {Bianchi}, {Forster}, {Friedman},
  {Heckman}, {Madore}, {Martin}, {Milliard}, {Morrissey}, {Neff},
  {Schiminovich}, {Small}, {Szalay}, \& {Welsh}}]{ree07}
{Ree}, C.~H., {Lee}, Y.-W., {Yi}, S.~K., {et~al.} 2007, \apjs, 173, 607

\bibitem[{{Sajina} {et~al.}(2012){Sajina}, {Yan}, {Fadda}, {Dasyra}, \&
  {Huynh}}]{saj12}
{Sajina}, A., {Yan}, L., {Fadda}, D., {Dasyra}, K., \& {Huynh}, M. 2012, \apj,
  757, 13

\bibitem[{{Salpeter}(1955)}]{sal55}
{Salpeter}, E.~E. 1955, \apj, 121, 161

\bibitem[{{Sargent} {et~al.}(2013){Sargent}, {Daddi}, {B{\'e}thermin},
  {et~al.}}]{sar13}
{Sargent}, M.~T., {Daddi}, E., {B{\'e}thermin}, M., {et~al.} 2013, \apj,
  submitted (arXiv:1303.4392)

\bibitem[{{Schawinski} {et~al.}(2012){Schawinski}, {Simmons}, {Urry},
  {Treister}, \& {Glikman}}]{sch12}
{Schawinski}, K., {Simmons}, B.~D., {Urry}, C.~M., {Treister}, E., \&
  {Glikman}, E. 2012, \mnras, 425, L61

\bibitem[{{Skrutskie} {et~al.}(2006){Skrutskie}, {Cutri}, {Stiening},
  {Weinberg}, {Schneider}, {Carpenter}, {Beichman}, {Capps}, {Chester},
  {Elias}, {Huchra}, {Liebert}, {Lonsdale}, {Monet}, {Price}, {Seitzer},
  {Jarrett}, {Kirkpatrick}, {Gizis}, {Howard}, {Evans}, {Fowler}, {Fullmer},
  {Hurt}, {Light}, {Kopan}, {Marsh}, {McCallon}, {Tam}, {Van Dyk}, \&
  {Wheelock}}]{skr06}
{Skrutskie}, M.~F., {Cutri}, R.~M., {Stiening}, R., {et~al.} 2006, \aj, 131,
  1163

\bibitem[{{Smith} {et~al.}(2012){Smith}, {Gomez}, {Eales}, {Ciesla}, {Boselli},
  {Cortese}, {Bendo}, {Baes}, {Bianchi}, {Clemens}, {Clements}, {Cooray},
  {Davies}, {de Looze}, {di Serego Alighieri}, {Fritz}, {Gavazzi}, {Gear},
  {Madden}, {Mentuch}, {Panuzzo}, {Pohlen}, {Spinoglio}, {Verstappen},
  {Vlahakis}, {Wilson}, \& {Xilouris}}]{smith12}
{Smith}, M.~W.~L., {Gomez}, H.~L., {Eales}, S.~A., {et~al.} 2012, \apj, 748,
  123

\bibitem[{{Stern} {et~al.}(2012){Stern}, {Assef}, {Benford}, {Blain}, {Cutri},
  {Dey}, {Eisenhardt}, {Griffith}, {Jarrett}, {Lake}, {Masci}, {Petty},
  {Stanford}, {Tsai}, {Wright}, {Yan}, {Harrison}, \& {Madsen}}]{stern12}
{Stern}, D., {Assef}, R.~J., {Benford}, D.~J., {et~al.} 2012, \apj, 753, 30

\bibitem[{{Swinbank} {et~al.}(2004){Swinbank}, {Smail}, {Chapman}, {Blain},
  {Ivison}, \& {Keel}}]{swi04}
{Swinbank}, A.~M., {Smail}, I., {Chapman}, S.~C., {et~al.} 2004, \apj, 617, 64

\bibitem[{{Symeonidis} {et~al.}(2013){Symeonidis}, {Vaccari}, {Berta}, {Page},
  {Lutz}, {Arumugam}, {Aussel}, {Bock}, {Boselli}, {Buat}, {Capak}, {Clements},
  {Conley}, {Conversi}, {Cooray}, {Dowell}, {Farrah}, {Franceschini},
  {Giovannoli}, {Glenn}, {Griffin}, {Hatziminaoglou}, {Hwang}, {Ibar},
  {Ilbert}, {Ivison}, {Le Floc'h}, {Lilly}, {Kartaltepe}, {Magnelli}, {Magdis},
  {Marchetti}, {Nguyen}, {Nordon}, {O'Halloran}, {Oliver}, {Omont},
  {Papageorgiou}, {Patel}, {Pearson}, {Perez-Fournon}, {Pohlen}, {Popesso},
  {Pozzi}, {Rigopoulou}, {Riguccini}, {Rosario}, {Roseboom}, {Rowan-Robinson},
  {Salvato}, {Schulz}, {Scott}, {Seymour}, {Shupe}, {Smith}, {Valtchanov},
  {Wang}, {Xu}, {Zemcov}, \& {Wuyts}}]{sym13}
{Symeonidis}, M., {Vaccari}, M., {Berta}, S., {et~al.} 2013, MNRAS, in press
  (arXiv:1302.4895)

\bibitem[{{Tacconi} {et~al.}(2010){Tacconi}, {Genzel}, {Neri}, {Cox}, {Cooper},
  {Shapiro}, {Bolatto}, {Bouch{\'e}}, {Bournaud}, {Burkert}, {Combes},
  {Comerford}, {Davis}, {Schreiber}, {Garcia-Burillo}, {Gracia-Carpio}, {Lutz},
  {Naab}, {Omont}, {Shapley}, {Sternberg}, \& {Weiner}}]{tac10}
{Tacconi}, L.~J., {Genzel}, R., {Neri}, R., {et~al.} 2010, \nat, 463, 781

\bibitem[{{Takata} {et~al.}(2006){Takata}, {Sekiguchi}, {Smail}, {Chapman},
  {Geach}, {Swinbank}, {Blain}, \& {Ivison}}]{tak06}
{Takata}, T., {Sekiguchi}, K., {Smail}, I., {et~al.} 2006, \apj, 651, 713

\bibitem[{{Tyler} {et~al.}(2009){Tyler}, {Le Floc'h}, {Rieke}, {Dey}, {Desai},
  {Brand}, {Borys}, {Jannuzi}, {Armus}, {Dole}, {Papovich}, {Brown},
  {Blaylock}, {Higdon}, {Higdon}, {Charmandaris}, {Ashby}, \&
  {Smith}}]{tyler09}
{Tyler}, K.~D., {Le Floc'h}, E., {Rieke}, G.~H., {et~al.} 2009, \apj, 691, 1846

\bibitem[{{Veilleux} \& {Osterbrock}(1987)}]{vo87}
{Veilleux}, S., \& {Osterbrock}, D.~E. 1987, \apjs, 63, 295

\bibitem[{{Veilleux} {et~al.}(2009){Veilleux}, {Rupke}, {Kim}, {Genzel},
  {Sturm}, {Lutz}, {Contursi}, {Schweitzer}, {Tacconi}, {Netzer}, {Sternberg},
  {Mihos}, {Baker}, {Mazzarella}, {Lord}, {Sanders}, {Stockton}, {Joseph}, \&
  {Barnes}}]{vei09}
{Veilleux}, S., {Rupke}, D.~S.~N., {Kim}, D., {et~al.} 2009, \apjs, 182, 628

\bibitem[{{Woods} \& {Geller}(2007)}]{woods07}
{Woods}, D.~F., \& {Geller}, M.~J. 2007, \aj, 134, 527

\bibitem[{{Wright} {et~al.}(2010){Wright}, {Eisenhardt}, {Mainzer}, {Ressler},
  {Cutri}, {Jarrett}, {Kirkpatrick}, {Padgett}, {McMillan}, {Skrutskie},
  {Stanford}, {Cohen}, {Walker}, {Mather}, {Leisawitz}, {Gautier}, {McLean},
  {Benford}, {Lonsdale}, {Blain}, {Mendez}, {Irace}, {Duval}, {Liu}, {Royer},
  {Heinrichsen}, {Howard}, {Shannon}, {Kendall}, {Walsh}, {Larsen}, {Cardon},
  {Schick}, {Schwalm}, {Abid}, {Fabinsky}, {Naes}, \& {Tsai}}]{wri10}
{Wright}, E.~L., {Eisenhardt}, P.~R.~M., {Mainzer}, A.~K., {et~al.} 2010, \aj,
  140, 1868

\bibitem[{{Wu} {et~al.}(2012){Wu}, {Tsai}, {Sayers}, {Benford}, {Bridge},
  {Blain}, {Eisenhardt}, {Stern}, {Petty}, {Assef}, {Bussmann}, {Comerford},
  {Cutri}, {Evans}, {Griffith}, {Jarrett}, {Lake}, {Lonsdale}, {Rho},
  {Stanford}, {Weiner}, {Wright}, \& {Yan}}]{wu12}
{Wu}, J., {Tsai}, C.-W., {Sayers}, J., {et~al.} 2012, \apj, 756, 96

\bibitem[{{Wyder} {et~al.}(2007){Wyder}, {Martin}, {Schiminovich}, {Seibert},
  {Budav{\'a}ri}, {Treyer}, {Barlow}, {Forster}, {Friedman}, {Morrissey},
  {Neff}, {Small}, {Bianchi}, {Donas}, {Heckman}, {Lee}, {Madore}, {Milliard},
  {Rich}, {Szalay}, {Welsh}, \& {Yi}}]{wyder07}
{Wyder}, T.~K., {Martin}, D.~C., {Schiminovich}, D., {et~al.} 2007, \apjs, 173,
  293

\bibitem[{{Yan} {et~al.}(2004){Yan}, {Helou}, {Fadda}, {Marleau}, {Lacy},
  {Wilson}, {Soifer}, {Drozdovsky}, {Masci}, {Armus}, {Teplitz}, {Frayer},
  {Surace}, {Storrie-Lombardi}, {Appleton}, {Chapman}, {Choi}, {Fan},
  {Heinrichsen}, {Im}, {Schmitz}, {Shupe}, \& {Squires}}]{yan04}
{Yan}, L., {Helou}, G., {Fadda}, D., {et~al.} 2004, \apjs, 154, 60

\bibitem[{{Yang} {et~al.}(2007){Yang}, {Greve}, {Dowell}, \& {Borys}}]{yang07}
{Yang}, M., {Greve}, T.~R., {Dowell}, C.~D., \& {Borys}, C. 2007, \apj, 660,
  1198

\bibitem[{{York} {et~al.}(2000){York}, {Adelman}, {Anderson}, {Anderson},
  {Annis}, {Bahcall}, {Bakken}, {Barkhouser}, {Bastian}, {Berman}, {Boroski},
  {Bracker}, {Briegel}, {Briggs}, {Brinkmann}, \& {SDSS
  Collaboration}}]{york00}
{York}, D.~G., {Adelman}, J., {Anderson}, Jr., J.~E., {et~al.} 2000, \aj, 120,
  1579

\bibitem[{{Younger} {et~al.}(2009){Younger}, {Omont}, {Fiolet}, {Huang},
  {Fazio}, {Lai}, {Polletta}, {Rigopoulou}, \& {Zylka}}]{you09td}
{Younger}, J.~D., {Omont}, A., {Fiolet}, N., {et~al.} 2009, \mnras, 394, 1685

\end{thebibliography}

\end{document}